\documentclass[aps, floatfix,nofootinbib,amsmath,amssymb,amsfonts,notitlepage, twocolumn,superscriptaddress]{revtex4-1} 
\usepackage{textgreek}
\usepackage{soul}

\usepackage[T1]{fontenc}
\usepackage[latin9]{inputenc}
\usepackage{lmodern} 
\usepackage{color}
\usepackage{bbold}
\usepackage{dsfont}
\usepackage{graphicx}
\usepackage[caption=false]{subfig}
\usepackage[colorlinks]{hyperref}
\usepackage[bottom]{footmisc}
\usepackage{enumerate}
\usepackage{float}
\usepackage{mathtools}

\usepackage{empheq}
\usepackage{tikz}
\usetikzlibrary{patterns,tikzmark, matrix,decorations.pathreplacing, calc, positioning,fit}

\usepackage{hf-tikz}

\usepackage{fancyhdr}

\usepackage[normalem]{ulem}
\DeclareMathAlphabet\mbc{OMS}{cmsy}{b}{n}
\usepackage{balance}

\pgfmathsetmacro{\flagscalefactor}{0.3}
\pgfmathsetmacro{\flagrotationdegree}{45}
\newcommand{\flagpolecolor}{brown!60!black}
\newcommand{\flagcolor}{yellow!67!red}

\newcommand{\flagadditionalcommands}{}
\usepackage{scalerel}

\newcommand{\overbar}[1]{\mkern 1.5mu\overline{\mkern-1.5mu#1\mkern-1.5mu}\mkern 1.5mu}
\newcommand\dbar[1]{\ThisStyle{%
  \setbox0=\hbox{$\SavedStyle\overbar{#1}$}%
  \ht0=\dimexpr\ht0-.15ex\relax
  \overbar{\copy0}%
}}
\newcommand{\tikzflag}[1][0,0]{
    \begin{tikzpicture}[scale=\flagscalefactor,rotate=\flagrotationdegree,shift={(#1)},overlay,remember picture]
        \draw[fill=\flagpolecolor,thick] (0,0) -- ++ (0,8) arc (180:0:0.4 and 0.1) -- ++ (0,-8) arc (360:180:0.4 and 0.1);
        \draw[thick] (0,8) arc (180:360:0.4 and 0.1);
        \draw[fill=\flagcolor,thick] (0.8,7.5) to[out=-30,in=210] ++(3,0) to[out=30,in=150] ++ (3,0) -- ++ (0,-4.5) to [out=150,in=30] ++(-3,0) to[out=210,in=-30] ++(-3,0) -- cycle;
        \flagadditionalcommands
    \end{tikzpicture}
}
\newcommand{\Rflag}[1]{
    \pgfmathsetmacro{\flagscalefactor}{0.1}
    \pgfmathsetmacro{\flagrotationdegree}{0}
    \renewcommand{\flagpolecolor}{black}
    \renewcommand{\flagcolor}{red}
    
    \marker{#1}{-1,-0.6}
    \tikzflag[m-#1]
}

\begin{document}
\pgfkeys{tikz/mymatrixenv/.style={decoration={brace},every left delimiter/.style={xshift=8pt},every right delimiter/.style={xshift=-8pt}}}
\pgfkeys{tikz/mymatrix/.style={matrix of math nodes,nodes in empty cells,left delimiter={(},right delimiter={)},inner sep=0.5pt,outer sep=4pt,column sep=6pt,row sep=6pt,nodes={minimum width=20pt,minimum height=10pt,anchor=center,inner sep=0pt,outer sep=0pt}}}
\pgfkeys{tikz/mymatrixbrace/.style={decorate,thick}}
 \newcommand\mymatrixbraceoffseth{0.5em}
 \newcommand\mymatrixbraceoffsetv{0.2em}

\newcommand*\mymatrixbraceright[4][m]{
\draw[mymatrixbrace] ($(#1.north west)!(#1-#3-1.south west)!(#1.south west)-    (\mymatrixbraceoffseth,0)$)
    -- node[left=2pt] {#4} 
    ($(#1.north west)!(#1-#2-1.north west)!(#1.south west)-(\mymatrixbraceoffseth,0)$);
}
 \newcommand*\mymatrixbraceleft[4][m]{
 \draw[mymatrixbrace] ($(#1.north east)!(#1-#2-1.north east)!(#1.south east)+    (\mymatrixbraceoffseth,0)$)
    -- node[right=2pt] {#4} 
     ($(#1.north east)!(#1-#3-1.south east)!(#1.south east)+    (\mymatrixbraceoffseth,0)$);
}
\newcommand*\mymatrixbracetop[4][m]{
\draw[mymatrixbrace] ($(#1.north west)!(#1-1-#2.north west)!(#1.north east)+(0,\mymatrixbraceoffsetv)$)
    -- node[above=2pt] {#4} 
    ($(#1.north west)!(#1-1-#3.north east)!(#1.north east)+(0,\mymatrixbraceoffsetv)$);
}
\newcommand*\mymatrixbracebottom[4][m]{
\draw[mymatrixbrace] ($(#1.south west)!(#1-1-#3.south east)!(#1.south east)-(0,\mymatrixbraceoffsetv)$)
    -- node[below=2pt] {#4} 
    ($(#1.south west)!(#1-1-#2.south west)!(#1.south east)-(0,\mymatrixbraceoffsetv)$);
}

\tikzset{style blue/.style={
    set fill color=blue!70!green!90,draw opacity=1,
    set border color=blue!70!green!90,fill opacity=0.08,
  },
  style teal/.style={
    set fill color=teal!90!, draw opacity=1,
    set border color=teal!90!,fill opacity=0.3,
  },
  style orange/.style={
    set fill color=orange!120, draw opacity=1,
    set border color=orange!120, fill opacity=0.3,
  },
  style yellow/.style={
    set fill color=yellow!80!green!100, draw opacity=1,
    set border color=yellow!80!green!100, fill opacity=0.3,
  },
  kwad/.style={
    above left offset={-0.15,0.2},
    below right offset={0.15,-0.2},
    #1
  },
  ,set fill color/.code={\pgfkeysalso{fill=#1}},
  set border color/.style={draw=#1}
}

\global\long\def\eqn#1{\begin{align}#1\end{align}}
\global\long\def\vec#1{\overrightarrow{#1}}
\global\long\def\ket#1{\left|#1\right\rangle }
\global\long\def\bra#1{\left\langle #1\right|}
\global\long\def\bkt#1{\left(#1\right)}
\global\long\def\sbkt#1{\left[#1\right]}
\global\long\def\cbkt#1{\left\{#1\right\}}
\global\long\def\abs#1{\left\vert#1\right\vert}
\global\long\def\cev#1{\overleftarrow{#1}}
\global\long\def\der#1#2{\frac{{d}#1}{{d}#2}}
\global\long\def\pard#1#2{\frac{{\partial}#1}{{\partial}#2}}
\global\long\def\re{\mathrm{Re}}
\global\long\def\im{\mathrm{Im}}
\global\long\def\dd{\mathrm{d}}
\global\long\def\ddd{\mathcal{D}}

\global\long\def\avg#1{\left\langle #1 \right\rangle}
\global\long\def\mr#1{\mathrm{#1}}
\global\long\def\mb#1{{\mathbf #1}}
\global\long\def\mc#1{\mathcal{#1}}
\global\long\def\tr{\mathrm{Tr}}

\global\long\def\nth{$n^{\mathrm{th}}$~}
\global\long\def\mth{$m^{\mathrm{th}}$~}
\global\long\def\non{\nonumber}

\newcommand{\orange}[1]{{\color{orange} {#1}}}
\newcommand{\cyan}[1]{{\color{cyan} {#1}}}
\newcommand{\teal}[1]{{\color{teal} {#1}}}
\newcommand{\blue}[1]{{\color{blue} {#1}}}
\newcommand{\yellow}[1]{{\color{yellow} {#1}}}
\newcommand{\green}[1]{{\color{green} {#1}}}
\newcommand{\red}[1]{{\color{red} {#1}}}

\global\long\def\todo#1{\cyan{{$\bigstar$ \orange{\bf\sc #1 }}$\bigstar$} }

\newcommand{\ks}[1]{{\textcolor{teal}{[KS: #1]}}}
\newcommand{\ec}[1]{{\textcolor{purple}{[EC: #1]}}}

\global\long\def\redflag#1{\Rflag{first} \red{\bf \sc #1}}

\title{Radiative Properties of an Artificial Atom coupled to a Josephson Junction Array}

\author{Kanu Sinha}
\affiliation{School of Electrical, Computer and Energy Engineering, Arizona State University, Tempe, AZ}
\affiliation{Department of Electrical and Computer Engineering, Princeton University, Princeton, NJ}
\author{Saeed A. Khan}
\affiliation{Department of Electrical and Computer Engineering, Princeton University, Princeton, NJ}
\author{Elif C\"uce}
\affiliation{Department of Electrical and Computer Engineering, Princeton University, Princeton, NJ}
\author{Hakan E. T\"{u}reci}

\affiliation{Department of Electrical and Computer Engineering, Princeton University, Princeton, NJ}
\begin{abstract}
 We study  the radiative properties---the Lamb shift, Purcell decay rate and the  spontaneous emission dynamics---of an artificial atom coupled to a long, multimode cavity formed by an array of Josephson junctions. Introducing a tunable coupling element between the atom and the array, we demonstrate that such a system can  exhibit a crossover from a perturbative to non-perturbative regime of light-matter interaction  as one strengthens the coupling between the atom and the Josephson junction array (JJA). 
 As a consequence, the concept of spontaneous emission as the occupation of the local atomic site  being governed by a single complex-valued exponent breaks down. 
  This breakdown, we show,  can be interpreted in terms of  formation of  hybrid atom-resonator modes with  radiative losses that are non-trivially related to the effective  coupling between individual modes. We develop a singular function expansion approach for the description of the open quantum system dynamics in such a  multimode non-perturbative regime. This modal framework generalizes the normal mode description of quantum fields in a finite volume, incorporating exact radiative losses and incident quantum noise at the delimiting surface. Our results are pertinent to recent experiments with Josephson atoms coupled to high impedance Josephson junction arrays.
\end{abstract}
\maketitle

\section{Introduction}

Progress in the fabrication and control of superconducting devices has provided a renewed impetus to reexamine some of the foundational problems of Quantum Electrodynamics (QED) \cite{Blais21} in the context of material systems. Research in the last two decades has spawned profound questions about radiative corrections  and the issue of the correct gauge-invariant description of the dynamics of artificial atoms in solid-state electromagnetic media. These questions are fueled in part by engineered superconducting electrodynamical systems that feature some of the strongest interactions between light and matter ever achieved in physics \cite{Kockum19, FornDiaz19},  opening up exciting possibilities for  applications in quantum information processing \cite{ Nataf11, Wang2017,  Stassi18, Kyaw15} while radically modifying various quantum optical phenomena, e.g., Purcell effect \cite{DeLiberato14}, Dicke physics \cite{Jaako16}, and ground state properties of atoms and the vacuum field \cite{Nataf11, Ashhab10, MartinezMartinez18, Hsiang19}.

In formulating a dynamical description of quantum electrodynamical systems, one must address the question of an appropriate basis of normal modes to express the problem efficiently. In Quantum Optics and more specifically in Cavity QED, one generally operates under the presumption that cavity normal modes have an existence that is independent of the atomic system they are coupled to \cite{Walther2006}. In atomic Cavity QED systems the characteristic weakness of light-matter interaction means this is generally a good starting point, as the hybridization of the atomic and cavity modes is weak except in a small spectral band. In that band the full quantum description of the atom coupled to one or few normal modes of the bare cavity is sufficient to capture the atomic dynamics accurately \cite{Kimble94}. Superconducting Cavity QED systems have brought forth conditions, however, where the atom-field hybridization can be substantial and give rise to significant renormalization of the atomic dynamics. It was therefore understood within the circuit QED framework that such hybridization has to be accurately captured, and appropriate theoretical and computational methods developed to do so \cite{Bourassa12, DiazCamacho16, SanchezBurillo14}.

Recent experiments have shown that an artificial atom embedded in a high-impedance Josephson metamaterial provides a setting where the strength of the coupling between the artificial atom and its environment can no longer be described via perturbation theory \cite{Leger19, Kuzmin19a}. Such Josephson junction arrays (JJAs) exhibit an array of interesting physical phenomena ranging from quantum phase transitions between a superconducting and insulating phase persisting at zero temperature \cite{Fisher90}, to synchronization \cite{Wiesenfeld96}, and  implementing low-loss  large impedances that can be comparable to the resistance quantum  \cite{Masluk12, Leger19, Kuzmin19a, Kuzmin19b, Pechenezhskiy19}. It has been experimentally observed that such high impedance environments can sustain large zero-point flux fluctuations, thereby resulting in  enhanced vacuum-induced Lamb shifts and spontaneous emission for artificial atoms coupled to JJAs \cite{Leger19, Kuzmin19a}.

\label{Sec:model}
\begin{figure*}[t]
    \centering
    \includegraphics[width = 6 in]{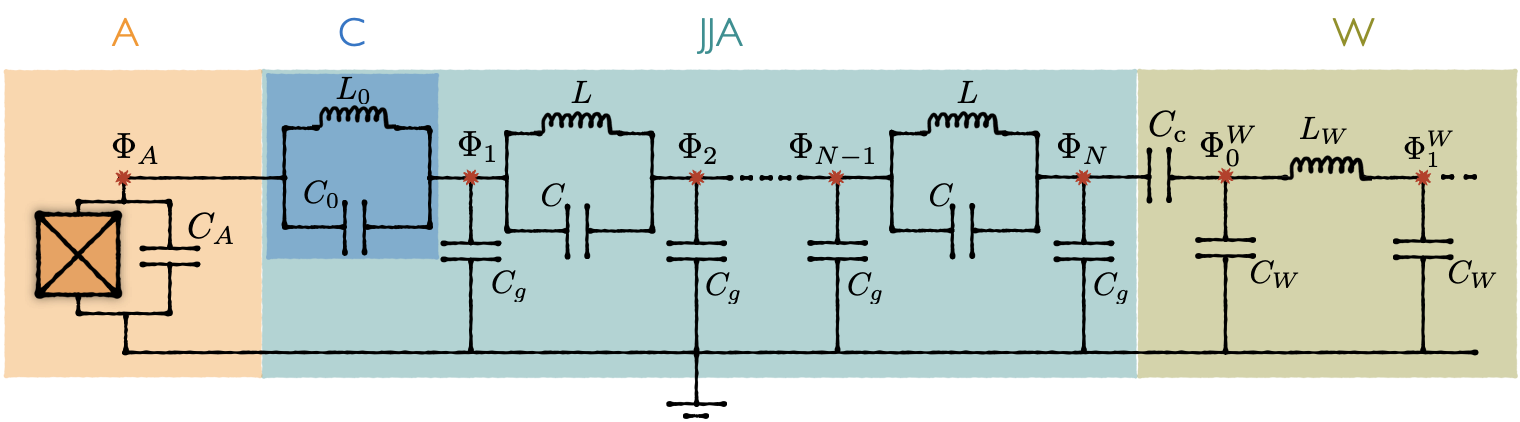}
    \caption{Schematic circuit for  an artificial atom coupled to a JJA.  The JJA  is  modeled as $N$ parallel LC-circuits with inductance $L$ and capacitance $C$  coupled to each other in series. Each node is connected to the ground via a capacitance $C_g$.  The waveguide is coupled to the JJA via a coupling capacitance $C_c$, and is described as a lumped element transmission line with inductance $L_W$ and ground capacitance $C_W $. The first element of the JJA has an inductance and capacitance value $\cbkt{L_0, C_0 }$ that can be different from the rest of the array. The parameter values assumed throughout all the calculations are (unless specified otherwise): $N = 1000$, $L = 1$~nH, $C = 150$~fF, $ C_g = 0.1$~fF, $C_c = 100$~fF. The atomic charging energy is assumed to be $E_C^A/\hbar =2\pi\times 15~$ GHz.  The capacitance and inductance values of the coupler are assumed to be $L_0 = L/\chi $ and $C_0 = \chi C$, where the coupling parameter $\chi $ can be varied. The transmission line is assumed to have an impedance of $Z_{W} = 50~\Omega$.
    }
    \label{Fig:Ckt}
\end{figure*}

In this paper we examine the  radiative corrections and open quantum system dynamics of an artificial atom coupled to such a high-impedance JJA,  with a view to address a subset of the aforementioned issues. We consider a  variable LC  coupler between the atom and the JJA that allows one to tune between different regimes of coupling strength and hybridization.  We demonstrate that  in a particular regime that we refer to here as "non-perturbative" (in atom-cavity coupling), one can no longer identify an eigenmode of the total system that is localized either (1) spatially  close to the atomic position or (2) spectrally close to the bare atomic frequency.  We analyze  the radiative properties of the atom using non-Hermitian eigenmodes of the finite system that bridge the perturbative and non-perturbative regimes and provides an interpretable unified description. We compare our results with those from second order perturbation theory, demonstrating a marked deviation of the obtained radiative corrections as the atom-JJA coupling strength is increased. The technical machinery to enable this dynamical description is based on the extension of the singular function expansion method to an open quantum system  description. Furthermore, such an approach allows one to derive the dynamics of the system in a multimode non-perturbative regime.

The rest of this paper is organized as follows. Section~\ref{Sec:Lagrangian} describes the model of the system in consideration, detailing the circuit Lagrangian and parameters. Section~\ref{Sec:EOM} discusses the equations of motion of the linear system, and a description of the open system dynamics in a reduced subspace. In Section~\ref{Sec:sfm} we describe the singular function expansion method that we use to determine the radiative properties and open system dynamics of the system. Section~\ref{Sec:Ham} discusses the effective Hamiltonian for the closed artificial atom+JJA system, defining the characteristic coupling strengths between the atomic and the JJA modes.  Section~\ref{Sec:Rad} details the radiative properties of the artificial atom, comparing the  atomic Lamb shifts and Purcell decay obtained via the modal analysis with those obtained via second order perturbation theory. The  dynamics of the open system is illustrated in Section~\ref{Sec:spem}, particularly considering the case of an initially excited atom. We  present our conclusions and outlook in Section~\ref{Sec:Diss}.

\section{Model}

\subsection{Lagrangian}
\label{Sec:Lagrangian}
Let us consider the system of an artificial atom coupled to the continuum of an infinite transmission line through an intervening JJA as shown in Fig.~\ref{Fig:Ckt}. The atom is modeled as a Josephson junction with a Josephson energy $E_A$, shunted by a capacitance with a charging energy $E_C^A \equiv e^2/(2C_A)$. In practice, the difference between the artificial atom junction and the individual junctions of the JJA resides in their junction areas, chosen such that the atomic junction has a much stronger anharmonicity. 

The JJA is constructed using $N$ nominally-identical component circuits coupled in series; each circuit has  a capacitance $C$, with an anharmonicity engineered to be substantially weaker than the artificial atom, allowing it to be treated as a linear inductor to lowest order.  A distinct coupling element, realized as a parallel $LC$ circuit with inductance and capacitance $\cbkt{L_0, C_0}$, connects the first element of the chain to the artificial atom, allowing for the possibility of having a weak or a strong coupling between the atom and the JJA. The waveguide is modeled as a lumped-element transmission line with inductance $L_W$ and a ground capacitance $C_W$.

The total system is described by the Lagrangian \eqn{\label{Ltot}\mc{L} = \mc{L}_A + \mc{L}_\mr{JJA} + \mc{L}_{A-\mr{JJA}} +  \mc{L}_{\mr{JJA}-W} + \mc{L}_W,}
where

\eqn{\label{LA}
\mc{L}_{A} = &\frac{1}{2} C_A \dot\Phi_A^2 -\frac{1}{2L_A}\Phi_A^2 - \mathcal{U}_A(\Phi_A) }
 stands for the bare atomic Lagrangian with $\Phi_A $ as the flux across the atom. Here $\mc{U}_A(\Phi_A)$ is the explicitly nonlinear part of the Josephson potential.

The bare JJA Lagrangian is given by
\begin{widetext}

\eqn{
\mc{L}_\mr{JJA} = & \frac{1}{2}\bkt{C+C_g} \dot{\Phi}_1 ^2 - \frac{1}{2L}\Phi_1^2 +  \sum _{n = 2}^{N-1}\sbkt{\frac{1}{2} \bkt{2C+ C_g}  \dot\Phi_n  ^2- \frac{1}{L }  \Phi_n  ^2} + \frac{1}{2}\bkt{C+C_g} \dot{\Phi}_N ^2 - \frac{1}{2L}\Phi_N^2 \non \\
&- \sum _{n = 1}^{N-1}\sbkt{ C \dot{\Phi}_n\dot{\Phi}_{n+1} - \frac{1}{L} \Phi_n \Phi_{n+1}} ,
}
\end{widetext}
where $\Phi_n $ corresponds to the (nodal) flux at the $n^\mr{th}$ node of the JJA measured with respect to the ground, as shown in Fig.~\ref{Fig:Ckt}. 

The  impedance of the array can be obtained by successively adding together the impedances of each unit of the JJA as shown in Appendix~\ref{App:imp}.  The impedance of an {\it infinite} JJA with the same parameters as assumed is 
\eqn{\label{zinf}
Z_\infty \approx \sqrt{Z_\mr{LC}Z_g},}  where  $ Z_\mr{LC} \equiv \frac{i \omega L }{1 - \omega^2/\Omega_0 ^2}$ corresponds to  the impedance of the individual $LC$-oscillator  units of the JJA and $ Z_g \equiv 1/\bkt{i\omega C_g}$ corresponds to the impedance of the capacitance to the ground, with  $\Omega_0 \equiv 1/\sqrt{LC} $ as the plasma frequency \cite{Pozar}. We note that for frequencies much lower than the cut-off frequency for the individual $LC$-oscillators  $(\omega\ll \Omega_0 )$ the array impedance can be approximated as $Z_\infty\approx \sqrt{L/C_g}\approx 3.16$~k$\Omega$, and for higher frequencies  $(\omega\gg \Omega_0 )$  $Z_\infty\approx i\sqrt{1/(\omega^2CC_g)}\approx i\bkt{\frac{\Omega_0}{\omega}} 3.16$~k$\Omega$, for the chosen set of parameter values as detailed in the caption of Fig.~\ref{Fig:Ckt}. Experimental systems with such large impedances  comparable to the resistance quantum $( R_Q \approx 6.15$~k$\Omega)$  have been instrumental in exploring quantum many-body effects  ($Z_\mr{\infty}\approx1.8 $~k$\Omega$) \cite{Leger19}, superconducting-insulator phase transitions ($Z_\mr{\infty}\approx0.7$--19~k$\Omega$) \cite{Kuzmin19b}, and `superstrong' coupling regimes wherein the atom-field coupling strength can be comparable to the mode-spacing of the environment ($Z_\mr{\infty}\approx5$--10~k$\Omega$)  \cite{Kuzmin19a, Meiser06}.

The interaction between the atom and the JJA has both capacitive and inductive contributions, and is given by the Lagrangian:
\eqn{
\mc{L}_{A-\mr{JJA}} = \frac{1}{2}C_0\left( \dot\Phi_A- \dot\Phi_1\right)^2 - \frac{1}{2L_0 }\left( \Phi_A -\Phi_1\right)^2.
}
In addition to explicit coupling contributions that arise as cross terms, the form of the coupling leads to a renormalization of the atom and JJA parameters.  To parametrize the coupling strength, we introduce the dimensionless  coupling parameter $\chi$ such that 
\eqn{\label{eq:chi}
L_0 = L/\chi\text{ and }
C_0 = \chi C,
}
such that the plasma frequency for the coupling circuit is equivalent to that of the rest of the chain. Particularly, we note that a value of $\chi = 1$ corresponds to the case of an artificial atom galvanically coupled to a high impedance JJA, similar to the experimental setups in \cite{Leger19, Kuzmin19a}. As we will demonstrate, the strength of the coupling and hybridization between atomic and JJA modes is determined by $\chi$, changing which allows us to observe a crossover from a perturbative to non-perturbative regime.  

The Lagrangians $\mc{L}_{W}$ and $\mc{L}_{\mr{JJA}-W}$ correspond to  the waveguide and the JJA-waveguide coupling respectively and are defined as 
\eqn{\mc{L}_W= & \sum_{n =0 }^{\infty}  \frac{1}{2} C_{W}  \bkt{\dot{\Phi}_n^{W}}^2 - \frac{1}{2 L_{W}} \bkt{ \Phi_{n }^{W} - \Phi_{n+1}^{W}}^2 \\
\mc{L}_{\mr{JJA}-W} =& \frac{1}{2} C_c\bkt{\dot \Phi_N -\dot \Phi_0 ^W}^2.
}
The coupling to an external waveguide renders the atom+JJA system open, leading to radiative losses.

\vspace{0.5 cm}

\subsection{Equations of motion}
\label{Sec:EOM}

While the inclusion of the nonlinear Josephson potential $\mathcal{U}_A(\Phi_A)$ ultimately enables the realization of an artificial atom, a number of important physical parameters of the joint system are already set at the linear level. These include strong hybridization effects renormalizing the frequency and dissipation rates of atomic and JJA modes. Understanding these linear effects is  crucial to the definition of an appropriate set of normal modes that can form the basis for describing the nonlinear quantum dynamics. We thus consider the linear chain by dropping at the first stage the nonlinear potential $\mathcal{U}_A(\Phi_A)$ from the bare artificial atom Lagrangian, Eq.~(\ref{LA}), hence considering the Josephson potential to linear order. We will later re-introduce the non-linearity through perturbation theory in Sec.~\ref{Sec:spem}.

The total Lagrangian for the linear system can be expressed in a matrix representation as follows:
\eqn{\label{eq:Ltot}\mc{L}_\mr{tot} = \frac{1}{2}\dot{\mb{\Phi}}^T_\mr{tot} \dbar{C}_\mr{tot}\dot{\mb{\Phi}}_\mr{tot} - \frac{1}{2}{\mb{\Phi}}^T_\mr{tot} \dbar{L}^{-1}_\mr{tot}{\mb{\Phi}}_\mr{tot},}
where $\mb{\Phi}_\mr{tot} = \cbkt{\Phi_A, \Phi_1, \dots, \Phi_N, \dots, \Phi_0^W, \dots}$ represents the flux variables for the various nodes, and the capacitance and inductance matrices $ \dbar {C}_\mr{tot}$ and $ \dbar {L}_\mr{tot}$ are given by:
\begin{widetext}
\begin{equation}
\label{Ctot}
        \dbar{C}_\mr{tot}=     \begin{tikzpicture}[baseline={-0.5ex},mymatrixenv]\matrix [mymatrix,inner sep=4pt] (m){  \tikzmarkin[kwad=style orange]{atC} \tikzmarkin[kwad=style blue]{cplg}  C_A + C_0 \tikzmarkend{atC}  &  -C_0 &  0   & \dots    & 0  & 0& 0&0&\dots \\
    -C_0 & \tikzmarkin[kwad=style teal]{jjaC}C_g + C + C_0\tikzmarkend{cplg} & -C & \dots & 0 & 0 & 0&0 &\dots \\
     0 & -C & C_g + 2C & \dots   &0 &0 &0 &0&\dots\\
    \vdots &\vdots &\vdots &  \ddots & \vdots & \vdots&\vdots&\vdots&\ddots \\
    0&0 &0 &  \dots & C_g + 2C & -C  &0 &0&\dots\\
    0 & 0 & 0 & \dots   & -C   &  ~\tikzmarkin[kwad=style yellow]{cc}C_g + C  + C_c \tikzmarkend{jjaC} &-C_c&0&\dots\\
    0 & 0 & 0 & \dots   & 0   &  -C_c  &C_c + C_W&0&\dots\\
    0 & 0 & 0 & \dots   & 0   &  0  &0&C_W&\dots\\
    \vdots & \vdots & \vdots & \ddots   & \vdots   &  \vdots &\vdots&\vdots&\ddots\tikzmarkend{cc}~\\
    };\mymatrixbraceright{1}{1}{$C_A'$}
  \mymatrixbraceleft{2}{6}{$\dbar{C}_\mr{JJA}$}
  \end{tikzpicture},
  \end{equation}
  
\begin{equation}
\label{Lintot}
\dbar L_\mr{tot}^{-1} =   \begin{tikzpicture}[baseline={-0.5ex},mymatrixenv]\matrix [mymatrix,inner sep=4pt] (m)          {  \tikzmarkin[kwad=style orange]{atL} \tikzmarkin[kwad=style blue]{cplL}  1/L_A + 1/L_0 \tikzmarkend{atL}  &  -1/L_0 &  0   & \dots    & 0  & 0&0&0&\dots  \\
    -1/L_0 & \tikzmarkin[kwad=style teal]{jjaL}1/L + 1/L_0\tikzmarkend{cplL} & -1/L & \dots & 0 & 0  &0&0&\dots \\
     0 & -1/L & 2/L & \dots   &0 &0  &0&0&\dots\\
    \vdots &\vdots &\vdots &  \ddots & \vdots & \vdots&\vdots&\vdots&\ddots\\
    0&0 &0 &  \dots & 2/L & -1/L   &0&0&\dots\\
    0 & 0 & 0 & \dots   & -1/L  &~\tikzmarkin[kwad=style yellow]{lw}  1/L  \tikzmarkend{jjaL}&0&0&\dots \\    
    0 & 0 & 0 & \dots   & 0  &  0 &1/L_W&-1/L_W&\dots \\    
    0 & 0 & 0 & \dots   & 0  &  0 &-1/L_W&2/L_W&\dots \\
    \vdots & \vdots & \vdots & \ddots   & \vdots  &  \vdots &\vdots&\vdots&\ddots\tikzmarkend{lw}~ \\    
    };\mymatrixbraceright{1}{1}{$1/L_A'$}
  \mymatrixbraceleft{2}{6}{$\dbar{L}^{-1}_\mr{JJA}$}
    \end{tikzpicture}.
\end{equation}
\end{widetext}
As indicated in the above matrices, the atom (orange) and JJA (green) subspaces are coupled capacitively and inductively via $C_0$ and $1/L_0$ respectively (the blue sector). The waveguide (shown in yellow) is coupled only capacitively to the JJA

Additionally, as discussed earlier, these terms lead to on-site contributions that renormalize the physically-coupled elements of the atom and JJA sectors respectively. For example, the `bare' atomic frequency is renormalized to $\omega_A' \equiv  \sqrt{\frac{1}{L_A'C_A'} }$, where $L_A ' \equiv \bkt{\frac{1}{L_A} + \frac{1}{L_0}}^{-1}$ and $ C_A' \equiv \bkt{C_A + C_0}$.

We define the total Hamiltonian for the system  via the standard Legendre transformation as
$ \mc{H}_\mr{tot} = \bkt{\sum_k Q_k \dot \Phi_k} - \mc L_\mr{tot}$, where the conjugate momenta to the node flux variables are given by $Q_n \equiv \frac{\delta \mc{L}_\mr{tot}}{\delta \dot\Phi_n}$, for the total Lagrangian $\mc{L}_\mr{tot}$ (Eq.~\eqref{eq:Ltot}). We next promote the flux and charge variables to quantum operators, satisfying the canonical commutation relation $ \sbkt{\hat\Phi_j , \hat Q_{j'}}= i\hbar\delta_{j,j'}$. The quantized Hamiltonian can then be written as:
\eqn{\mc{H}_\mr{tot} = \frac{1}{2}\hat{\mb{Q}}_{\mr{tot}}^T \dbar{C}_\mr{tot}^{-1} \hat{\mb{Q}}_{\mr{tot}} + \frac{1}{2}\hat{\mb{\Phi}}_{\mr{tot}}^T \dbar{L}_\mr{tot}^{-1} \hat{\mb{\Phi}}_{\mr{tot}}.
}
 The Heisenberg equations of motion for the flux and charge dynamical variables are:
\eqn{
\label{eq:dphitot}
\der{}{t}\hat{\mb{\Phi}}_{\mr{tot}}  =& \dbar{C}_\mr{tot}^{-1} \hat{\mb{Q}}_{\mr{tot}}\\
\label{eq:dqtot}
\der{}{t}\hat{\mb{Q}}_{\mr{tot}} =&- \dbar{L}_\mr{tot}^{-1} \hat{\mb{\Phi}}_{\mr{tot}}.
}
Formally, the Heisenberg equations of motion include a discrete but infinite set of equations for the waveguide nodes coupled to the atom+JJA system. While the output waveguide is itself a multimode transmission line, its {primary} role is to serve as a uniform
environment allowing observers to direct inputs to, and extract outputs from, the system of interest. An analysis of the waveguide field in terms of incoming and outgoing modes with respect to the resonator (defined as $-$ and $+$ respectively) allows the separation of the incoming noise component of the waveguide modes from the outgoing component carrying information about the atom and the resonator. We show in Appendix~\ref{App:wgin} that it is then possible to place a transparent boundary after the first waveguide node. The remaining finite set of equations for the atom+JJA+first waveguide node (the latter denoted here by index $0$) includes then a dissipation and noise term, given by:

\begin{widetext}
\eqn{
\label{eq:dphiexdt}
\der{}{\tilde t} \hat{\mb{\Phi}}_\mr{red} &= \dbar{C}_{\mr{red}}^{-1}\sbkt{Z_0 \hat{\mb{Q}}_\mr{red}}\\
\label{eq:dqexdt}
\der{}{\tilde t}\sbkt{Z_0\hat{\mb{Q}}_\mr{red}} &= -  \dbar{L}_{\mr{red}}^{-1} \hat{\mb{\Phi}}_\mr{red}-\underbrace{ \frac{Z_0}{Z_W} \dbar{\delta}_\mr{N+2}\dbar{C}_\mr{red}^{-1} \sbkt{Z_0\hat{\mb{Q}}_\mr{red}}}_{\text{Dissipation}} +\underbrace{ \frac{2}{\Omega_0} \sbkt{Z_0 \hat{Q}_\mr{in} \bkt{t}} \textbf{\textdelta}_{N+2} }_{\text{Noise}},
}
\end{widetext}
where we have defined  $Z_0\equiv \sqrt{L/C}$,    $\Omega_0\equiv 1/\sqrt{LC}$, $\tilde t \equiv \Omega_0 t$ and  the reduced subspace vectors $\hat{\mb{\Phi}}_\mr{red} = \cbkt{\hat\Phi_A, \hat\Phi_1, \dots, \hat\Phi_N, \hat\Phi_0^W}$ and $\hat{\mb{Q}}_\mr{red} = \cbkt{\hat Q_A, \hat Q_1, \dots, \hat Q_N,\hat Q_0^W}$ that encompass the atom+JJA+first waveguide node. The second and third terms in the  equation of motion for the charge variables indicate dissipation and noise due to coupling to the waveguide, with $\textbf{\textdelta}_{N+2}\equiv \cbkt{0,0,\dots,1}$, $\dbar{\delta}_\mr{N+2}\equiv \textbf{\textdelta}^T_{N+2}\textbf{\textdelta}_{N+2}$, and $\hat{{Q}}_\mr{in}\bkt{\tilde t}$ representing the input noise  at the zeroth waveguide node (see Appendix~\ref{App:SFM} for details).   The dimensionless capacitance and inductance matrices in the reduced subspace   are defined as:
\begin{widetext}

\begin{equation}
\label{Cex}
        \dbar{C}_\mr{red}=\frac{1}{C}     \begin{tikzpicture}[baseline={-0.5ex},mymatrixenv]\matrix [mymatrix,inner sep=4pt] (m){  \tikzmarkin[kwad=style orange]{cat} \tikzmarkin[kwad=style blue]{couple}  C_A + C_0 \tikzmarkend{cat}  &  -C_0 &  0   & \dots    & 0  & 0& 0\\
    -C_0 & \tikzmarkin[kwad=style teal]{jja}C_g + C + C_0\tikzmarkend{couple} & -C & \dots & 0 & 0 & 0\\
     0 & -C & C_g + 2C & \dots   &0 &0 &0 \\
    \vdots &\vdots &\vdots &  \ddots & \vdots & \vdots&\vdots\\
    0&0 &0 &  \dots & C_g + 2C & -C  &0 \\
    0 & 0 & 0 & \dots   & -C   &  ~\tikzmarkin[kwad=style yellow]{wg}C_g + C  + C_c \tikzmarkend{jja} &-C_c\\
    0 & 0 & 0 & \dots   & 0   &  -C_c  &C_c + C_W\tikzmarkend{wg}~\\
    };
  \end{tikzpicture},
  \end{equation}
  
\begin{equation}
\label{Linex}
\dbar L_\mr{red}^{-1} = L   \begin{tikzpicture}[baseline={-0.5ex},mymatrixenv]\matrix [mymatrix,inner sep=4pt] (m)          {  \tikzmarkin[kwad=style orange]{lat} \tikzmarkin[kwad=style blue]{couplel}  1/L_A + 1/L_0 \tikzmarkend{lat}  &  -1/L_0 &  0   & \dots    & 0  & 0&0  \\
    -1/L_0 & \tikzmarkin[kwad=style teal]{Ljja}1/L + 1/L_0\tikzmarkend{couplel} & -1/L & \dots & 0 & 0  &0 \\
     0 & -1/L & 2/L & \dots   &0 &0  &0\\
    \vdots &\vdots &\vdots &  \ddots & \vdots & \vdots&\vdots\\
    0&0 &0 &  \dots & 2/L & -1/L   &0\\
    0 & 0 & 0 & \dots   & -1/L  &~\tikzmarkin[kwad=style yellow]{wgl}  1/L  \tikzmarkend{Ljja}&0 \\    
    0 & 0 & 0 & \dots   & 0  &  0 &0 \tikzmarkend{wgl}~ \\    
    };
    \end{tikzpicture}.
\end{equation}
\end{widetext}

We remark that such an approach for eliminating the environmental modes to describe the dynamics of an open quantum system does not require one to take into account the full Hilbert space of the bath. The effects of the system-bath interaction are captured  in terms of the appropriate effective dissipation and noise terms at the boundary node, thereby reducing the computational resources needed for a numerical solution significantly.

\subsection{Singular function expansion}
\label{Sec:sfm}

 We now turn to  describing the open quantum system dynamics in terms of  a singular function expansion of the propagator for the system dynamics.  The equations of motion Eqs.~\eqref{eq:dphiexdt} and \eqref{eq:dqexdt} can be  solved  by taking a Laplace transform to obtain the linear dynamics of node flux variables as (see Appendix~\ref{App:Laplace} for details of the derivation):
\eqn{\label{phiext}
 \hat {\mb{\Phi} }^{(0)} _\mr{red}\bkt{\tilde t} = \frac{1}{2\pi i } \int_{-i \infty }^{i\infty } \dd\tilde s \, e^{\tilde s \tilde t} \, \, \dbar{G} \bkt{\tilde s} \hat{\mb{Y}}\bkt{\tilde s}
}
where  the superscript $(0)$ denotes the linear dynamics of the system and \eqn{\dbar G\bkt{\tilde s} \equiv \sbkt{ \tilde s^2\dbar C_\mr{red} +\frac{Z}{Z_W}  \tilde s \dbar{\delta}_{N+2} + \dbar{ L}_\mr{red}^{-1}}^{-1},}
corresponds to the propagator for the open system dynamics. The initial conditions and the input noise from the waveguide are represented by the operator:
\eqn{\label{eq:Ys}
\hat{\mb{Y}}\bkt{\tilde s} \equiv &\sbkt{\tilde s \dbar C_\mr{red} +\frac{Z_0}{Z_W} \dbar{\delta}_{N+2} }\hat{\mb{\Phi}}_\mr{red}\bkt{0} +  Z_0 \hat{\mb{Q}}_\mr{red}\bkt{0} \non\\
&+ 2  \sbkt{Z_0 \tilde Q_\mr{in}\bkt{\tilde s}}\textbf{\textdelta}_{N+2},
}
with $ \tilde Q _\mr{in} \bkt{\tilde s}\equiv \int _0 ^\infty \dd \tilde t \, e^{- \tilde s \tilde t} \, \hat{Q}_\mr{in}\bkt{\tilde t} $, the Laplace transform of the input quantum field at the first waveguide node\footnote{For a coherently driven resonator, the spectral form the input field $\tilde {Q}_\mr{in, tot} (\tilde s) = \sum_n \frac{Q_\mr{dn}}{\tilde s - i\tilde \omega_\mr{dn}} + \tilde{Q}_\mr{in} (\tilde s)$ is analytic except at a discrete set of real poles $\omega_\mr{dn}$ corresponding to the frequencies of the  driving field,  with $Q_\text{dn}$ as complex-valued scalars.}

We next decompose the propagator using a  singular function expansion as $ \dbar G ^{-1}\bkt{\tilde s} = \dbar {\alpha}\bkt{\tilde s} \dbar \gamma\bkt{\tilde s}  \dbar {\beta}\bkt{\tilde s} $, where $\dbar \gamma\bkt{\tilde s}$ is a diagonal matrix with $\gamma_{jj}\bkt{\tilde s}$ as the singular values of the propagator and $\dbar \alpha \bkt{\tilde s}$ and $\sbkt{\dbar \beta \bkt{\tilde s}}^T$ as orthogonal matrices.

We express the quantized flux field of the system as $\mb{\Phi}^{(0)}_\mr{red}\bkt{\tilde t} = \sum_p e^{-i\tilde\omega_p \tilde t}\vec{\varphi}_p$, where $\tilde\omega_p = -i \tilde s_p$ are a set of generally complex-valued poles of the system propagator $\dbar G \bkt{\tilde s_p}=0$. The poles and the associated  set of system  eigenmodes $\vec{\varphi}_p$ are determined by the generalized, quadratic eigenvalue problem
\eqn{\label{eigtot}
\sbkt{\tilde\omega_p^2 \dbar{C}_\mr{red}- i\tilde \omega_p\frac{Z_0}{Z_W} \dbar{\delta}_{N+2} - \dbar{L}_\mr{red}^{-1}}\vec{\varphi}_p=0.}
The transient quantum dynamics of the linear problem (e.g. the spontaneous emission dynamics) is encoded entirely in the eigenmodes and complex eigenfrequencies obtained by solving Eq.~(\ref{eigtot}). This approach makes no assumptions on the strength of coupling between the atom and JJA sectors, and is thus expected to be valid across regimes of varying atom-field coupling strengths. As a consequence, there is no formal distinction between the JJA modes and the atomic mode; the obtained eigenmodes $\vec{\varphi}_p$ are not necessarily restricted to either atom or JJA sectors.

Considering the transient oscillations in an initially excited system (which, for cavity-like modes is referred to as ringdown and for the qubit-like mode as spontaneous emission) we obtain the dynamics of the flux at node $j $ as:
\eqn{
\label{Eq:phiext}
 \hat{\Phi}^{(0)} _{\mr{red}, j} \bkt{\tilde t}
 =&\sum_{p,q} e^{\tilde s_p \tilde t} \eta_{p,q}\bkt{\tilde s_p} \hat{Y}_{q}\bkt{\tilde s_p},
}
where 
\eqn{\label{Eq:eta}\eta_{p,q}\bkt{\tilde s_p}\equiv \frac{1}{\bkt{\frac{\delta\gamma_{pp} \bkt{\tilde s_p}}{\delta \tilde s}}}\beta^{-1}_{j,p}\bkt{\tilde s_p}  \alpha^{-1}_{p,q}\bkt{\tilde s_p}  ,}
with $ \dbar \alpha \bkt{\tilde s _p}$ and  $\dbar \beta \bkt{\tilde s _p}$ representing  the singular vectors evaluated at the poles  $\tilde s_p (= i \tilde \omega_p )$  of the propagator evaluated using Eq.~\eqref{eigtot}.

Substituting Eq.~\eqref{Eq:phiext} in Eq.~\eqref{eq:dphiexdt}, we obtain the linear dynamics of the charge variables as follows:

\eqn{\label{Eq:qext}
Z_0 \hat{Q}^{(0)} _{\mr{red},j} \bkt{\tilde t} =\sum_{p,q}  e^{\tilde s_p \tilde t} \zeta_{p,q}\bkt{\tilde s_p} \hat{Y}_{q}\bkt{\tilde s_p},
}
where 
\eqn{\label{Eq:zeta}\zeta_{p,q}\bkt{\tilde s_p} \equiv \sum_r\tilde s_p \frac{1}{\bkt{\frac{\delta\gamma_{pp} \bkt{\tilde s_p}}{\delta \tilde s}}}C_{\mr{red},j,r}\beta^{-1}_{r,p}\bkt{\tilde s_p}  \alpha^{-1}_{p,q}\bkt{\tilde s_p}.} 

Thus, the linear dynamics of the open atom+JJA  system is described exactly by Eqs.~\eqref{Eq:phiext} and \eqref{Eq:qext}. However, an often-used  perspective is to treat the artificial atom as the system of interest and the JJA as its environment. Next, we show how this perspective can be obtained from our approach, with the eventual aim of comparing how perturbation theory results using the standard perspective compare to exact results obtained using Eq.~(\ref{eigtot}). The atom+JJA description typically centers around a Hamiltonian formulation, assuming the combined artificial atom+JJA system to be only weakly-coupled to the lossy waveguide.  We will thus begin by deriving the exact  Hamiltonian of the artificial atom+JJA system.

\section{Hamiltonian for the closed system: Atom-JJA interaction}
\label{Sec:Ham}

In this section we consider the atom+JJA as a closed system without the external coupling to the waveguide ($C_c\to 0$) to describe the interaction between the atom and the array in terms of an effective Hamiltonian. It proves useful to separate the flux variables into atomic and JJA sectors, $\mb{\Phi}_\mr{closed} = \{\Phi_A;\mb{\Phi}_{\mr{JJA}}\}$, where $\mb{\Phi}_{\mr{JJA}} = \cbkt{\Phi_1,\dots, \Phi_N}$. Then, the full closed system Lagrangian can be equivalently written as:
\eqn{
\mc{L} &\equiv \frac{1}{2} \dot{\mb{\Phi}}_{\mr{JJA}} ^T\dbar{C}_\mr{JJA}\dot{\mb{\Phi}}_{\mr{JJA}} - \frac{1}{2} \mb{\Phi}_{\mr{JJA}} ^T \ \dbar{L}^{-1}_\mr{JJA}\mb{\Phi}_{\mr{JJA}} \non\\
&+\frac{1}{2}(C_A+C_0)\dot{\Phi}_A^2 - \frac{1}{2}\left(\frac{1}{L}+\frac{1}{L_0}\right)\Phi_A^2  \non \\
&- C_0 \dot\Phi_A \dot\Phi_1 + \frac{1}{L_0 }\Phi_A \Phi_1.
}

We use the Lagrangian of the JJA as given by the first line, thus including on-site renormalization due to the coupling to the artificial atom, to define an appropriate set of JJA modes. More precisely, we calculate the Euler-Lagrange equations of motion for the JJA nodes, and obtain a generalized eigenvalue problem for eigenmodes of the JJA introduced via $\mb{\Phi}_{\mr{JJA}} = e^{-i\omega_{k,\mr{JJA}} t}\mb{\Phi}_{k,\mr{JJA}}$, and eigenfrequencies $\omega_{k,\mr{JJA}}$ (real in the closed system limit), as detailed in Appendix \ref{App:JJAeig}: 
\eqn{\label{jjaep}
\omega_{k, \mr{JJA}}^2 \dbar{C}_\mr{JJA}\mb{\Phi}_{k, \mr{JJA}} = \dbar{L}_\mr{JJA}^{-1} \mb{\Phi}_{k, \mr{JJA}},
}
assuming $C_c\rightarrow 0 $ for a closed JJA. The obtained eigenvalues form a photonic band as shown in Fig.~\ref{Fig:gk}, with a band edge at $ \Omega_0/(2\pi) \approx12.95$~GHz (see also Fig.~\ref{Fig:disp}~(a) in the Appendix). We emphasize here that the modes $\mb{\Phi}_{k, \mr{JJA}}$ of the JJA  are calculated by including the coupling capacitance ($C_0$) and inductance ($L_0$)  between the JJA and the atom in the $\dbar C_\mr{JJA}$ and $\dbar L_{\mr{JJA}}$ matrices. Thus, the  JJA spatial eigenmodes follow the appropriate boundary conditions determined self-consistently by the strength of the coupling element (see Appendix~\ref{App:JJAeig} for details).

Having defined the JJA modes, we will now rewrite the \textit{total} closed system Lagrangian in this basis. To this end,  we can define the vector $ \mb{\Psi}\equiv \cbkt{\Phi_A; \mb{\Phi}_{k, \mr{JJA}}}$ as the composite of the bare atomic mode $\Phi_A$ and the JJA eigenmodes $\mb{\Phi}_{k, \mr{JJA}}$. Rewriting the closed system Lagrangian in this composite basis, we obtain:
\eqn{\label{Ltilde}
\mc{L}= \frac{1}{2} \dot{\mb{\Psi}}^T \tilde C \dot{\mb{\Psi}}- \frac{1}{2} \mb{\Psi}^T \tilde L ^{-1} \mb{\Psi}.}
The transformed capacitance and inductance  matrices are  defined as $\tilde C \equiv U ^T \dbar{C} U $  and $\tilde L^{-1} \equiv U ^T \dbar{L}^{-1} U $, where the matrix $U $ relates the flux vector in the spatial basis to that in the partially diagonalized basis $  \mb{\Phi }_\mr{closed} = U \mb{\Psi}$. 

\begin{figure}[t]
    \centering
    \subfloat[]{
 \includegraphics[width = 3.3 in]{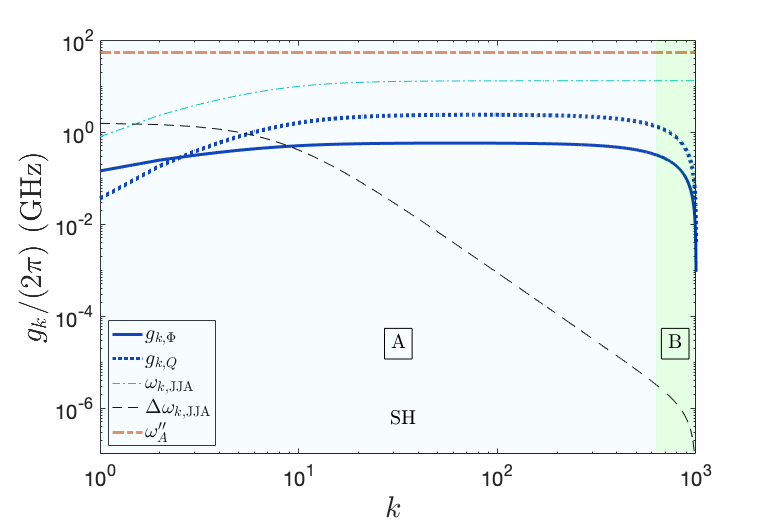}}\\
     \subfloat[]{
 \includegraphics[width = 3.3 in]{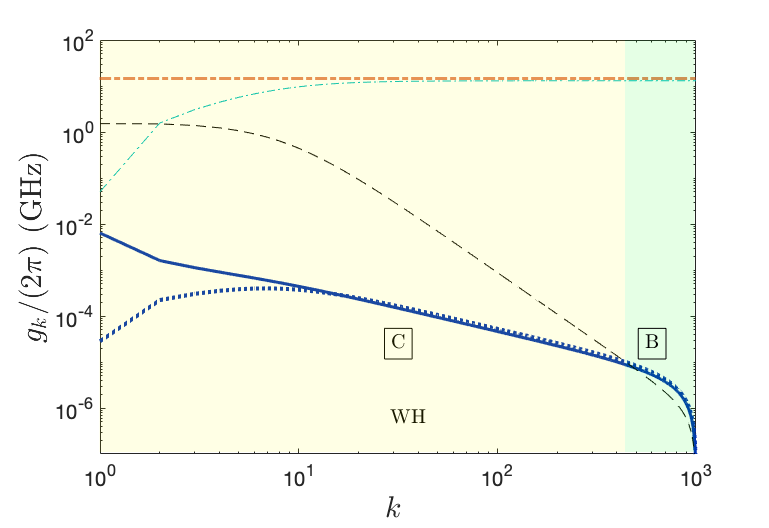}}
    \caption{Coupling coefficients $ g_{k,\Phi}$ (solid blue) and $ g_{k,Q}$ (dotted blue),  as obtained numerically from Eq.~\eqref{gkphisa} and Eq.~\eqref{gkqsa}, corresponding to  (a) $\chi = 1$  and (b) $\chi = 10^{-5}$. The bare and renormalized atomic frequencies ($\omega_A$  and $\omega_A''$) are denoted by the dashed orange horizontal lines for each case. The dashed-dotted curves correspond to the JJA eigenfrequencies $\omega_{k, \mr{JJA}}$,  and the dashed curve corresponds to the free spectral range $ \Delta\omega_{k,\mr{JJA}}$. The shaded regions represent the different coupling regimes (A),  (B) and (C) denoted by the  blue,  green and yellow shaded areas, respectively. We have chosen the bare atomic frequency to be $\omega_A /(2\pi )\approx15 $~GHz.}
    \label{Fig:gk}
\end{figure}

We define the conjugate momenta corresponding to the eigenmodes of the uncoupled JJA and atomic flux variables as  $\mb{Q}_{k,\mr{JJA}} = \frac{\delta \mc{L}}{\delta \dot{\mb{\Phi}}_{k, \mr{JJA}}}$, and $Q_A = \frac{\delta \mc{L}}{\delta \dot{\Phi}_{A}}$ respectively. This yields $\mb{Q} = \tilde C \dot{\mb{\Psi}}$, where $\mb{Q}\equiv \cbkt{Q_A ; \mb{Q}_{k,\mr{JJA}} }$. The Hamiltonian is thus obtained via the Legendre transformation as
\eqn{\label{Ham2}
H = \frac{1}{2} \mb{Q}^T \tilde C ^{-1} \mb{Q} + \frac{1}{2} \mb{\Psi}^T \tilde{L}^{-1}\mb{\Psi}.
}

Promoting the flux and charge variables $\cbkt{\mb{\Psi}, \mb{Q}}$ to quantum observables, one can express those in terms of bosonic operators as

\eqn{
\label{hatphik}
\hat \Phi_{k, \mr{JJA} (A)} =& \sqrt{\frac{\hbar Z_{k (A)}}{2}} \bkt{\hat{a}_{k (A)}  + \hat{a}_{k (A)} ^\dagger}\\
\label{hatqk}
\hat Q_{k, \mr{JJA} (A)} =& -i \sqrt{\frac{\hbar}{2Z_{k (A)}}} \bkt{\hat{a}_{k (A)}  - \hat{a}_{k(A)} ^\dagger},}
where the creation and annihilation operators corresponding to the array modes (atom) satisfy the canonical commutation relations $\sbkt{\hat{a}_{k(A) }^\dagger, \hat{a}_{k (A) }} = 1$. The impedances associated with the atom and the $k^{\mr{th}}$ JJA mode are defined as $ Z_A \equiv  \sqrt{L_A'/C_A''}$ and $ Z_{k} \equiv \frac{1}{\bkt{C_g + 2C}\omega_{k, \mr{JJA}}}$.  The renormalized atomic  capacitance is given by
\eqn{\label{capp}
C_A'' \equiv& C_A ' - \frac{C_0 ^2}{C_g + 2C}  \sum_k \Phi_{k,\mr{JJA}}^2(1)
.}

This allows one to rewrite the Hamiltonian in Eq.~\eqref{Ham2} as (see Appendix~\ref{App:Ham} for details)
\begin{widetext}
\eqn{
\label{Ham}
H = &\hbar \omega_A'' \hat{a}^\dagger_A \hat{a}_A + \sum_k\sbkt{\hbar  \omega'_{k,\mr{JJA}} \hat{a}_k ^\dagger \hat{a}_k+  \hbar g_{k,\Phi} \bkt{\hat a_A + \hat{a}_A ^\dagger}\bkt{\hat a_k + \hat{a}_k ^\dagger}+ \hbar  g_{k,Q} \bkt{\hat a_A - \hat{a}_A ^\dagger}\bkt{\hat a_k - \hat{a}_k ^\dagger} }\non\\
&+\sum_{k\neq k'}\sbkt{\hbar  \xi_{k,k'} \bkt{\hat a_k  - \hat a_k ^\dagger}\bkt{\hat a_{k'}  - \hat a_{k'} ^\dagger}},
}
\end{widetext}

We emphasize that the Hamiltonian of Eq.~(\ref{Ham}) is \textit{exact} for the closed system: there is no truncation in the number of JJA modes retained, or any weak-coupling approximations. We have introduced the renormalized atomic and array mode frequencies, defined respectively as
\eqn{\label{waren}
\omega_A '' = & \omega_A' \sbkt{1 - \frac{C_0 ^2}{C_A' \bkt{C_ g + 2C}}\sum _ k \Phi_{k,\mr{JJA}}^2 \bkt{1}} ^{-1/2}\\
\label{wkren}
\omega'_{k, \mr{JJA}} = &\omega_{k, \mr{JJA}} \sbkt{1 + \frac{C_0 ^2 }{C_A'' \bkt{C_ g + 2C} }\Phi_{k,\mr{JJA}} ^2(1)}^{-1/2}.
}
We remark that the  renormalized atomic frequency $\omega_A'' $ can be drastically different from the bare atomic frequency, as well as the physical eigenfrequency corresponding to the atomic mode in the non-perturbative regimes, as analyzed in Appendix~\ref{App:Ham}. 

The flux and charge coupling coefficients between the atom and the JJA modes are $g_{k,\Phi}$ and  $g_{k,Q}$,   are given, respectively,  by
\eqn{\label{gkphisa}g_{k,\Phi} =& - \frac{\chi \sqrt{Z_A Z_k}}{2L} \Phi_{k,\mr{JJA}}(1)\\
\label{gkqsa}
g_{k,Q} =& - \frac{\chi C }{2 \bkt{C_g + 2C} C_A''\sqrt{Z_A Z_k }}\Phi_{k,\mr{JJA}} (1).}
We note from the above  that the coupling strength between the atom and the JJA modes goes linearly as the coupling parameter $\chi$. It is also pertinent to note here that the amplitude of the JJA eigenmodes at the atomic position $ \Phi_{k,\mr{JJA}}$ also varies as one changes $\chi$. 

\eqn{\label{Eq:xikk}
\xi_{k,k'} = \frac{C_0^2}{4\bkt{C_g + 2C }^2 C_A'' \sqrt{ Z_kZ_{k'}}}\Phi_{k,\mr{JJA}}\bkt{1}\Phi_{k',\mr{JJA}} \bkt{1}} corresponds to the strength of coupling between the $k$ and $k'$ JJA modes, mediated by the atom.

We plot the coupling coefficients between the atom and the JJA, $g_{k\Phi}$ and  $g_{kQ}$, in  Fig.~\ref{Fig:gk}~(a) and (b) for $\chi = 1$ and $\chi = 10^{-5}$, respectively, with increasing $\chi$ indicating stronger coupling values (see Eq.~\eqref{eq:chi}).
One can approximately identify the following  coupling regimes as indicated by the different shaded regions in the plots:
\begin{enumerate} [A.]
\item{Region (A) is identified by the condition \eqn{\label{eq:rega} g_k/\omega_k \gtrsim 0.1,} where $g_k \equiv \mr{max}\cbkt{g_{k,\Phi}, g_{k,Q}}$ is defined as the maximum of the two coupling coefficients  and $ \omega_k = \mr{min}\cbkt{\omega'_{k,\mr{JJA}}, \omega''_A }$ is defined as the minimum of the bare excitation frequencies  for a given $k $ value. In such a regime perturbation theory does not apply anymore and non-RWA terms become important in describing the radiative properties and dynamics of the atomic system. Such a regime is often referred to as the `ultrastrong' or `deep strong' coupling regime \cite{Kockum19,  FornDiaz19}. However, we note that contrary to these regimes that rely on a single environmental mode,  the coupling strength in the present system can be greater than or comparable to the free spectral range of the environment $(g_k \gtrsim \Delta\omega_{k, \mr{JJA}}$),  necessitating a consideration of multiple environmental modes.}
\item{ Region (B) corresponds to the case  where the coupling strength is comparable to or greater than the free spectral range  but smaller than the bare excitation frequencies of the atomic and JJA modes \eqn{\label{eq:regb}\Delta\omega_{k,\mr{JJA}}\lesssim g_k\ll \omega_k,}
for a specific $k$ value. Such a regime is also referred to as the multimode strong coupling or the `superstrong' coupling  regime, wherein the atomic mode couples strongly to multiple modes of the environment \cite{Sundaresan15, Kuzmin19a, Meiser06}.}
\item{Region (C) corresponds to the case where the coupling strength is smaller than both the free spectral range and the individual mode frequencies.}
\end{enumerate}

 Our analysis indicates that a careful treatment of the coupling terms, incorporating the spatial dependence of the array modes, is necessary to model the interaction between the artificial atom and a high-impedance resonator.
The atom in such experimental setups~\cite{Kuzmin19a, Leger19} generally corresponds to a junction with a distinctly different non-linearity than the rest of the junctions. Thus, in practice, it is convenient to identify the atom by its locality.  However, its ``coupling'' to the environment must be physically defined. 
One approach  used in the literature to arrive at a Hamiltonian description of such a system is to carry out an equivalent circuit analysis via the Foster theorem to derive a Caldeira-Leggett model for the atom-environment interaction \cite{Manucharyan2017, Kuzmin19a,  DevoretLesHouches}.  In particular, while performing an equivalent circuit transformation the coupling term between the atom and the JJA must be appropriately transformed.  The resulting Caldeira-Leggett model, Eq.~\eqref{Ham}, generally contains both a flux and a charge coupling term, even when a further diagonalization of the JJA sector is performed to remove the $\xi_{k, k'}$ term.

Having analyzed the closed atom-JJA interaction, we now return to the open system description.

\section{Radiative Properties of the atom}
\label{Sec:Rad}

\subsection{Complex eigenfrequencies in the linear regime}

We will now consider the radiative properties of the atom, namely the radiative frequency shift, a.k.a. the Lamb shift, and the spontaneous emission decay rate, a.k.a the Purcell decay rate. In calculating the exact atomic mode eigenfrequency, we must first address the question of how the atomic mode is identified when solving Eq.~(\ref{eigtot}) for eigenmodes that are defined over the joint atom+JJA system. We note that in the absence of the atom-JJA coupling, $\cbkt{C_{0}, 1/L_{0} } \rightarrow0$, Eq.~(\ref{eigtot}) becomes block-diagonal, reducing to two independent eigenproblems for the separate atom and JJA modes respectively. The atomic mode eigenproblem yields an eigenmode $\vec{\varphi}_A \propto \cbkt{1;\mb{0}}$ that is completely localized at the spatial index corresponding to the atom, with bare frequency $\omega_A = \frac{1}{\sqrt{C_AL_A}}$. 
\begin{figure}[t]
    \centering
    {\includegraphics[width = 3.4 in]{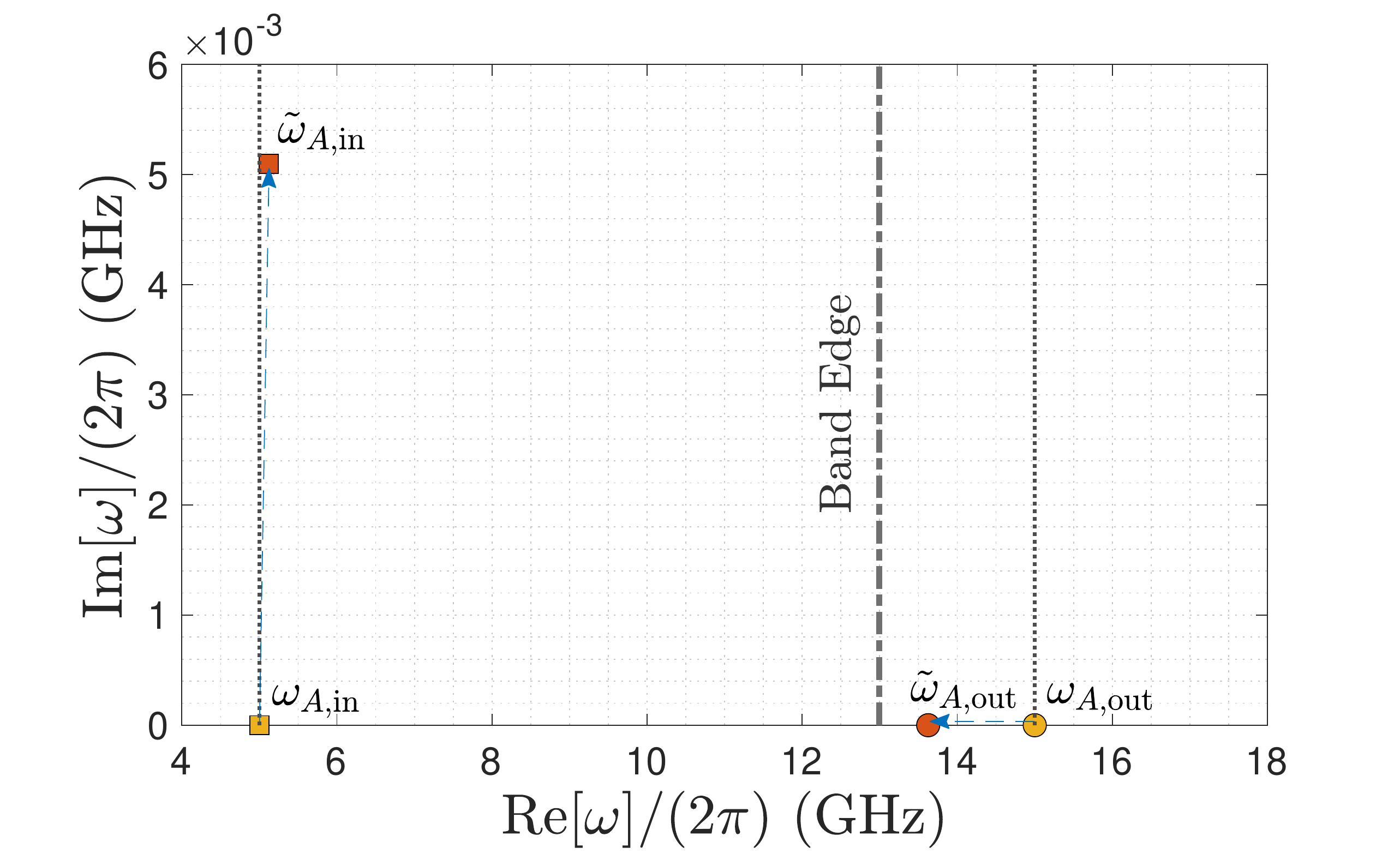}}
    \caption{The change in the complex atomic frequency $\omega_\text{A}$ when $\chi=0 \rightarrow \chi=1$. Plotted is this shift for two cases: for the bare atomic frequency in the band $\omega_{A,\mr{in}}/(2\pi) =5$~GHz (squares) and outside the band $\omega_{A, \mr{out}}/(2\pi) =$15~GHz (circles). The photonic band edge frequency is shown as the vertical dashed-dotted line around $ \Omega_0/(2\pi) \approx12.95$~GHz.} 
    \label{Fig:poles}
\end{figure}

This uncoupled regime forms the starting point for an adiabatic procedure that allows us to track the evolution of the atomic frequency and spatial eigenmode as the coupling strength is increased to a desired nonzero value, determined by $C_0,L_0$. More precisely, we consider an iterative procedure $\cbkt{C_{0}^{(n)}, 1/L_{0}^{(n)}}\rightarrow \cbkt{n C_0/N_a, N_a/(n L_0)} $, for $n=0,\ldots,N_a$, yielding eigenmode $\vec{\varphi}_A^{(n)}$ and complex eigenfrequency $\omega_A^{(n)}$ for the $n^\mr{th}$ step, such that $\vec{\varphi}^{(0)}_A \propto \cbkt{1;\mb{0}}$ and $\omega_A^{(0)} = \frac{1}{\sqrt{C_AL_A}}$. At each step we identify the atomic mode as the one that has maximum overlap with the atomic mode at the previous step. For a sufficiently large number of steps such that the change in the eigenmode frequency between  iteration steps is smaller than other energy scales, we obtain a convergence to a specific eigenmode that we identify as the atomic mode with eigenfrequency $\tilde\omega_A$, and the corresponding spatial eigenfunction $\vec \varphi_A$.

One may wonder why such an adiabatic limiting procedure is needed to identify the atomic mode. In the perturbative limit ($\chi \ll 1$), the atomic mode can easily be identified by the local nature of the associated eigenmode. Our analysis in Sec.~\ref{Sec:Atommode} will show that in the non-perturbative regime such an identification is not possible. This is also the reason why in recent experiments accessing this regime~\cite{Leger19}, the Lamb shift could not be directly measured. Instead, it was inferred via an indirect measurement of the splitting between various modes in the system.

The adiabatic procedure described here is applicable to both closed and open system cases. As shown in Fig.~\ref{Fig:poles}, one can plot the atomic eigenfrequency on the complex plane as one increases the coupling strength between the atom and the JJA.  The Lamb shift of the atom is identified as the  difference between the real part of the  eigenvalue and the bare atomic frequency for the uncoupled atom. The spontaneous emission rate of the atom is given by the imaginary part of the complex atomic eigenfrequency. We note that the atom here is assumed to not be coupled to any other bath than the waveguide. It can be seen from Fig.~\ref{Fig:poles} that for the case where the bare atomic frequency is outside the band (at $\chi=0$), the atomic mode incurs a substantial Lamb shift while the spontaneous emission rate remains negligible, as a result of the lack of density of modes.

\begin{figure*}[t]
    \centering
\subfloat[]{    \includegraphics[width = 2.5 in]{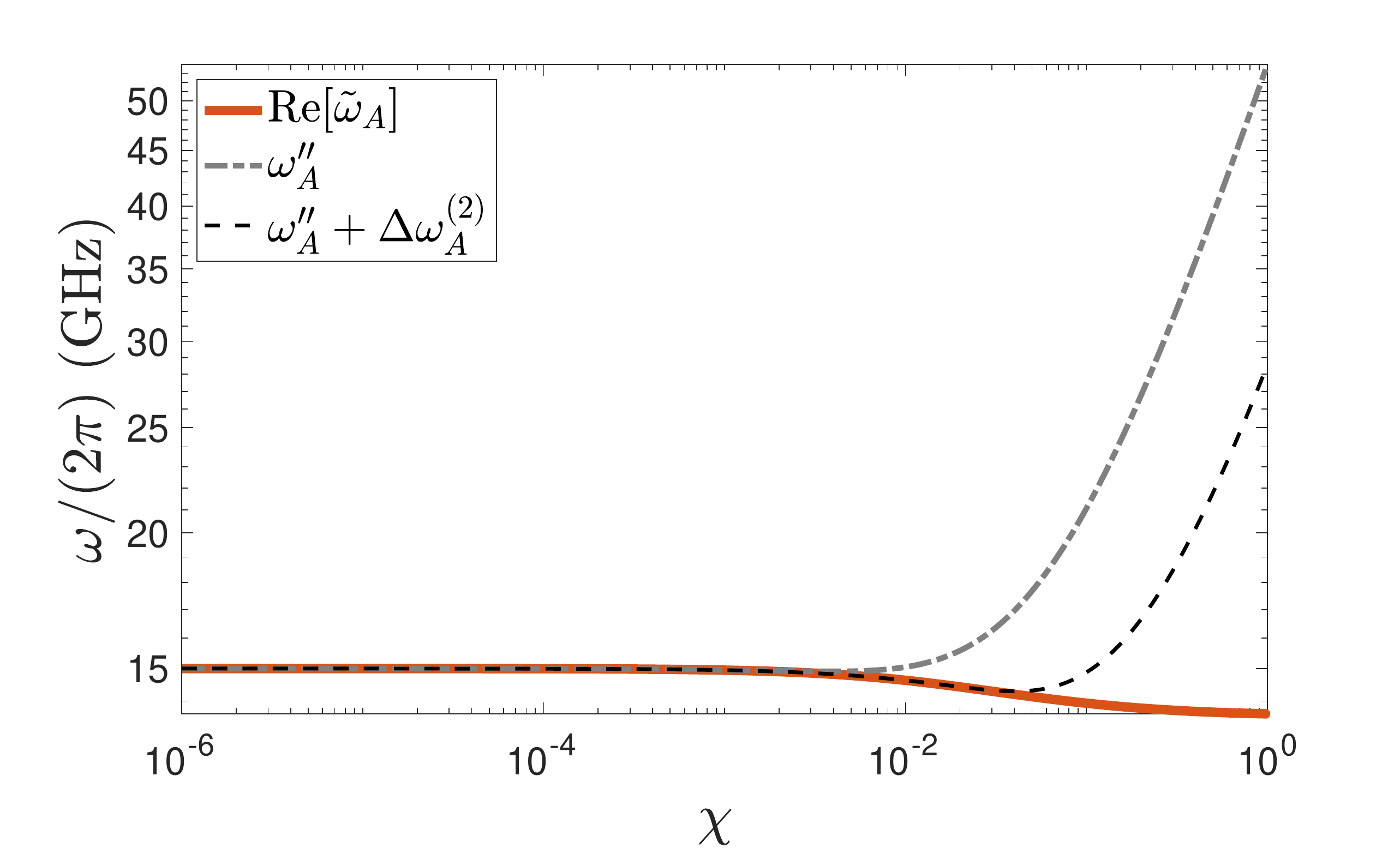}}
\subfloat[]{\includegraphics[width = 2.5 in]{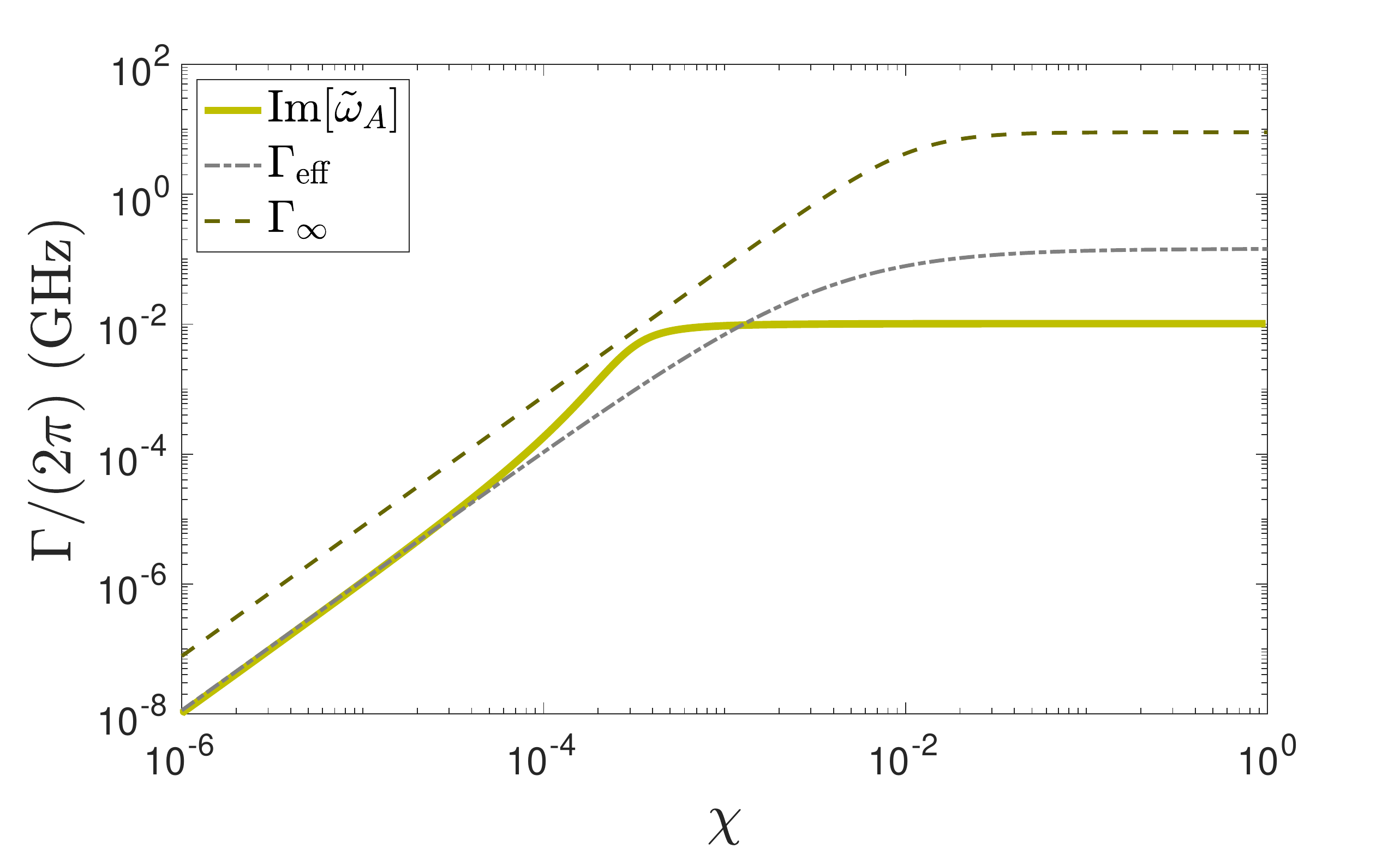}}\\
\subfloat[]{\includegraphics[width = 2.5 in]{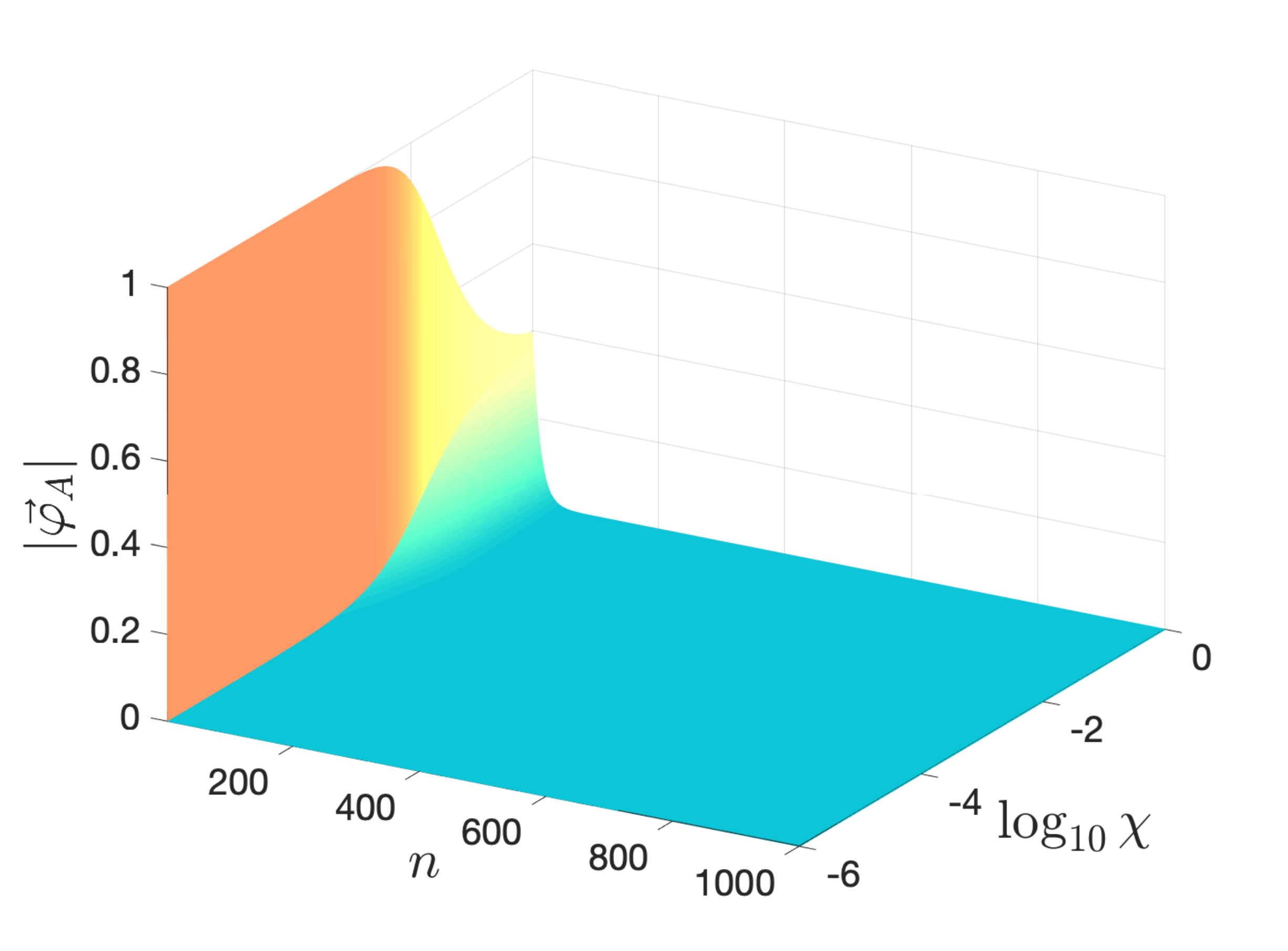}}
\subfloat[]{\includegraphics[width = 2.5 in]{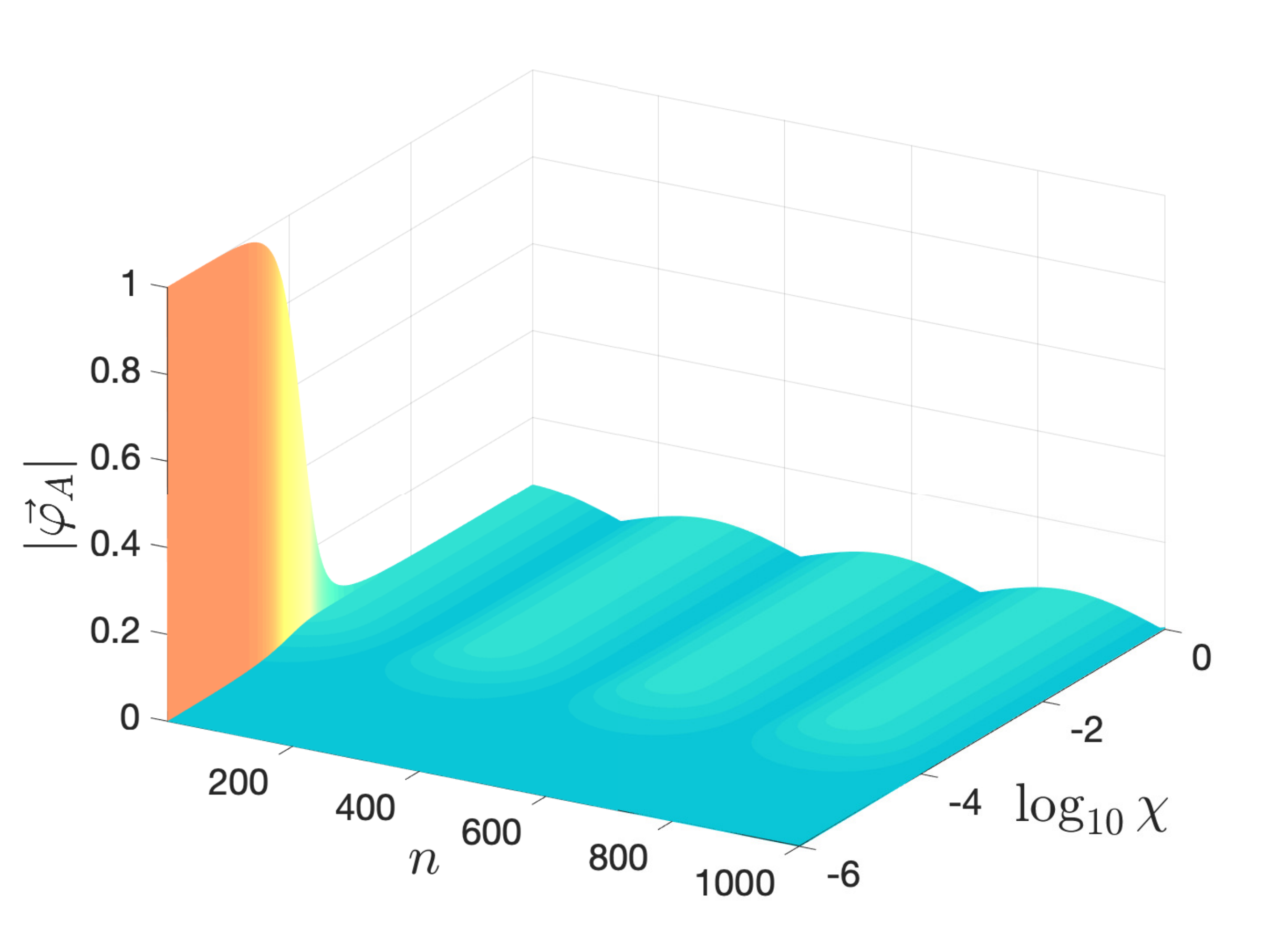}}
    \caption{(a) Atomic frequency as a function of the coupling parameter $\chi$ for $\omega_A /(2\pi)= 15 $~GHz. The solid curve stands for the real part of the  numerically obtained atomic eigenmode frequency $\mr{Re}\sbkt{\tilde \omega _A}$.  The dashed-dotted curve corresponds to the renormalized atomic frequency $\omega_A'' $ (Eq.~\eqref{waren}), and the  dashed curve represents the renormalized atomic frequency including second-order perturbative corrections as given by Eq.~\eqref{deltawa2}.  (b) Atomic dissipation as a function of the coupling parameter $\chi$ for $\omega_A /(2\pi)= 5 $~GHz. The solid curve denotes the numerically obtained atomic dissipation $\bkt{\mr{Im}[\tilde \omega_A]}$, the dash-dotted curve represents the decay obtained via the  perturbative expression $\Gamma_\mr{eff}$ given by Eq.~\eqref{gammaeff}, and the dashed curve corresponds to the perturbative decay rate  $\Gamma_\infty$ obtained with considering an infinite JJA impedance. Atomic eigenmode as a function of spatial position and coupling strength for (c) $\omega_A /(2\pi) = 15$~GHz and (d) $\omega_A /(2\pi) = 5$~GHz. }
    \label{Fig:pert}
\end{figure*}

\subsection{Comparison with perturbative approach}
\label{Sec:pert}
\begin{figure*}[t]
    \centering
     \subfloat[]{\includegraphics[width = 2.25 in]{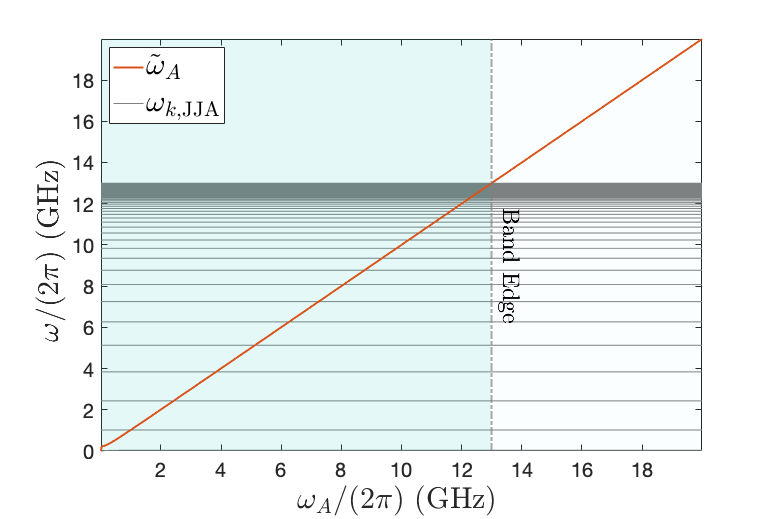}}
\subfloat[]{\includegraphics[width = 2.25 in]{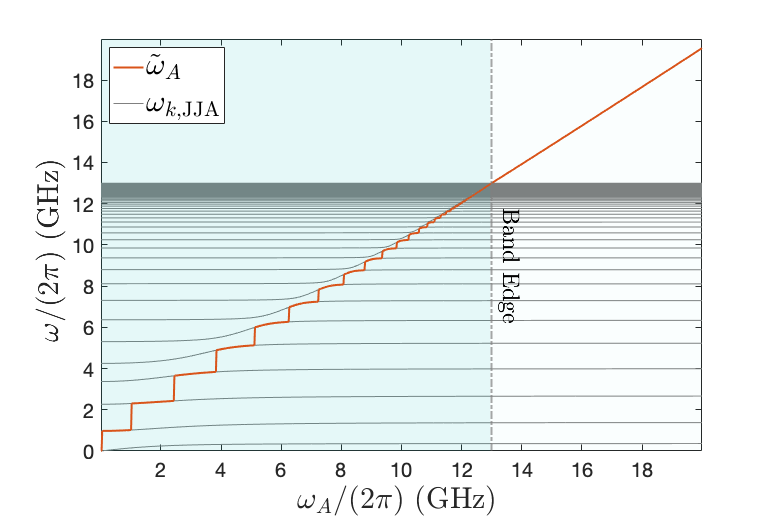}}
    \subfloat[]{\includegraphics[width = 2.3 in]{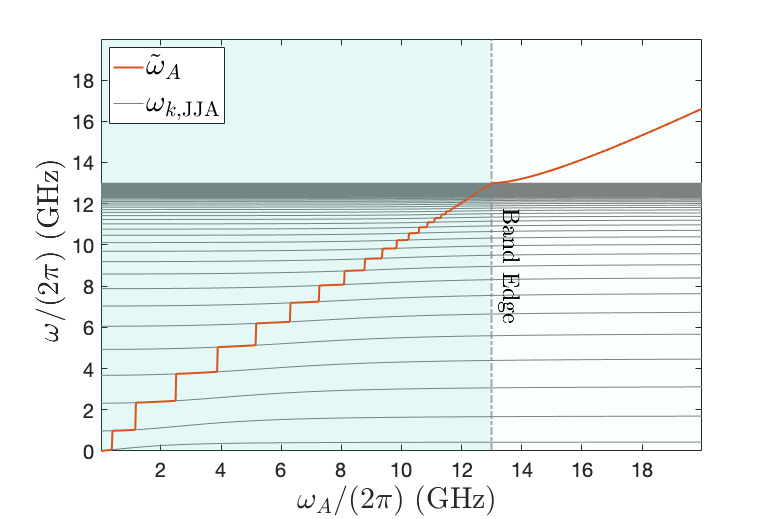}}
    \\
    \subfloat[]{\includegraphics[width = 2.25 in]{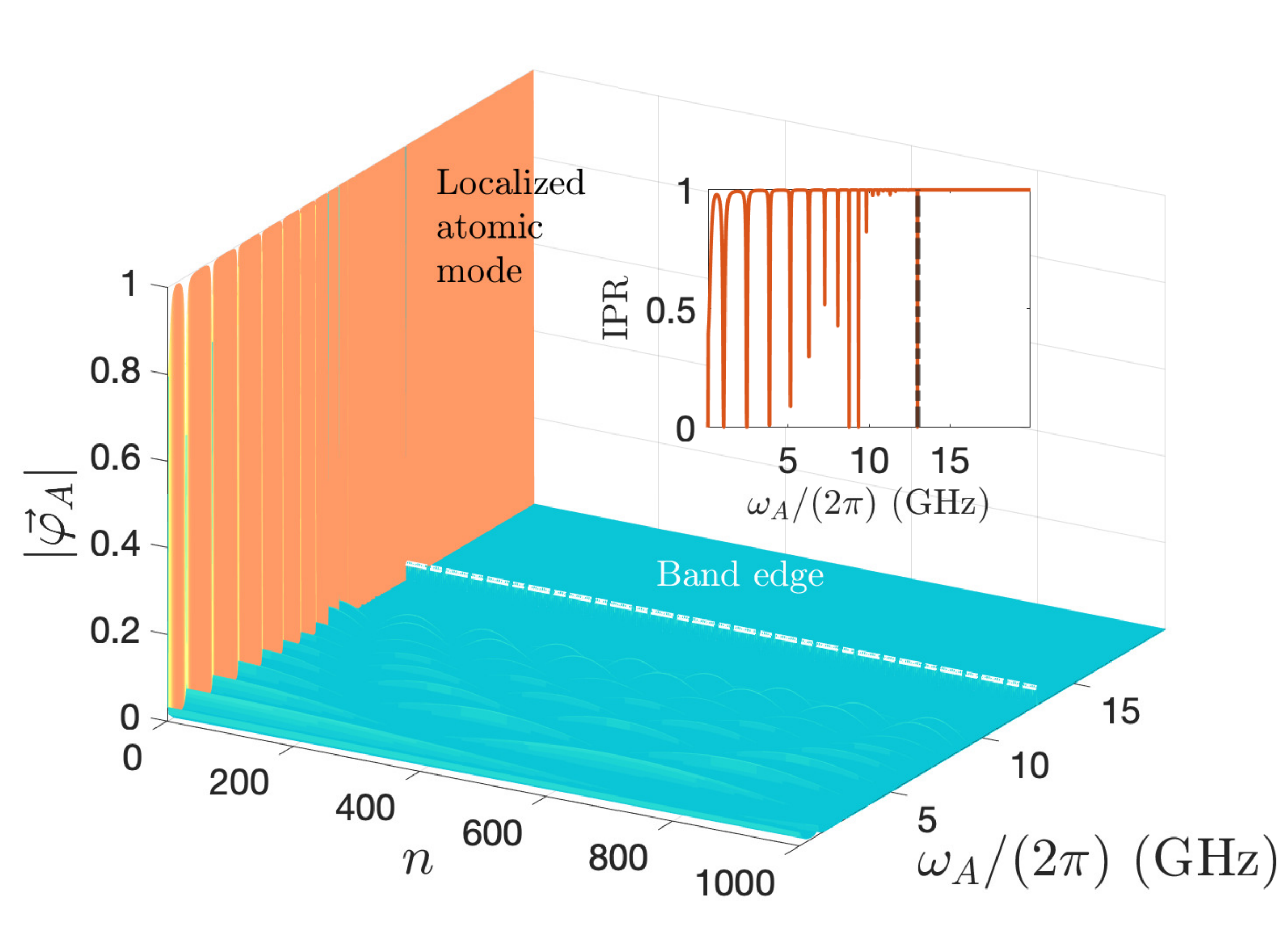}}     \subfloat[]{\includegraphics[width = 2.25 in]{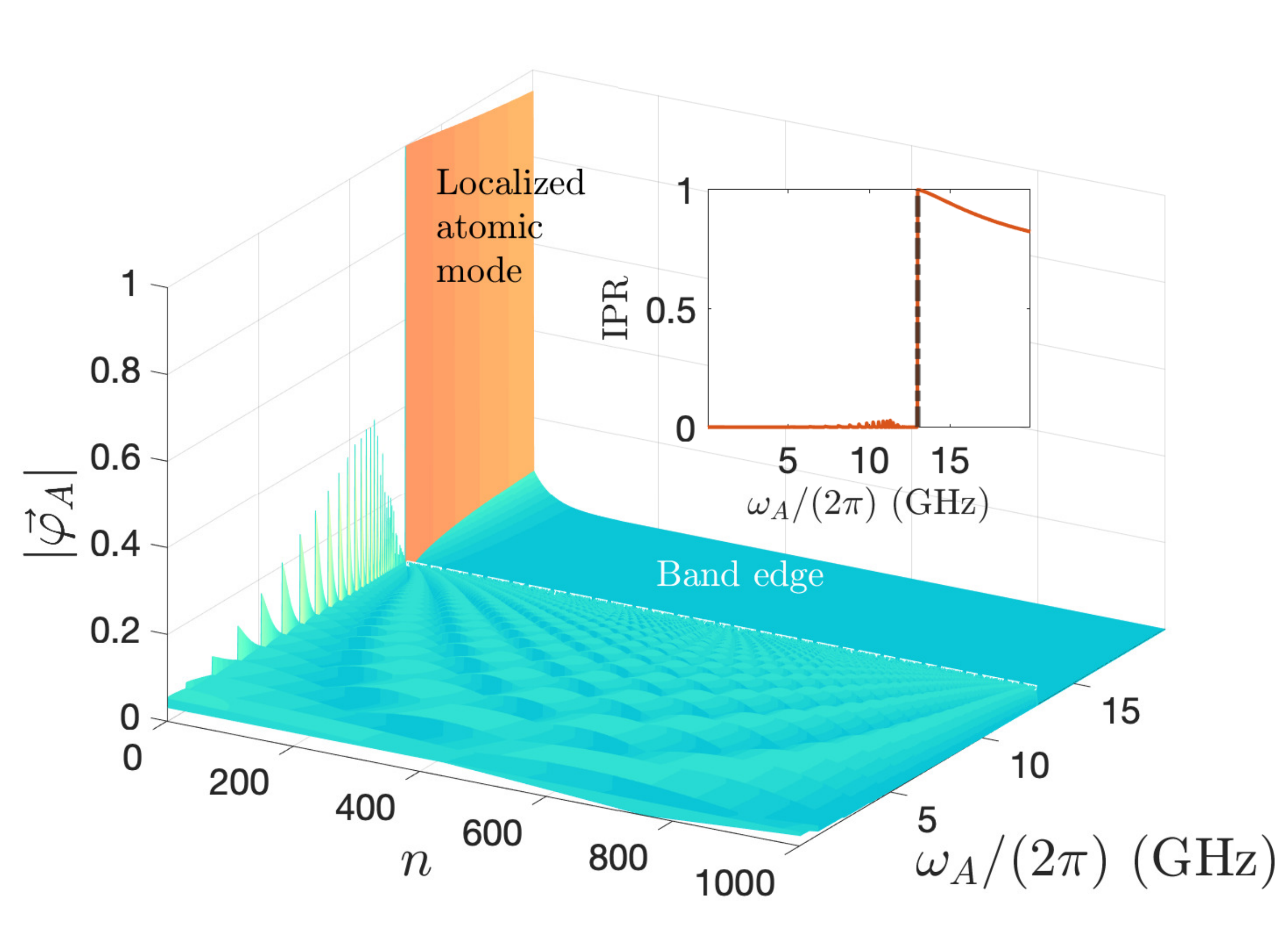}} \subfloat[]{\includegraphics[width = 2.25 in]{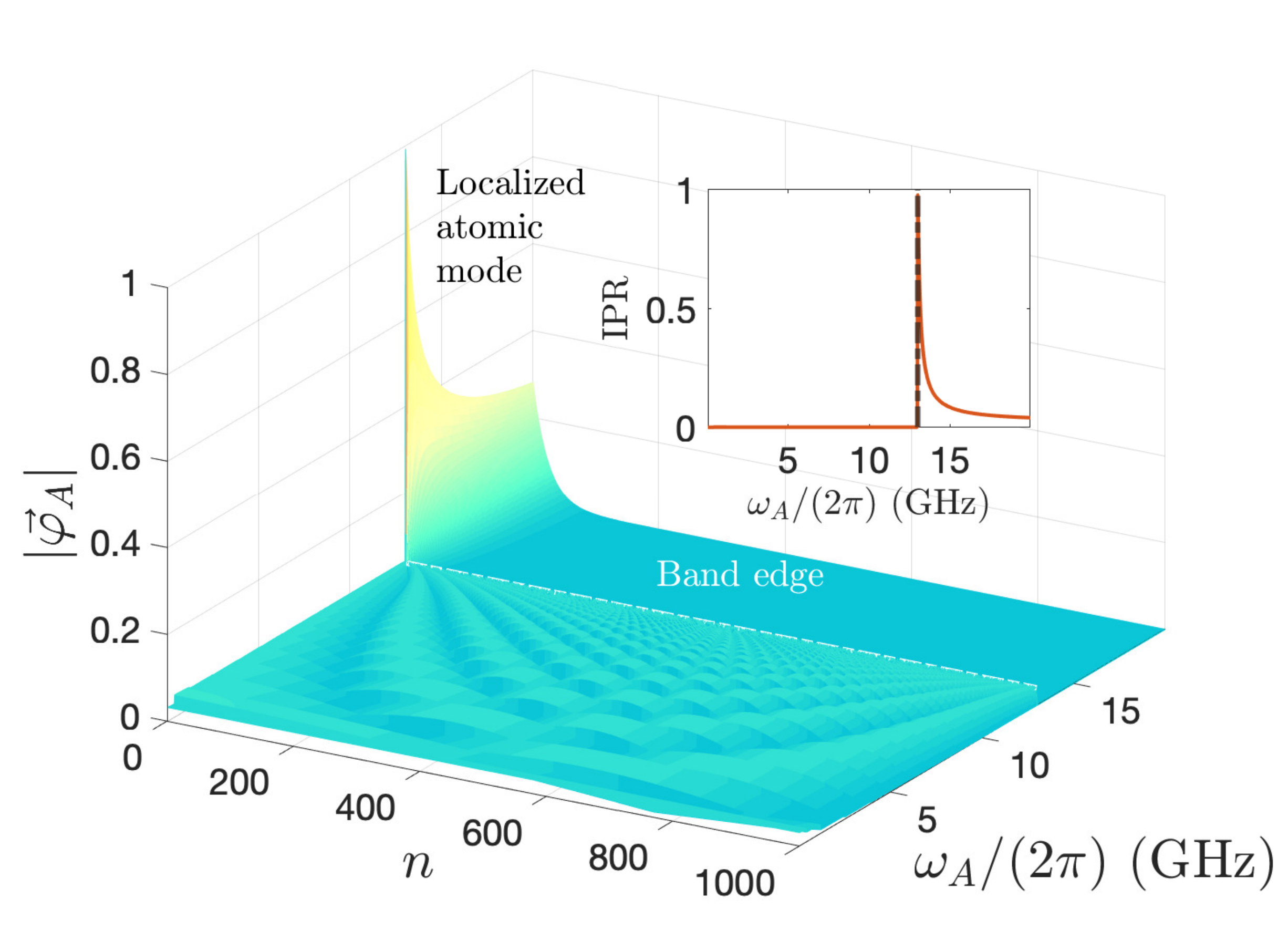}}  
    \caption{Eigenfrequencies of the coupled atom+JJA system as a function  of the bare atomic frequency  for (a) $\chi = 10^{-5}$, (b) $\chi = 10^{-2.5}$ and (c)  $\chi = 1$.   The atomic mode frequency is denoted by the solid orange curve in each plot and the gray  curves corresponds to the JJA normal modes. The atomic eigenmodes corresponding to (a), (b) and (c) are shown in (d), (e) and (f), respectively. The inset  shows the Inverse Participation Ratio (IPR) for the atomic mode as a function of the bare atomic frequency as determined by Eq.~\eqref{Eq:IPR}. }
    \label{Fig:local}
\end{figure*}
A standard approach to calculating the field-induced modification of atomic properties is via perturbation theory. Reintroducing the Josephson potential $\mathcal{U}_A(\Phi_A)$ into the linear Hamiltonian, Eq.~(\ref{Ham}), furnishes the nonlinearity necessary to render the bare atomic spectrum anharmonic, thus allowing the addressability of two individual energy levels {under coherent monochromatic input}. We can thus  derive the  frequency shift of the atomic mode  as the second order perturbative correction  to the difference between the energies of these lowest two states. The correction to the ground and first-excited  states $ \ket{0 }_A \ket{\cbkt{0}_k }$ and $ \ket{1 }_A \ket{\cbkt{0}_k }$ are given by $ \delta \omega_{A,0} ^{(2) }= -\hbar \sum_k \frac{\bkt{g_{k, \Phi} + g_{k,Q}}^2}{\omega_A '' + \omega_{k, \mr{JJA}}'}$ and $\delta \omega_{A,1} ^{(2) }= \hbar \sum_k \frac{\bkt{g_{k, \Phi} - g_{k,Q}}^2}{\omega_A '' - \omega_{k, \mr{JJA}}'}-2\hbar \sum_k \frac{\bkt{g_{k, \Phi} + g_{k,Q}}^2}{\omega_A '' + \omega_{k, \mr{JJA}}'}$, respectively. This yields the perturbative shift to the atomic mode as:

\eqn{
\label{deltawa2}
\Delta\omega_A^{(2)} =  \sum_{k }\sbkt{ \frac{\bkt{g_{k,\Phi}- g_{k,Q} }^2}{ \omega_A''- \omega'_{k, \mr{JJA}}}-\frac{\bkt{g_{k,\Phi}+ g_{k,Q} }^2}{ \omega_A''+\omega'_{k, \mr{JJA}} }}.
}

We remark here that the perturbation theory is performed by segregating the total Hamiltonian Eq.~\eqref{Ham} into   $H_0 \equiv \hbar \omega_A'' \hat{a}_A^\dagger \hat{a}_A$ as the unperturbed atomic Hamiltonian, and $ H_I \equiv \sum_k\sbkt{ \hbar g_{k,\Phi} \bkt{\hat a_A + \hat{a}_A ^\dagger}\bkt{\hat a_k + \hat{a}_k ^\dagger}\right.$ $\left.+ \hbar  g_{k,Q} \bkt{\hat a_A - \hat{a}_A ^\dagger}\bkt{\hat a_k - \hat{a}_k ^\dagger} }$ as the perturbative correction. Fig.~\ref{Fig:pert}~(a) shows the renormalized frequency with the above perturbative corrections as a function of the coupling parameter $\chi$. It can be seen that including  the perturbative corrections brings the renormalized frequency closer to the exact eigenmode frequency, and we see an agreement between $\omega_A'' + \Delta\omega_A ^{(2)}$ and the exact normal mode frequency for $\chi\lesssim 0.05$.

 Fig.~\ref{Fig:pert}~(c) shows the corresponding atomic eigenmode as a function of the array position and coupling parameter $\chi $. It can be seen that the atomic mode is  localized at the atomic position for coupling parameter $\chi\lesssim 0.05$, at which point we see a concomitant breakdown of perturbation theory in Fig.~\ref{Fig:pert}~(a) and (c).

The perturbative expression for the spontaneous emission decay rate of an atom coupled to continuum is given by~\cite{Nigg12}
\eqn{\label{gammaeff}
\Gamma_\mr{eff} = \frac{1}{2\pi C_A}\re\sbkt{\frac{1}{Z_\mr{eff} \bkt{\omega_A}}},
}
where $ Z_\mr{eff}\bkt{\omega_A}$ corresponds to the effective impedance of the environment at the atomic frequency (See Appendix~\ref{App:imp}  for a detailed derivation).  We remark that the effective impedance of the environment seen by the atom is different from that of an infinite JJA $Z_\infty$  (see Appendix~\ref{App:imp} for comparison), which yields a spontaneous emission rate of $\Gamma_\infty \equiv \frac{ 1}{2 \pi C_A} \re\sbkt{ \frac{1}{Z_\infty \bkt{\omega_A}}}.$

It can be seen from Fig.~\ref{Fig:pert}~(b) that for coupling parameter $\chi \lesssim 10^{-4}$, there is an agreement between the perturbative $\Gamma_\mr{eff}$ and the numerically obtained atomic decay $ \bkt{\im\sbkt{\tilde\omega_A}}$, though for larger coupling strengths the two differ significantly. We further note that the perturbative $\Gamma_\infty$ calculated with the infinite array impedance is appreciably different from both the exact decay rate as well as $\Gamma_\mr{eff}$. Fig.~\ref{Fig:pert}~(d) shows the atomic eigenmode for the bare atomic frequency $\omega_A/(2\pi) =  5$~GHz. We find that the atomic mode is mostly localized at the atomic position for $\chi \lesssim 10^{-4}$, and is delocalized over the entire array for larger coupling parameters. Specifically, the atomic mode is pinned to the spectrally-closest JJA mode corresponding to $k = 4$, as we will illustrate in the following subsection, and the decay rate saturates to the loss for that array mode for large $\chi$.

\subsection{Atomic mode in the spatial and spectral domain}
\label{Sec:Atommode}

\begin{figure}[t]
    \centering
    \subfloat{  \includegraphics[width = 3.3 in]{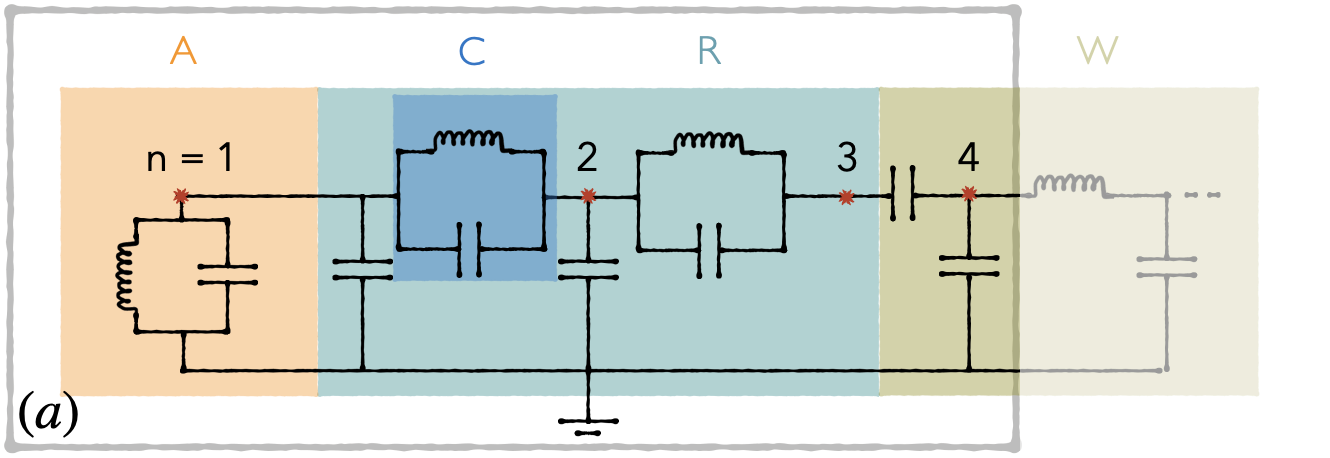}}\\
    \subfloat{\includegraphics[width = 3.4 in]{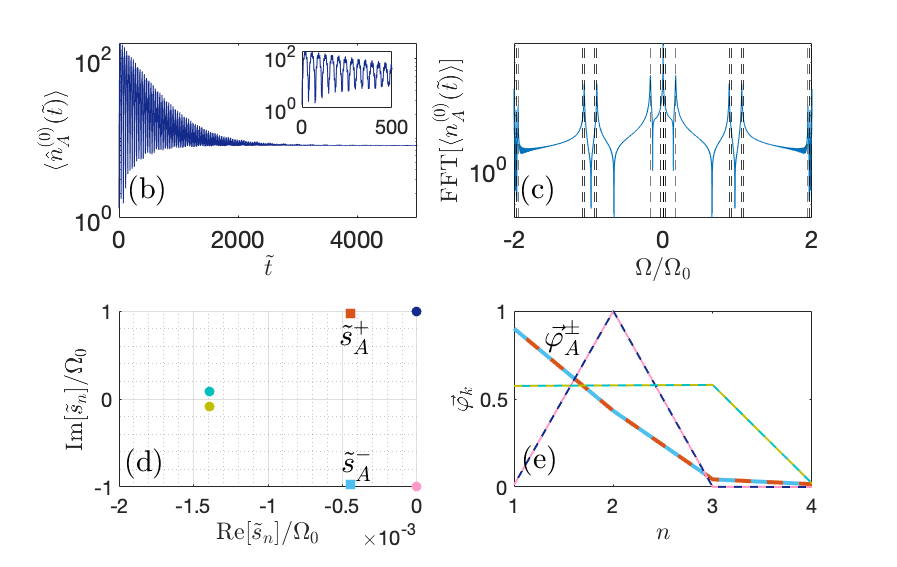}}
    \caption{(a) Schematic circuit for a toy model with an artificial atom (A)  coupled to an open single LC-resonator (R) via a coupler (C). The resonator is in turn coupled to a transmission line (W). (b) Dynamics of expected number of excitations at the atomic node for couplings $ \chi = 1$.  (c) The Fourier transform of $ n_A\bkt{\tilde t}$ exhibits resonances at the beat frequencies between the various poles in the system, as indicated by the dashed vertical lines. The pole values and the corresponding eigenmodes are shown in (d) and (e) respectively. The 
    bare atomic frequency is taken to be $\omega_A/\bkt{2\pi} \approx 5 $~GHz, and temperature $T = 50 $~mK.
    } 
    \label{Fig:fft}
\end{figure}
As seen in the previous section, atomic properties depend not only on the coupling parameter $\chi$ but also on the atomic frequency, and in particular whether it is within or outside the photonic frequency band. Realizing an artificial atom in cQED using a SQUID loop provides the advantage of being able to tune the atomic mode frequency \textit{in-situ}; we thus explore in this section how the spatial and spectral properties of the atomic mode vary as its bare frequency is tuned across the photonic band.

One can observe from Figs.~\ref{Fig:local}~(a)--(c) that for small coupling parameter $\chi$  as one tunes  the bare atomic frequency through the photonic band, the atomic mode goes through a series of avoided crossings, while for larger  $\chi$, the atomic frequency appears `pinned' to those of the array.  Outside  the photonic band, for small $\chi$ the atomic frequency is close to the bare atomic frequency with a negligible shift, while for  $\chi\gtrsim 1$ the atomic frequency exhibits a significant Lamb shift. Furthermore, it can be seen from Fig.~\ref{Fig:local}~(c) that for a galvanic coupling $(\chi = 1)$ the eigenvalues of the array modes change as we vary the bare atomic frequency within the band \cite{Kuzmin21}. For smaller values of $\chi$ (e.g., Fig.~\ref{Fig:local}~(a)),  while the atomic mode goes through a series of avoided crossings,  there is negligible effect of changing the atomic frequency on the eigenvalues of the JJA.

The corresponding atomic eigenmode is shown is Figs.~\ref{Fig:local}~(d)--(f), which shows the localization of the atomic mode.  To quantify  the hybridization of the atomic eigenmode with the modes of the JJA, we define the Inverse Participation  Ratio (IPR) as a measure of atomic mode localization as \cite{Shrinivasan03}
\eqn{\label{Eq:IPR}
\mr{IPR} = \sum_n \abs{\vec \varphi_A (n)}^4.
} 
We see from Fig.~\ref{Fig:local}~(d)  that for small $\chi $,  the atomic mode is spatially localized at the position of the atom except at the points of avoided crossings where it hybridizes strongly with the near-resonant modes of the JJA. For $\chi =1$, for the atomic frequency within the band the atomic mode is delocalized over the entire array, with the mode function corresponding to the near resonant eigenmodes of the array. For the bare atomic frequency  outside of the photonic band the atomic mode is  spatially localized at the atomic position as seen from Fig.~\ref{Fig:local}~(f). The localization increases near the band edge as the atom hybridizes with a larger number of modes close to the edge which can help create an effective localized mode near the position of the atom.

Having considered the radiative properties of the atom, we know turn to the open quantum system dynamics in the following section. 

\vspace{0.5 cm}
\section{Spontaneous Emission  Dynamics}
\label{Sec:spem}

We now consider the evolution of the number of excitations at the atomic node starting with the initial state $\rho(0) = \ket{1}_A\bra{1}_A \otimes \ket{\cbkt{0}}\bra{\cbkt{0}}$, such that the atomic node contains one excitation to start with and remainder of the nodes are in vacuum. Such an initial state can be prepared by driving the atomic node locally via an external drive to the first excited state and, if necessary, switching on the coupling between the atom and the JJA quickly compared to the time scales of the system dynamics. The number of excitations at the atomic node at any given time are obtained as:

\eqn{\label{Eq:nat}
&\avg{\hat n_A^{(0)}\bkt{\tilde t}} =\non\\
& \frac{1}{2\hbar Z_A}\sbkt{\avg{:\bkt{\hat\Phi_A^{(0)}\bkt{\tilde t}}^2:} + Z_A^2 \avg{:\bkt{\hat Q_A^{0}\bkt{\tilde t}}^2:}}.
}

Substituting  Eqs.~\eqref{Eq:phiext} and \eqref{Eq:qext}  in Eq.~\eqref{Eq:nat}, one can obtain the dynamics of the excitation number expectation value at the atomic node.
as follows (see Appendix~\ref{App:nAt} for details of the derivation):

\begin{widetext}
\eqn{\label{Eq:nAt}
&\avg{\hat {n}_A^{(0)}\bkt{\tilde t}} =  \frac{1}{2} \sum_{p,q,m,n} e^{\bkt{\tilde s_p + \tilde s_m}\tilde t}\tilde s_p \tilde s_m  C_{\mr{red}, q,1} C_{\mr{red}, n,1}   \sbkt{  \eta_{p,q}\bkt{\tilde s_p}  \eta_{m,n}\bkt{\tilde s_m}+\bkt{ \frac{Z_A}{Z_0}}^2\zeta_{p,q}\bkt{\tilde s_p}  \zeta_{m,n}\bkt{\tilde s_m}}\non\\
&+\frac{1}{2} \sum_{p,m} e^{\bkt{\tilde s_p + \tilde s_m}\tilde t}  \bkt{\frac{Z_0}{Z_A}}^2\sbkt{ \eta_{p,1}\bkt{\tilde s_p}  \eta_{m,1}\bkt{\tilde s_m}+\zeta_{p,1}\bkt{\tilde s_p}  \zeta_{m,1}\bkt{\tilde s_m}}\non\\
&+\frac{k_B TZ_0^2}{\hbar \Omega_0 Z_AZ_W}\sum_{p,m}\frac{ 1 }{\bkt{\tilde s_p + \tilde s_m }}\sbkt{  \eta_{p,N+2}\bkt{\tilde s_p}  \eta_{m,N+2}\bkt{\tilde s_m}+\bkt{ \frac{Z_A}{Z_0}}^2\zeta_{p,N+2}\bkt{\tilde s_p}  \zeta_{m,N+2}\bkt{\tilde s_m}} 
 \bkt{1 -e^{\bkt{\tilde s_p + \tilde s_m}\tilde t}  },
}
\end{widetext}

\begin{figure}[b]
    \centering
    \subfloat[]{\includegraphics[width = 3.4 in]{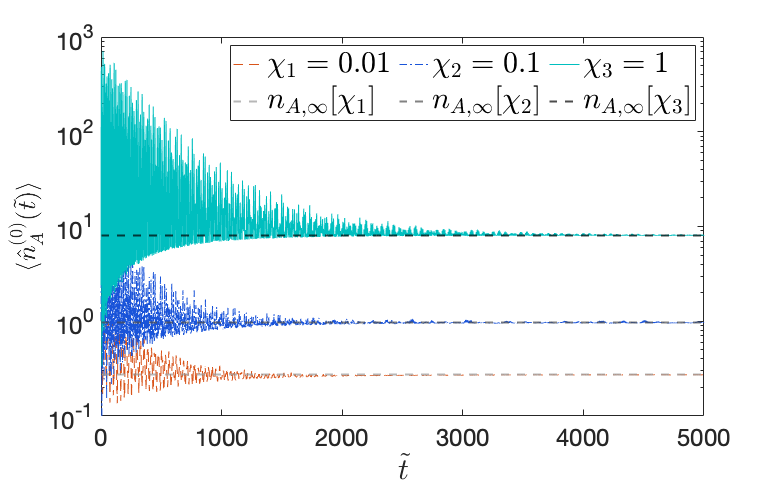}}\\
    \subfloat[]{\includegraphics[width = 3.4 in]{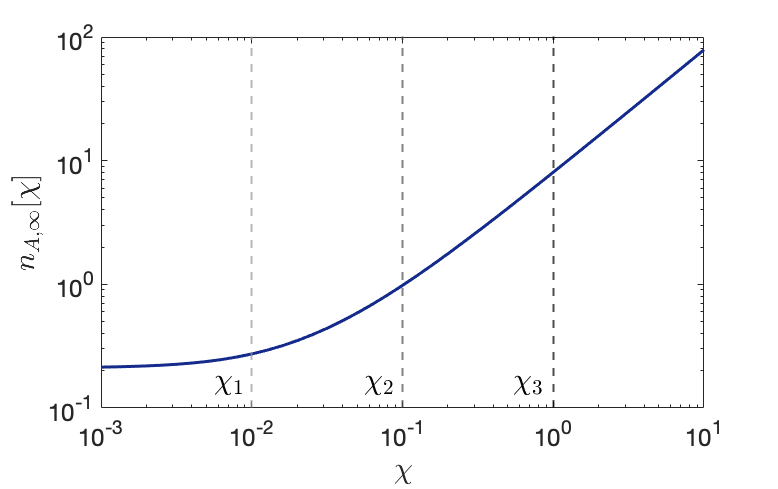}}
    \caption{(a) Dynamics of the number expectation value at the atomic node for  various coupling coefficients $\chi$ for an array with $N = 100$ junctions, with $\omega_A/\bkt{2\pi} = 5 $~GHz, and temperature $T = 50 $~mK. (b) Steady state excitation number values at the atomic node $n_{A,\infty}[\chi]$ as a function of the coupling strength $\chi$.
    } 
    \label{Fig:spem}
\end{figure}
with  $\eta_{p,q}$ and $\zeta_{m,n}$ defined by Eqs.~\eqref{Eq:eta} and \eqref{Eq:zeta}. The first two lines in the above equation represent the contribution from initial conditions, and the last line corresponds to the input noise from the waveguide.  The indices $p,m$ indicate the sum over  the various eigenmodes of the system.  We have  assumed here that the system is in the high temperature limit, such that  the noise correlation time is much smaller compared to the characteristic system relaxation time scale $\bkt{\frac{\hbar }{k_B T} \ll \Gamma^{-1} }$. The input noise from the transmission line can thus be approximated to be delta-correlated:

\eqn{\label{Eq:deltanoise}
\avg{:\hat Q_\mr{in}\bkt{t_1} \hat Q_\mr{in}\bkt{t_2}:} \rightarrow
 \frac{ k_B T }{2Z_W } \delta \bkt{t_1 - t_2}.}
We illustrate the dynamics of the number expectation value at the atomic position for in a simple system with an atom coupled to an open resonator, as shown in Fig.~\ref{Fig:fft}(a), corresponding to the $N =1$ limit of the JJA.  The dominant frequencies in the dynamics as obtained via the Fourier transform of the time domain signal in Fig.~\ref{Fig:fft}(c). The vertical dashed lines in Fig.~\ref{Fig:fft}(c) represent various beat frequencies obtained as $\im\tilde s_p + \im\tilde s_m$ for different poles $\tilde s_{p,m}$ in the system, shown in  Fig.~\ref{Fig:fft}(d). The eigenmodes corresponding to the various poles are shown in Fig.~\ref{Fig:fft}(e).

Fig.~\ref{Fig:spem}(a) shows the dynamics of the number expectation value at the atomic position for different coupling coefficients in the non-perturbative regime for a larger array with N = 100 junctions. We note from Fig.~\ref{Fig:spem}(b)   that as the coupling strength is increased, the system decays into a steady state with increasingly  larger number of  excitations. The steady state excitation number $n_{A,\infty}\sbkt{\chi}\equiv \lim_{\tilde t\rightarrow\infty} \avg{\hat n_A^{(0) }\bkt{\tilde t}}$ can be obtained from Eq.~\eqref{Eq:nAt} as:
\eqn{\label{Eq:ninf}
n_{A,\infty}\sbkt{\chi}\rightarrow &\frac{k_B TZ_0^2}{\hbar \Omega_0 Z_AZ_W}\sum_{p,m}\frac{ 1 }{\bkt{\tilde s_p + \tilde s_m }}\non\\
&\sbkt{\eta_{p,N+2}\bkt{\tilde s_p}  \eta_{m,N+2}\bkt{\tilde s_m}\right.\non\\
&\left.+\bkt{ \frac{Z_A}{Z_0}}^2\zeta_{p,N+2}\bkt{\tilde s_p}  \zeta_{m,N+2}\bkt{\tilde s_m}} .
}
The increase in the steady state occupation of the excitation number at atomic node with $\chi$ can be attributed to the fact that the number non-conserving non-RWA terms become prominent in the non-perturbative regime.

\subsection{Including atomic nonlinearity}

We rewrite the equations of motion for the reduced system subspace including the nonlinear potential for the atomic junction, using \eqref{eq:dphiexdt} and \eqref{eq:dqexdt}, as follows:

\eqn{\label{eq:phinl}
\der{}{\tilde t} \hat{\mb{\Phi}}_\mr{red} =& \dbar{C}_{\mr{red}}^{-1}\sbkt{Z_0 \hat{\mb{Q}}_\mr{red}}\\
\label{eq:qnl}
\der{}{\tilde t}\sbkt{Z_0\hat{\mb{Q}}_\mr{red}} =& -  \dbar{L}_{\mr{red}}^{-1} \hat{\mb{\Phi}}_\mr{red}-  \frac{Z_0}{Z_W} \dbar{\delta}_{N+2}\dbar{C}_\mr{red}^{-1} \sbkt{Z_0\hat{\mb{Q}}_\mr{red}} \non\\
&+ \frac{2}{\Omega_0^2} \sbkt{Z_0\hat Q_\mr{in} }\textbf{\textdelta}_{N+2} +  \frac{1}{\Omega_0}\underbrace{ \pard{\mc{U}_A\bkt{\Phi_A}}{{\Phi}_A}}_\text{Nonlinear source}.
}

We assume the perturbative solutions up to first order in atomic nonlinearity to be:  \eqn{ \hat{\Phi}_{\mr{red},j}\bkt{t} &= \hat {\Phi}_{\mr{red},j}^{(0)}\bkt{t} + \lambda  \hat {\Phi}_{\mr{red},j}^{(1)}\bkt{t}\\
\hat{Q}_{\mr{red},j}\bkt{t} &= \hat {Q}_{\mr{red},j}^{(0)}\bkt{t} +\lambda  \hat {Q}_{\mr{red},j}^{(1)}\bkt{t},} 
where $\cbkt{\hat{\mb{\Phi}}_\mr{red}^{(0)},\hat{\mb{Q}}_\mr{red}^{(0) }}$ represents the solution to the linear problem as derived in Eqs.~\eqref{Eq:phiext} and \eqref{Eq:qext},   $\cbkt{\hat{\mb{\Phi}}_\mr{red}^{(1)},\hat{\mb{Q}}_\mr{red}^{(1) }}$ represents the first order perturbative correction and $\lambda $ is the perturbative parameter that scales with the strength of nonlinearity.   Substituting the perturbative ansatz into the nonlinear equations of motion (Eqs.~\eqref{eq:phinl} and \eqref{eq:qnl}), one can obtain the dynamics of the perturbative corrections to the flux and charge dynamical variables as:
\eqn{
\der{}{\tilde t} \hat{\mb{\Phi}}_\mr{red}^{(1)} &=  \dbar{C}_{\mr{red}}^{-1}\sbkt{Z_0 \hat{\mb{Q}}_\mr{red}^{(1)}}\\
\der{}{\tilde t}\sbkt{Z_0\hat{\mb{Q}}_\mr{red}^{(1)}} &= - \dbar{L}_{\mr{red}}^{-1} \hat{\mb{\Phi}}_\mr{red}^{(1)}-  \frac{Z_0}{Z_W} \dbar{\delta}_{N+2}\dbar{C}_\mr{red}^{-1} \sbkt{Z_0\hat{\mb{Q}}^{(1)}_\mr{red}}\non\\
&+  \frac{1}{\lambda\Omega_0} \dbar{\delta}_1 \bkt{\pard{\mc{U}_A\bkt{\Phi_A}}{\Phi_A}}.
}
We note that the homogeneous part of the above equations of motion is the same as Eqs.\,\eqref{eq:dphiexdt} and \eqref{eq:dqexdt}, thus corresponding to the same propagator $\bkt{\dbar{G} \bkt{\tilde s}}$ as in the linear problem.  It is pertinent to remark here that a Kerr-type nonlinearity gives rise to secular terms, which can be addressed by a multi-scale perturbation theory~\cite{Malekakhlagh16}. For an odd-order nonlinearity, such as a cubic potential realizable e.g. with a SNAIL-based atom \cite{Frattini2017}, one can obtain lowest order nonlinear  corrections to the dynamics as follows:

\eqn{
\label{phi1t}
&\hat{\Phi}^{(1)} _{\mr{red}, j} \bkt{\tilde t}
 =\non\\
 &\sum_{p,q} e^{\tilde s_p \tilde t}  \frac{1}{\bkt{\frac{\delta\gamma_{pp} \bkt{\tilde s_p}}{\delta \tilde s}}}\beta^{-1}_{j,p}\bkt{\tilde s_p}  \alpha^{-1}_{p,q}\bkt{\tilde s_p} \hat{Y}_{q}^{(1)}\bkt{\tilde s_p},\\
 \label{q1t}
 &Z_0\hat{{Q}}^{(1)}_{\mr{red},j} \bkt{\tilde t} =\non\\
 &\sum_{p,q,r} \tilde s_p e^{\tilde s_p \tilde t}  \frac{1}{\bkt{\frac{\delta\gamma_{pp} \bkt{\tilde s_p}}{\delta \tilde s}}}C_{\mr{red},j,q}\beta^{-1}_{q,p}\bkt{\tilde s_p}  \alpha^{-1}_{p,r}\bkt{\tilde s_p} \hat{Y}_{r}^{(1)}\bkt{\tilde s_p}.
}
where $ \hat{\mb{Y}}^{(1)}\bkt{\tilde s} \equiv \int \dd\tilde t~ e^{- \tilde s \tilde t}\sbkt{\frac{1}{\lambda} \dbar{\delta}_1 \bkt{\pard{\mc{U}_A\bkt{\Phi_A}}{\Phi_A}}}$ corresponds to the nonlinear source term.

We can thus obtain the nonlinear corrections to the number expectation value at the atomic node as:

\eqn{\label{nat1}
&\avg{\hat n_A^{(1)}\bkt{\tilde t}} = \non\\
&\frac{\lambda^2}{2\hbar Z_A}\sbkt{\avg{:\bkt{\hat \Phi_A^{(1)}\bkt{\tilde t}}^2:} + Z_A^2 \avg{:\bkt{\hat Q_A^{(1)}\bkt{\tilde t}}^2:}},
}
The lowest order contribution is at the second order in nonlinearity. For a cubic nonlinearity, the above expression can be evaluated as described in Appendix~\ref{App:NL}.

\section{Discussion}
\label{Sec:Diss}
We have analyzed the  radiative properties and open system  dynamics of an artificial atom coupled to a high impedance JJA. We study the crossover from a perturbative to a non-perturbative regime of light-matter interaction,  considering a tunable  coupler between the atom and the JJA that allows one to isolate the atom from the array modes. We develop  a singular function expansion approach to describe the  atom+JJA system in Sec.~\ref{Sec:EOM}, which allows one to analyze the properties and dynamics of the system in terms of its non-Hermitian eigenmodes. The dissipation and noise from  system-bath interaction are  accounted for by eliminating the waveguide modes via appropriate boundary conditions without requiring a full consideration of the waveguide Hilbert space, thereby making the approach computationally efficient.  We derive an effective Hamiltonian to describe the  closed  atom+JJA system in Sec.~\ref{Sec:Ham}, delineating different regimes of coupling strengths realizable in the system. It is shown that the system can exhibit multimode nonperturbative coupling strengths between the atomic and the resonator modes  (Fig.~\ref{Fig:gk}).  Sec.~\ref{Sec:Rad} discusses a scenario of large coupling where the Lamb shift and Purcell decay in such a system can no longer be described via a perturbative approach (Fig.~\ref{Fig:pert}). We  define and identify the atomic eigenmode across different coupling regimes, and discuss its qualitative behavior in terms of the spatial and spectral  properties in  Sec.~\ref{Sec:Atommode}. In multimode non-perturbative coupling regimes, it can be seen that, the  atomic mode is no longer spatially or spectrally localized due to a strong hybridization between the atomic and field modes (Fig.~\ref{Fig:local}). Finally we illustrate the spontaneous emission dynamics of such a system in the non-perturbative regime in Sec.\,\ref{Sec:spem}. It is found that there is a significant contribution from the non-RWA terms to the steady state occupation of the atomic node as the light-matter coupling becomes non-perturbatively strong (Fig.~\ref{Fig:spem}).

This work opens several new directions to explore with regard to  fluctuation phenomena in high impedance environments in non-perturbative regimes of 1+1 dimensional QED. In the presence of strongly hybridized matter and field degrees of freedom, the quantum vacuum fluctuations are also hybridized, and can lead to non-perturbatively strong dispersive and dissipative effects as we have shown in this work.  Previous experiments have explored Lamb shifts and dynamical Casimir effects in cQED setups  \cite{Wen19, Nation12}; it would be interesting to extend such studies to strongly hybridized regimes and analyze the non-perturbative effects therein.

The dynamics of the atom in such a system can be highly non-Markovian as a result of several factors coming into play \cite{Breuer16, deVega17, Malekakhlagh16b} -- particularly, a nonperturbative multimode strong coupling between the atom and its environment, going beyond the multimode strong coupling regime of cavity QED \cite{Meiser06, Krimer14}.  We show that such non-Markovian effects manifest themselves as a multi-exponential ocillatory decay of the atomic mode, where the individual exponents can be related to a set of discrete complex-valued poles that correspond to the eigenfrequencies of the non-Hermitian modes of the system. Understanding the non-Markovian dynamics in terms of these eigenmodes can offer insights into how excitations and coherences evolve in such a non-perturbative multimode regime of light-matter interaction.

In the presence of a drive, the non-linearity of the atom can result in rich dynamical behavior such as bistability and self-oscillations, leading to generation of frequency combs~\cite{Khan18}. While this dynamical instability has been observed for a single mode environment~\cite{Lu2021}, JJA arrays provide an ideal platform for its study in a multimode setting, where quantum features such as multipartite entanglement and soliton formation may prevail. It has also been discussed that in the presence of strong hybridization the atomic non-linearity can be transferred to the JJA modes \cite{Leger19}, diluting the effect of the atomic non-linearity as predicted in Ref.~\cite{Malekakhlagh16b} (see Sec.~\ref{Sec:pert}), an effect that requires further careful theoretical analysis.

Additionally, JJAs exhibit several interesting properties as an optical medium. Through their strong  non-linearities and large or negative refractive index \cite{Zueco12}, they can serve as a platform to study electromagnetic phenomena such as slow and stopped light \cite{Shen07} in new regimes. It has also been proposed that the optical properties of the JJAs can be dynamically controlled by quantum coherent states of qubits coupled to them \cite{Rakhmanov08}. The tunability of the individual junctions forming the JJA can be used to design the spectral properties of the JJA modes and engineer band gaps \cite{Hutter11}. Thus coupling an artificial atom to an optical medium that possesses a great degree of tunability and inherent quantum non-linearity provides for several opportunities for exploring and understanding novel quantum optical phenomena.

\section{Acknowledgments}
We are grateful to  Kaan G\"{u}ven, Archana Kamal, Alicia Koll\'{a}r, Roman Kuzmin,  S\'{e}bastien L\'{e}ger, Vladimir Manucharyan and Nicolas Roch for insightful discussions. We acknowledge support from the US Department of Energy, Office of Basic Energy Sciences, Division of Materials Sciences and Engineering, under Award No. DESC0016011. The simulations presented in this article were performed on computational resources managed and supported by Princeton Research Computing, a consortium of groups including the Princeton Institute for Computational Science and Engineering (PICSciE) and the Office of Information Technology's High Performance Computing Center and Visualization Laboratory at Princeton University.\\

\appendix

\section{JJA impedance}
\label{App:imp}
One can write the effective impedance seen by the atom as a cumulative sum of the impedance of all junctions of the array plus the external waveguide \cite{Pozar}.  The external impedance is given as
\eqn{
Z_\mr{ext}\bkt{\omega}  = Z_\mr{TL} + \frac{1}{i \omega C_c},
}
where $Z_\mr{TL} = 50~\Omega$ is the impedance of the transmission line. We add the impedances of the junctions in the array successively by defining the effective impedance after adding $n$ junctions as $ Z_{\mr{eff}  }^{(n)}(\omega)$, such that
\eqn{\label{zeff}
Z_{\mr{eff}}^{(n)}\bkt{ \omega} = Z_\mr{LC} + \sbkt{\frac{1}{Z_g} +\frac{ 1}{Z_{\mr{eff}}^{(n-1)}\bkt{ \omega}}}^{-1},
}
where we start with $Z_{\mr{eff}}^{(0)} (\omega) \equiv Z_\mr{ext}(\omega)$. The impedance seen by the atom is \eqn{Z_{\mr{eff}}^{(N)} \bkt{\omega}= \frac{Z_\mr{LC} }{\chi} + \sbkt{ \frac{1}{Z_g} + \frac{1}{Z_{\mr{eff}}^{(N-1)} ( \omega)}}^{-1} , } 
where we note that the impedance of the coupler is $ Z_\mr{LC}/\chi$.

\begin{figure}[t]
    \centering
    {\includegraphics[width = 3.4 in]{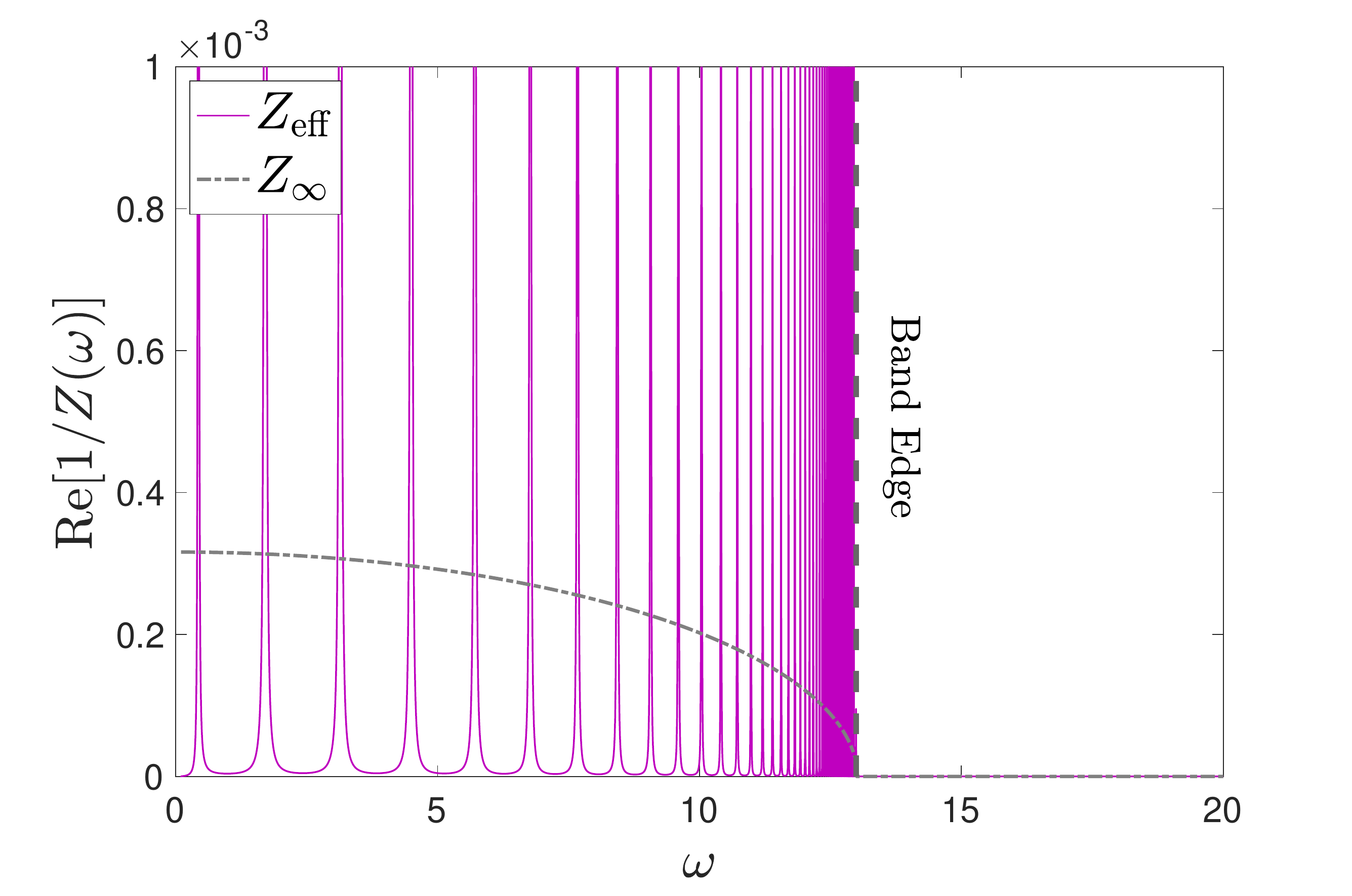}}
    \caption{Real part of the admittance of the array as a function of the atomic frequency. The solid line represents the cumulative impedance of the JJA $Z_\mr{eff}^{(n)}$ as given by Eq.~\eqref{zeff}. The gray dashed-dotted line represents the impedance of the infinite array $Z_\infty$ as given by Eq.~\eqref{zinf}.
    }
    \label{Fig:zinf}
\end{figure}

Fig.~\ref{Fig:zinf} compares the real part of the effective admittance $1/Z_{\mr{eff}}^{(N)} ( \omega)$ with that of an infinite JJA.  It can be seen that for the atomic frequency within the photonic band the effective admittance seen by the atom exhibits an oscillatory behavior owing to the resonances of the multimode cavity. Outside of the photonic band we see that both $1/Z_\mr{eff}^{N}$ and $ Z_\infty$ vanish.

\section{Waveguide dissipation and noise}
\label{App:SFM}

\subsection{Incoming and Outgoing waveguide modes}
\label{App:wgin}
We define the bosonic operators associated with the waveguide modes as \cite{ClerkRMP}:
\eqn{
\hat{b}_k \bkt{t} \equiv &\frac{1}{\sqrt{2 \hbar N_W}} \sum_{n = 0}^{N_W} e^{-i k n}\sbkt{ i \sqrt{Z_W} \hat Q_n^W + \frac{\abs{k}}{\sqrt{Z_W}} \hat{\Phi}_n^W}\\
\hat{b}_k^\dagger \bkt{t}\equiv &\frac{1}{\sqrt{2 \hbar N_W}} \sum_{n = 0}^{N_W} e^{i k n}\sbkt{ -i \sqrt{Z_W} \hat Q_n^W + \frac{\abs{k}}{\sqrt{Z_W}} \hat{\Phi}_n^W}.
}
One can note from the above that the bosonic operators $\hat b_k$ and $\hat b_{k'}^\dagger$ satisfy the commutation relations:
\eqn{
\sbkt{\hat b_k, \hat b_{k'}^\dagger } = \abs{k}\delta_{k,k'}
}
The waveguide node flux and charge variables can be defined in terms of  incoming and outgoing set of modes as follows:

\eqn{
\hat \Phi_n^W\bkt{t} =&\hat  \Phi_n^{W+}\bkt{t} +\hat  \Phi_n^{W-} \bkt{t}\\
\hat Q_n^W\bkt{t} =& \hat Q_n^{W+}\bkt{t} + \hat Q_n^{W-} \bkt{t},
}
where
\eqn{
\label{eq:phiwpm}
&\hat \Phi_n^{W\pm}\bkt{t} \non\\
&\equiv\sqrt{\frac{\hbar Z_W}{2N_W}}\sum_{k>0} \frac{1}{\abs{k}} \sbkt{\hat b_{\pm k}\bkt{0} e^{-i \bkt{\omega t \mp kn }} + \hat b_{\pm k}^\dagger\bkt{0} e^{i \bkt{\omega t \mp kn }}}\\
\label{eq:qwpm}
&\hat Q_n^{W\pm}\bkt{t} \non\\
&\equiv-i\sqrt{\frac{\hbar }{2Z_WN_W}}\sum_{k>0}  \sbkt{\hat b_{\pm k}\bkt{0} e^{-i \bkt{\omega t \mp kn }} - \hat b_{\pm k}^\dagger \bkt{0} e^{i \bkt{\omega t \mp kn }}},
}
such that $\omega $ and $k$ are related by the waveguide dispersion relation:  \eqn{\label{Eq:WGdisp}
\frac{\omega^2}{\Omega_{W}^2}  =  2 (1-\cos k),
}
where $\Omega_W = 1/\sqrt{L_WC_W}$ corresponds to the plasma frequency of the transmission line.

We write the Heisenberg equations of motion at the first  waveguide node as follows:
\eqn{\label{dq0wdt}\der{\hat Q_0 ^W}{t} = & -\frac{1}{L_W} \bkt{\hat \Phi_0 ^W - \hat \Phi_1 ^W}\\
  \bkt{C_W + C_c} \der{\hat \Phi_0 ^W}{t} - C_c \der{\hat \Phi_N}{t}=&\hat Q_0 ^W.
  }
  
\begin{widetext}
  We note that the only coupling between the first waveguide node and the rest of the waveguide is inductive. In order to eliminate the remainder of waveguide, we expand the RHS of  Eq.~\eqref{dq0wdt} in terms of the incoming and outgoing modes (Eq.~\eqref{eq:phiwpm}) as follows:
  \eqn{
  \der{\hat Q_0 ^W}{t} =  \frac{1}{L_W} \sqrt{\frac{\hbar Z_W }{2N_W}}&\sbkt{\sum_{k>0}\frac{1}{\abs{k}} \cbkt{\hat{b}_k e^{-i \bkt{\omega t -k}} - \hat{b}_k^\dagger e^{i \bkt{\omega t -k} }}+\sum_{k>0}\frac{1}{\abs{k}} \cbkt{\hat{b}_{-k} e^{-i \bkt{\omega t +k}} - \hat{b}_{-k}^\dagger e^{i \bkt{\omega t +k} }}\right.\non\\
  &\left. -\sum_{k>0}\frac{1}{\abs{k}} \cbkt{\hat{b}_k e^{-i \omega t } - \hat{b}_k^\dagger e^{i \omega t  }}-\sum_{k>0}\frac{1}{\abs{k}} \cbkt{\hat{b}_{-k} e^{-i \omega t } - \hat{b}_{-k}^\dagger e^{i \omega t } }}\\
  =  \frac{1}{L_W} \sqrt{\frac{\hbar Z_W }{2N_W}}&\sbkt{\sum_{k>0}\frac{1}{\abs{k}} \cbkt{\hat{b}_k e^{-i \omega t }\bkt{e^{ik}-1} - \hat{b}_k^\dagger e^{i \omega t } \bkt{e^{-ik} -1}}\right.\non\\
  &\left.+\sum_{k>0}\frac{1}{\abs{k}} \cbkt{\hat{b}_{-k} e^{-i \omega t } \bkt{e^{-ik}-1} - \hat{b}_{-k}^\dagger e^{i \omega t  } \bkt{e^{ik}-1}}}\non\\
    = \frac{1}{L_W} \sqrt{\frac{\hbar Z_W }{2N_W}}&\sbkt{\sum_{k>0}\frac{1}{\abs{k}} \cbkt{\hat{b}_k e^{-i \omega t }\bkt{\cos k-1 +i \sin k} - \hat{b}_k^\dagger e^{i \omega t } \bkt{\cos k -1 - i \sin k}}\right.\non\\
 & \left.+\sum_{k>0}\frac{1}{\abs{k}} \cbkt{\hat{b}_{-k} e^{-i \omega t } \bkt{\cos k-1 - i \sin k} - \hat{b}_{-k}^\dagger e^{i \omega t  } \bkt{\cos k-1 + i \sin k}}}
  }
  We use the dispersion relation for the waveguide (Eq.~\eqref{Eq:WGdisp}), keeping terms up to lowest order in $\omega/\Omega_W$ in the continuum limit of the waveguide,  to simplify the above as:
  \eqn{
  \der{\hat Q_0 ^W}{t} = & \frac{1}{L_W} \sqrt{\frac{\hbar Z_W }{2N_W}}\sbkt{\sum_{k>0}\frac{1}{\abs{k}}\bkt{\frac{i\omega}{\Omega_W}} \cbkt{\hat{b}_k e^{-i \omega t } + \hat{b}_k^\dagger e^{i \omega t } }+\sum_{k>0}\frac{1}{\abs{k}}\bkt{-\frac{i\omega}{\Omega_W}} \cbkt{\hat{b}_{-k} e^{-i \omega t }  + \hat{b}_{-k}^\dagger e^{i \omega t  } }}\\
  = &-\frac{1}{Z_W} \bkt{\der{\hat \Phi_0^{W+}}{t} - \der{\hat \Phi_0^{W-}}{t}}\\
  =& \underbrace{- \frac{1}{Z_W} \der{\hat\Phi_0 ^W}{t }}_{\mr{Dissipation}} +\underbrace{2\hat Q_\mr{in}}_{\mr{Noise}},
  }
where we have defined $\hat{Q}_\mr{in}\bkt{t} \equiv \Omega_W  \hat Q_0^{W-}\bkt{t}$. The first and the second terms in the above equation correspond to the dissipation and noise, respectively. We thus obtain the equations of motion in the reduced subspace as in Eqs.~\eqref{eq:dphiexdt} and \eqref{eq:dqexdt}.

\subsection{Noise correlation}
\label{App:noise}
We evaluate the noise correlation function between the input noise at two times as follows:
\eqn{
\avg{:\hat Q_\mr{in}\bkt{t_1} \hat Q_\mr{in}\bkt{t_2}:} =& -\frac{\hbar\Omega_W^2 }{2Z_W N_W}\sum_{k,k'}\avg{:\sbkt{ \hat b_{-k }e^{-i\omega_k t_1} - \hat b_{-k }^\dagger e^{i\omega_k t_1} } \sbkt{ \hat b_{-k' }e^{-i\omega_k' t_2} - \hat b_{-k' }^\dagger e^{i\omega'_k t_2} }:}\\
= &\frac{\hbar\Omega_W }{Z_W N_W}\sum_k \frac{\omega_k}{e^{\hbar  \omega_k/\bkt{k_B T}} - 1} \cos\sbkt{\omega_k \bkt{t_1 - t_2}}\\
\approx&  \frac{\hbar }{2\pi Z_W} \int
_0 ^\infty \dd \omega_k \frac{\omega_k}{e^{\hbar \omega_k /\bkt{k_B T}} -1} \cos \sbkt{ \omega_k \bkt{t_1 - t_2}}\\
= &\frac{\hbar}{4\pi Z_W } \sbkt{\frac{1}{\bkt{t_1 - t_2}^2} - \frac{\pi^2\text{cosech}^2\bkt{\frac{\pi k_B T \bkt{t_1 - t_2}}{\hbar}} }{\bkt{\hbar /(k_B T)}^2 }}
}
where we have assumed  continuum limit for the waveguide. This can be rewritten as Eq.~\eqref{eq:q1q2}.

\section{Solving equations of motion in reduced subspace}
\label{App:Laplace}

Let us consider the Laplace transform of the equations of motion in Eq.~\eqref{eq:dphiexdt} and \eqref{eq:dqexdt} as follows:

\eqn{
\label{eq:lapdphidt}
\tilde s \tilde{\mb{\Phi}}_\mr{red}\bkt{\tilde s} &= \hat{\mb{\Phi}}_\mr{red}(0) +  \dbar{C}_{\mr{red}}^{-1}\sbkt{Z_0 \tilde{\mb{Q}}_\mr{red}\bkt{\tilde s}}\\
\label{eq:lapdqdt}
\tilde s\sbkt{Z_0\tilde{\mb{Q}}_\mr{red}\bkt{\tilde s}} &=Z_0\hat{\mb{Q}}_\mr{red}\bkt{0} -  \dbar{L}_{\mr{red}}^{-1} \tilde{\mb{\Phi}}_\mr{red}\bkt{\tilde s}- \frac{Z_0}{Z_W} \dbar{\delta}_\mr{N+2}\dbar{C}_\mr{red}^{-1} \sbkt{Z_0\tilde{\mb{Q}}_\mr{red}\bkt{\tilde s}} + \frac{2}{\Omega_0} \sbkt{Z_0 \tilde{Q}_\mr{in} \bkt{\tilde s}} \textbf{\textdelta}_{N+2},
}
where we have defined $ \tilde {\mc{O}}\bkt{\tilde s}\equiv \int _0 ^\infty \dd \tilde t e^{- \tilde s \tilde t} \hat {\mc{O}}\bkt{\tilde t}$ as the Laplace transform of the operator $\hat {\mc{O}}\bkt{\tilde t}$.

One can use Eq.~\eqref{eq:lapdqdt} to express $Z_0\tilde {\mb{Q}}_\mr{red}\bkt{\tilde s}$ in terms of $ \tilde {\mb{\Phi}}_\mr{red}\bkt{\tilde s}$ and the initial conditions and noise as:
\eqn{
\label{Eq:tildeQs}
Z_0\tilde {\mb{Q}}_\mr{red}\bkt{s} = \sbkt{ \tilde s + \frac{Z_0}{Z_W} \dbar{\delta}_{N+2} \dbar{C} _{\mr{red}}^{-1}}^{-1} \sbkt{ Z_0 \hat{\mb{Q}}_\mr{red} \bkt{0} - \dbar{L}^{-1}_\mr{red} \tilde {\mb\Phi}_\mr{red} \bkt{\tilde s}+\frac{2}{\Omega_0}\sbkt{ Z_0  \tilde {Q}_\mr{in}\bkt{\tilde s}}\textbf{\textdelta}_{N+2}}.
}
Substituting above in Eq.~\eqref{eq:lapdphidt}, we obtain:

\eqn{
\label{Eq:Phis}
\tilde {\mb{\Phi}}_\mr{red}\bkt{\tilde s} = \sbkt{\tilde s^2 \dbar{C}_\mr{red} + \tilde s \frac{Z_0}{Z_W} \dbar{\delta }_{N+2} + \dbar{L}_\mr{red}^{-1}}^{-1}\sbkt{\cbkt{\tilde s \dbar{C}_\mr{red} + \frac{Z_0}{Z_W} \dbar \delta_{N+2}}\hat{\mb{\Phi}}_\mr{red}\bkt{0} + Z_0 \hat{\mb{Q}}_\mr{red}\bkt{0}+ \frac{2}{\Omega_0 } \sbkt{Z_0 \tilde Q_\mr{in}\bkt{\tilde s}} \textbf{\textdelta}_{N+2}}.
}
\end{widetext}

Taking the inverse Laplace transform of the above equation we obtain Eq.~\eqref{phiext}.

\section{JJA Eigenvalues and Eigenmodes}
\label{App:JJAeig}
\subsection{Plane-wave basis}
Let us consider the eigenvalue problem Eq.~\eqref{jjaep} for partial diagonalization of the JJA, expressing the eigenmodes in a plane-wave basis $\mb{\Phi}_{k, \mr{JJA}} = \sum_ {p = - \bkt{N-1}/2 }^{\bkt{N-1}/2} c_{kp} \mb{\phi}_p,$
such that  $ \phi_p (n) = \frac{1}{\sqrt{N}} e^{2\pi i n p /N}$. One can rewrite such a plane-wave expansion of the eigenmodes in a matrix representation  as 
\eqn{\mb{\Phi}_{k, \mr{JJA}} = \dbar\phi \mb{c}_k,} where $\dbar \phi \equiv \cbkt{ \mb \phi _{-\bkt{N-1}/2}\dots \mb \phi _{\bkt{N-1}/2}}$ and $\mb{c}_k \equiv \cbkt{c_{k\bkt{-\bkt{N-1}/2}},\dots c_{k\bkt{-\bkt{N-1}/2}}}^T $. One can thus substitute the plane-wave expansion in the eigenvalue problem Eq.~\eqref{jjaep} to obtain
\eqn{\omega_{k, \mr{JJA}}^2 \bar{C}_\mr{JJA} \mb{c}_k =  \bar{L}^{-1}_\mr{JJA} \mb{c}_k,}
where we have defined $ \bar M\equiv \dbar \phi ^\dagger \dbar{M} \dbar{\phi}$.

It can be shown that
\eqn{
\bkt{\bar{L}^{-1}_\mr{JJA}}_{k,k'} =& \frac{1}{L_k}\delta_{k,k'}  + \mc{B}_L,\\
\bkt{\bar C_{\mr{JJA}}}_{k,k'} = &C_k\delta_{k,k'} + \mc{B}_C,
}
where the terms $\mc{B}_{L, C}\sim 1/N$  correspond to the boundary contributions and  the effective inductance and capacitance values  corresponding to the JJA modes are given as 
\eqn{ \label{ck}
C_{k} \equiv& {C_g} + 2C \sbkt{1- \cos \bkt{\frac{2\pi k }{N}}}\\
\label{lk}
L_{k}\equiv &\frac{L}{2\bkt{1- \cos \bkt{\frac{2\pi k }{N}}}}.}
 We can thus arrive at the dispersion relation in the limit of large $N$
\eqn{\label{wkjja}
\omega_{k, \mr{JJA}} \approx \frac{1}{\sqrt{L_k C_k} } = \Omega_0 \sqrt{\frac{1 - \cos\bkt{\frac{2 k \pi}{N}}}{\frac{C_g}{2C} + 1 - \cos\bkt{\frac{2 k \pi}{N}}}}.
}

Furthermore, considering the diagonal forms of the matrices  $\bar {L}^{-1}_\mr{JJA}$ and  $\bar {C}_\mr{JJA}$, it can be shown that  
\eqn{c_{kk' } \begin{cases}
    \neq 0 ,& \text{if } k' = \pm k\\
    =0,              & \text{otherwise}.
\end{cases}
}
This allows us to write the eigenmodes as
\eqn{ \label{planewaveapp}
\Phi_{k,\mr{JJA}} (n) \approx c_{+k } \phi_{+k}(n)+ c_{-k } \phi_{-k}(n)
}

We now apply the boundary conditions
\eqn{\label{leftBC}
&\omega_{k, \mr{JJA}}^2 \sbkt{ \bkt{C + C_g + C_0 } \Phi_{k, \mr{JJA}} (1) - C \Phi_{k, \mr{JJA}} (2)}\non\\
&=\bkt{\frac{1}{L } + \frac{1}{L_0}}\Phi_{k, \mr{JJA}} \bkt{1}-\frac{1}{L}\Phi_{k, \mr{JJA}} \bkt{2}}
\eqn{
\label{rightBC}
&\omega_{k, \mr{JJA}}^2 \sbkt{ \bkt{C + C_g + C_c } \Phi_{k, \mr{JJA}} (N) - C \Phi_{k, \mr{JJA}} (N-1)}\non\\
&= \frac{1}{L } \Phi_{k, \mr{JJA}} \bkt{N}-\frac{1}{L}\Phi_{k, \mr{JJA}} \bkt{N-1}.
}

Substituting the plane-wave decomposition of the flux vector in Eq.~\eqref{leftBC} we obtain
\begin{widetext}
\eqn{
\frac{c_{-k}}{c_{+k}} = - \frac{ \bkt{C + C_g + C_0 }\omega_{k, \mr{JJA}}^2 e^{2\pi i k /N} - C\omega_{k, \mr{JJA}}^2 e^{4\pi i k /N} - \bkt{\frac{1}{L} + \frac{1}{L_0 }} e^{2\pi i k /N} +\frac{1}{L} e^{4\pi i k /N}}{ \bkt{C + C_g + C_0 }\omega_{k, \mr{JJA}}^2 e^{-2\pi i k /N} - C\omega_{k, \mr{JJA}}^2 e^{-4\pi i k /N} - \bkt{\frac{1}{L} + \frac{1}{L_0 }} e^{-2\pi i k /N} +\frac{1}{L} e^{-4\pi i k /N}}
}.

We note from the above that $\abs{c_{+k }} = \abs{c_{-k}}$. From the right boundary condition Eq.~\eqref{rightBC} one further has that
\eqn{
\frac{c_{-k}}{c_{+k}}= - \frac{ \bkt{C + C_g + C_c }\omega_{k, \mr{JJA}}^2 e^{2\pi i k } - C\omega_{k, \mr{JJA}}^2 e^{2\pi i k (N-1)/N }  - \frac{1}{L} e^{2\pi i k }  +\frac{1}{L} e^{2\pi i k (N-1)/N } }{ \bkt{C + C_g + C_c}\omega_{k, \mr{JJA}}^2 e^{-2\pi i k } - C\omega_{k, \mr{JJA}}^2 e^{-2\pi i k (N-1)/N } - \frac{1}{L} e^{-2\pi i k }  +\frac{1}{L}e^{-2\pi i k (N-1)/N } },
\label{cmkcpk}
}
\end{widetext}
which provides the allowed values of $k$. We consider the following approximate parameter values to determine the allowed values of $k$: $C_g\ll{C}$, $C_c \approx C$, $N\gg1$, and (a) $ \cbkt{1/L_0, C_0} = \cbkt{1/L,C}$ and (b) $ \cbkt{1/L_0, C_0}  \ll\cbkt{1/L,C}$. We look at the two cases of homogeneous array and weakly coupled atom as follows:

\begin{enumerate}[(a)]
    \item {For a  homogeneous array with  $ \cbkt{1/L_0, C_0} = \cbkt{1/L,C}$ one obtains from left BC  (Eq.\eqref{leftBC}) $\frac{c_{-k}}{c_{+k}} \approx - \frac{ 1 - 2\pi ik /N }{1 + 2\pi ik /N }\approx-1$. Together with the right BC (Eq.~\eqref{rightBC}), this yields $\sin \bkt{2\pi k }\approx 0 $, such that the allowed $k$ values are $2k = q$, with $ q\in \mathbb I$.

    Using the normalization condition $\mb\Phi_{k, \mr{JJA} }^T \bar{C}_\mr{JJA} \mb\Phi_{k', \mr{JJA}} = \delta_{k,k'}$ we can determine the eigenmodes of the JJA as
\eqn{\label{phika}
\Phi_{k,\mr{JJA}} (n) \approx \frac{2i}{\sqrt{N}}\sqrt{\frac{2C+ C_g}{2 C_k}}\sin \bkt{\frac{2\pi k n}{N}}
.}
}
\item{For a weakly coupled atom with $ \cbkt{1/L_0, C_0}  \ll\cbkt{1/L,C}$, we obtain from the left BC $\frac{c_{+k}}{c_{-k}}\approx 1$, which taken together with the right BC gives $\cos \bkt{2\pi k } =0$. Thus for a weakly coupled atom one obtains that $ k = q + \frac{1}{2} $, with $q\in \mathbb I$.

The eigenmodes are thus given as 
\eqn{\label{phikb}
\Phi_{k,\mr{JJA}} (n) \approx \frac{2}{\sqrt{N}}\sqrt{\frac{2C+ C_g}{2C_k}}\cos \bkt{\frac{2\pi k n}{N}}
.}
}
\end{enumerate}

\subsection{JJA  Modes}

\begin{figure}[t]
    \centering
    \subfloat{\includegraphics[width = 3.35 in]{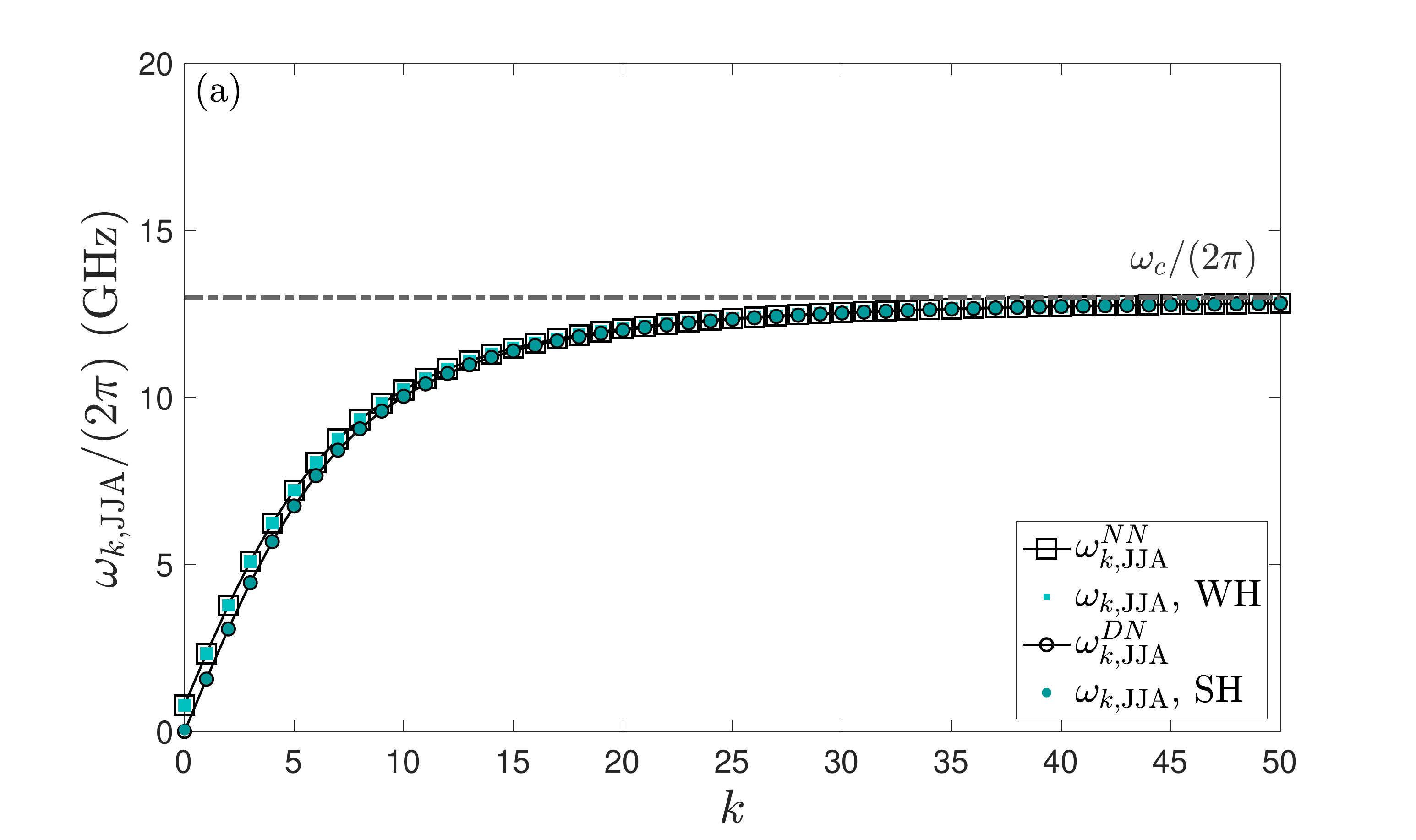}}\\
    \subfloat{\includegraphics[width = 3.35 in]{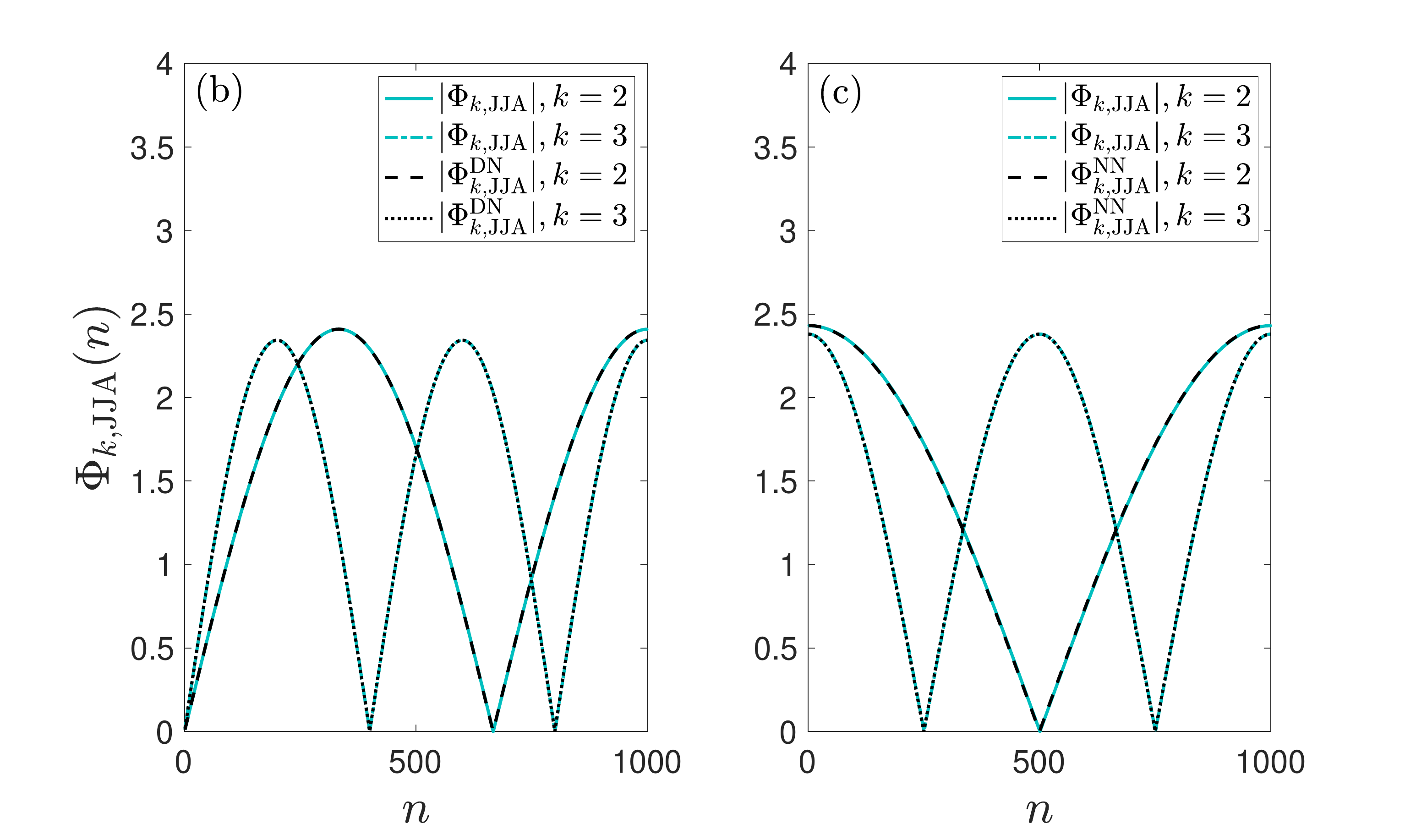}}
    \caption{ Eigenfrequencies  and eigenmodes for the uncoupled JJA. (a)  The black solid curve represents the analytical dispersion relation for the JJA assuming Neumann BC at both ends of the JJA (see Eq.~\eqref{wkjjann}) and the gray curve represents the JJA dispersion assuming Dirichlet BC at the atomic end and Neumann at the waveguide end (see Eq.~\eqref{wkjjadn}). The  squares and circles  represent the numerically obtained JJA eigenfrequencies for $\chi = 10^{-5}$  and $\chi = 1$ respectively. The band edge frequency is $\omega_{c}/(2\pi)\approx 12.99$~GHz, as depicted by the horizontal dashed-dotted line. The JJA eigenmodes corresponding to  (b) $\chi = 1$ and (c) $\chi = 10^{-5}$. The numerically obtained eigenmodes are depicted in blue solid and dashed-dotted curves for $k = 2$ and $k = 3 $ respectively. For $\chi = 1$ ($\chi = 10^{-5}$), the black dashed and dotted curves represent  the $k = 2 $ and $k = 3$   eigenmodes corresponding to the approximate analytical solution with Dirichlet (Neumann) BC at the atomic end given by  Eq.~\eqref{phikdn} (Eq.~\eqref{phiknn}), respectively.}
    \label{Fig:disp}
\end{figure}
Let us consider the  equations of motion for the JJA subspace as follows \eqn{\label{eomjja}\dbar {C} _\mr{JJA} \ddot{\mb{\Phi}}_\mr{JJA} = - \dbar{L}^{-1}_\mr{JJA} \mb{\Phi}_\mr{JJA},}
where the matrices $\dbar {C}_\mr{JJA}$ and $\dbar {L}_\mr{JJA}^{-1}$ are the capacitance and inverse inductance matrices as indicated in Eqs.~\eqref{Ctot} and \eqref{Lintot}, and $\mb{\Phi}_\mr{JJA} = \cbkt{\Phi_1, \dots, \Phi_N}$ represents the flux on the JJA nodes. The generalized eigenvalue problem for the JJA subspace is thus given by Eq.~\eqref{jjaep}.

The  generalized eigenvalue problem can be solved numerically to obtain the eigenfrequencies $\omega_{k,\mr{JJA}}$ and eigenmodes $\mb{\Phi}_{k,\mr{JJA}}$ of the JJA. We choose the normalization of the eigenmodes such that
\eqn{\label{norm}\mb{\Phi}^T _{k, \mr{JJA}} \dbar C _\mr{JJA} \mb{\Phi} _{k', \mr{JJA}}& =\bkt{C_g + 2C} \delta_{k,k'}\non\\
\mb{\Phi}^T _{k, \mr{JJA}} \dbar L ^{-1}_\mr{JJA} \mb{\Phi} _{k', \mr{JJA}} &=\frac{\omega_{k, \mr{JJA}}^2}{L} \delta_{k,k'}.
}

In the limit of a large JJA, the eigenmodes of the array can be well-approximated as two counterpropagating plane wave solutions which yields approximate analytical expressions for the eigenvalues and eigenvectors. While we are exclusively considering the JJA subspace here, the first element of the inductance and capacitance matrices includes the coupling term $\cbkt{L_0 , C_0 } $, which is crucial in determining the boundary condition (BC) at the atomic end. We consider the two hybridization regimes corresponding to the two coupler values as follows:
\begin{itemize}
    \item{\textit{Strongly hybridized (SH) regime}: The SH regime corresponds to having a Dirichlet  BC at the atomic end, with the dispersion relation given by  substituting the allowed values of $k$ in the dispersion relation Eq.~\eqref{wkjja} as
    \eqn{
\label{wkjjadn}\omega_{k, \mr{JJA}} ^{DN} =& \Omega_{0} \sqrt{\frac{1 - \cos\bkt{\bkt{k + 1/2} \pi/N}}{C_g/(2C) + 1 - \cos\bkt{\bkt{k + 1/2} \pi/N}}},}
with the band edge frequency $\omega_c \equiv 1/\sqrt{L(C_g/2 + C)}$.

The corresponding eigenmodes of the JJA can be obtained as (see Eq.~\eqref{phika})
\eqn{\label{phikdn}
{\Phi}^{DN}_{k, \mr{JJA}}(n) \approx& \sqrt{\frac{ C_g + 2C }{N\bkt{C \sbkt{ 1 - \cos\bkt{\frac{\pi \bkt{k + \frac{1}{2}} }{N}}} +\frac{C_g}{2}} }} \non\\
&\quad\quad\quad\quad\quad\quad\quad\quad\cos\bkt{\frac{\pi (k+ \frac{1}{2})n}{N}}.
}
\begin{figure*}[t]
    \centering \subfloat{\includegraphics[width = 3.5 in]{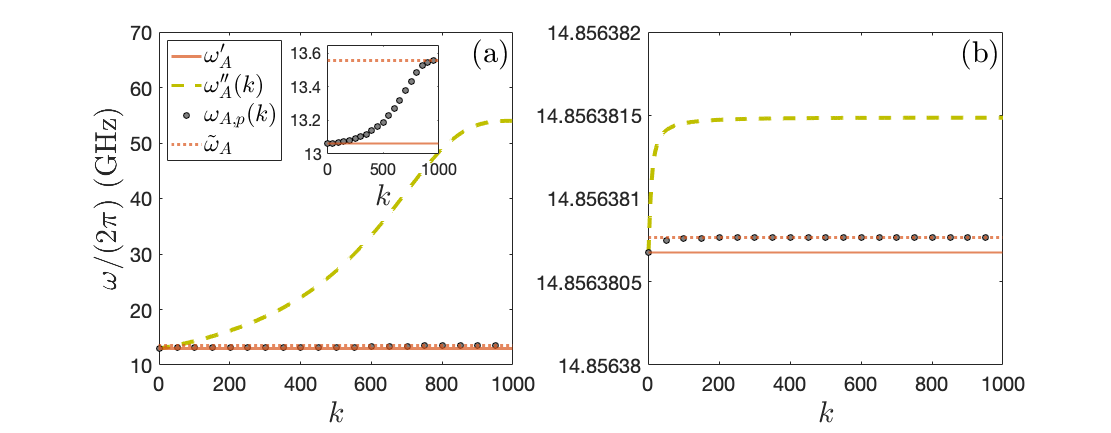}} 
\subfloat{ \includegraphics[width = 3.5 in]{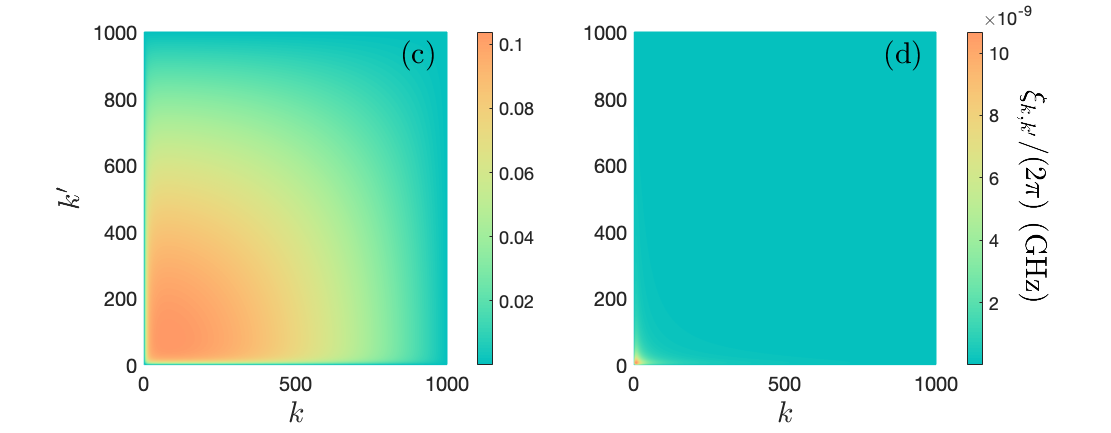}}
    \caption{Renormalized atomic frequency $\omega_A''$  (dashed curve) as a function of the number of modes incorporated, for (a) $\chi = 1$ and (b) $\chi = 10^{-5}$ . The bare atomic frequency is $\omega_A/(2\pi)\approx15$ GHz, the atomic frequency in the presence of on-site coupling  $\omega_{A}'$ is denoted by the solid horizontal line. The gray circles denote the eigenfrequencies $\omega_{A,p}(k)$, corresponding to the eigenmode of the Hamiltonian $H(k)$ with the highest amplitude at the atomic position and the dotted horizontal line corresponds to the adiabatically obtained atomic mode frequency $\tilde \omega_A$. The inset illustrates the convergence of $\omega_{A,p}(k)$ to $\tilde \omega_A$. The coupling coefficient $ \xi_{k, k'}$ between the JJA modes $k$ and $k'$ for (c) $\chi = 1$ and (d) $\chi = 10 ^{-5}$. }
    \label{Fig:waa}
\end{figure*}
We note that in the SH regime, the array eigenmodes have a minimum at the atomic end.
}
    \item{\textit{Weakly hybridized (WH) regime}: For $ \cbkt{1/L_0, C_0 }\ll \cbkt{1/L, C }$,  we obtain Neumann BC at the atomic end,  substituting the allowed values of $k$ in the dispersion relation Eq.~\eqref{wkjja}, we obtain   \eqn{\label{wkjjann}\omega_{k, \mr{JJA}} ^{NN} =& \Omega_{0} \sqrt{\frac{1 - \cos\bkt{k \pi/N}}{C_g/(2C) + 1 - \cos\bkt{k \pi/N}}}.}
    The corresponding eigenvectors for the Neumann-Neumann BCs are (see Eq.~\eqref{phikb})
\eqn{\label{phiknn}
{\Phi}^{NN}_{k, \mr{JJA}}(n) \approx&i\sqrt{\frac{ C_g + 2C }{N \bkt{C \sbkt{ 1 - \cos\bkt{\frac{\pi k }{N}}} +\frac{C_g}{2}}} } \sin\bkt{\frac{\pi kn}{N}}.}
The array modes have a maximum at the atomic end in the WH regime.
}
\end{itemize}
We note that having a general  strength of the coupler corresponds to mixed BCs at the atomic end, as can be determined from Eq.~\eqref{cmkcpk}.  The eigenvalues for $\chi = 1$ (SH) and $\chi = 10^{-5}$ (WH) are plotted in Fig.~\ref{Fig:disp}~(a).   It can be seen from Fig.~\ref{Fig:disp}~(a) that the eigenvalues for the SH case  agree with the approximate analytical dispersion relation for Dirichlet BC at left end and Neumann on the right (Eq.~\eqref{wkjjadn})  and those for WH case agree with the dispersion relation for Neumann  BC at both ends (Eq.~\eqref{wkjjann}).  Similarly, the eigenmodes corresponding to Dirichlet-Neumann (Neumann-Neumann) BC agree well with the numerically obtained eigenmodes for the SH (WH) case, as shown in Fig.~\ref{Fig:disp}~(b) and (c).

\vspace{1 cm}

\section{Hamiltonian derivation }
\label{App:Ham}

Using the partially diagonalized basis for the JJA, we can write the atom+JJA Lagrangian in Eq.~\eqref{Ltilde} as
\eqn{
\mc{L} =& \mc{L}_A  +  \sum_k \bkt{\frac{1}{2}\tilde C_{k} \dot{\mb{\Phi}}_{k,\mr{JJA}}^2 - \frac{\tilde L^{-1}_{k}}{2 }{\mb{\Phi}}_{k,\mr{JJA}}^2} \non\\
&- C_0 \dot{\Phi}_A \sum_{k}{\textbf{\textdelta}}_1^T\dot{\mb{\Phi}}_{k,\mr{JJA}} - \frac{1}{L_0} {\Phi}_A \sum_{k}{\textbf{\textdelta}}_1^T{\mb{\Phi}}_{k,\mr{JJA}},
}
where ${\textbf{\textdelta}}_1^T{\mb{\Phi}}_{k,\mr{JJA}} = \Phi_{k,\mr{JJA}}(1)$, and  $\tilde C_{k}$ and $\tilde L^{-1}_{k}$ represent the $(k+1)^\mr{th}$ diagonal element of the matrices $\tilde C$ and   $\tilde L^{-1}$. From the normalization conditions (Eq.~\eqref{norm}) we note that $ \tilde C_k = C_g + 2C$ and $ \tilde L^{-1}_k = \omega_{k, \mr{JJA}}^2 /L$.

Let us define the conjugate momenta corresponding to the flux variables for the atom $\Phi_A $ and the uncoupled JJA $ \mb\Phi_{k,\mr{JJA}}$ as
\eqn{
Q_A \equiv& \frac{\delta \mc{L}}{\delta \dot \Phi_{A}} =C_A' \dot{\Phi}_A - C_0  \sum_k  {\textbf{\textdelta}}_1^T\dot{\mb{\Phi}}_{k,\mr{JJA}}\\
\mb{Q}_{k,\mr{JJA}}\equiv &  \frac{\delta \mc{L}}{\delta \dot{ \mb{\Phi}}_{k,\mr{JJA}}} = \bkt{C_g + 2C} \mb{\Phi}_{k,\mr{JJA}} - C_0 \dot{\Phi}_A  {\textbf{\textdelta}}_1.
}
This allows us to write the Hamiltonian as 
\eqn{
H = Q_A \dot{\Phi}_A + \sum_k \mb{Q}_{k,\mr{JJA}}^T \dot{\mb{\Phi}}_{k,\mr{JJA}} - \mc{L}.
}

We can simplify the above Hamiltonian to obtain
\begin{widetext}
\eqn{\label{Ham0}
H = & \frac{1}{2} \frac{Q_A^2 }{C_A''} + \frac{1}{2} \frac{\Phi_A^2 }{L_A'} + \sum_{k}\sbkt{\frac{ Q^2_{k, \mr{JJA}} }{2\bkt{C_g + 2C} } + \omega_{k, \mr{JJA}}^2\frac{ \Phi^2_{k, \mr{JJA}} }{2L } + \frac{C_0}{C_A''\bkt{C_g + 2C} }\Phi_{k} (1) Q_A Q_{k, \mr{JJA}}-\frac{1}{L_0 } \Phi_{k} (1)  \Phi_A \Phi_{k, \mr{JJA}}} \non\\
& + \frac{C_0 ^2}{2C_A''\bkt{C_g + 2C}^2} \bkt{\sum_k\Phi_k (1) Q_{k, \mr{JJA}} }^2,
}
\end{widetext}
where  the renormalized  capacitance $C_A''$ is defined in Eq.~\eqref{capp}. Defining the renormalized mode frequency to include the self-interaction terms between the JJA modes as  (see Eq.~\eqref{wkren}), and promoting the flux and charge variables  to quantum operators as defined in  Eq.~\eqref{hatphik} and  Eq.~\eqref{hatqk} yields the Hamiltonian in Eq.~\eqref{Ham}.

It can be seen from Fig.~\ref{Fig:waa} (a) and (b)  that for a galvanically coupled atom ($\chi = 1$) the renormalized atomic frequency  can be drastically different from the bare atomic frequency, and, does not actually correspond to a physical eigenfrequency of the total system. As a benchmark for comparison, one can define the frequency $\omega_{A,p}(k)$ of the eigenmode of the Hamiltonian that has the largest amplitude at the atomic position. The eigenmode corresponding to $\omega_{A,p}(k)$ is obtained by numerically diagonalizing the Hamiltonian $H(k)$ defined as
\eqn{ H(k) = &\frac{1}{2} \bkt{\mb{Q}^{(k)}}^T \bkt{\tilde C^{(k)}}^{-1}\mb{Q}^{(k)} \non\\
&+ \frac{1}{2} \bkt{\mb{\Psi}^{(k)}}^T\bkt{\tilde L^{(k)}}^{-1}\mb{\Psi}^{(k)},
}
where $\mb{\Psi}^{(k)} \equiv \cbkt{\Phi_A; \mb{\Phi}_{j, \mr{JJA}}\vert j\in 1\dots k}$, $\mb{Q}^{(k)} \equiv \cbkt{Q_A; \mb{Q}_{j, \mr{JJA}}\vert j\in 1\dots k}$, $\bkt{\tilde C^{(k)} }_{mn}= \bkt{\tilde C}_{mn}$ and $\bkt{\tilde L^{(k)} }_{mn}= \bkt{\tilde L}_{mn}( m,n\in 1\dots k)$. Thus diagonalizing the Hamiltonian after including the coupling terms, we find that the mode frequencies of the total system do
not see as large a renormalization as indicated by $\omega_A''$.

It can also be seen from Fig.~\ref{Fig:waa}(c) and (d) that the coupling $ \xi_{k, k'}$ between the modes $k$ and $k'$ of the JJA can be as large as $\sim 0.1 $~GHz for $\chi = 1$, while it remains negligibly small for $\chi = 10^{-5}$. 

\section{Average number of excitations at the atomic node}
\label{App:nAt}
The average number of excitations at the atomic node is given by Eq.~\eqref{Eq:nat}, which can be simplified by substituting Eqs.~\eqref{Eq:phiext} and \eqref{Eq:qext} as follows:
\begin{widetext}
\eqn{
\avg{\hat n_A^{(0)}\bkt{\tilde t}} =& \frac{1}{2\hbar Z_A}\sbkt{\avg{:\bkt{\hat\Phi_A^{(0)}\bkt{\tilde t}}^2:} + Z_A^2 \avg{:\bkt{\hat Q_A^{(0)}\bkt{\tilde t}}^2:}}\\
= &  \frac{1}{2\hbar Z_A}\sbkt{\avg{:\bkt{\sum_{p,q} e^{\tilde s_p \tilde t} \eta_{p,q}\bkt{\tilde s_p} \hat{Y}_{q}\bkt{\tilde s_p}}^2:} +\bkt{ \frac{Z_A}{Z_0}}^2 \avg{:\bkt{\sum_{p,q}  e^{\tilde s_p \tilde t} \zeta_{p,q}\bkt{\tilde s_p} \hat{Y}_{q}\bkt{\tilde s_p}}^2:}}\\
= &  \frac{1}{2\hbar Z_A}\sbkt{\sum_{p,q}\sum_{m,n}e^{\bkt{\tilde s_p +\tilde s_m}\tilde t}\cbkt{  \eta_{p,q}\bkt{\tilde s_p}  \eta_{m,n}\bkt{\tilde s_m}+\bkt{ \frac{Z_A}{Z_0}}^2\zeta_{p,q}\bkt{\tilde s_p}  \zeta_{m,n}\bkt{\tilde s_m}} \avg{:\hat{Y}_{q}\bkt{\tilde s_p}\hat{Y}_{n}\bkt{\tilde s_m}:}}.
\label{Eq:nat0}
}
We can now evaluate the correlation $ \avg{:\hat{Y}_{q}\bkt{\tilde s_p}\hat{Y}_{n}\bkt{\tilde s_m}:}$ explicitly for an initial state of the system $\rho(0) = \ket{1}_A\bra{1}_A \otimes \ket{\cbkt{0}}\bra{\cbkt{0}}$ as:

\eqn{\label{Eq:Ycorr}
\avg{:\hat{Y}_{q}\bkt{\tilde s_p}\hat{Y}_{n}\bkt{\tilde s_m}:} =& \avg{:\cbkt{\tilde s_p  \sum_{r}C_{\mr{red},q,r}\hat{\Phi}_{\mr{red},r}\bkt{0} +\frac{Z_0}{Z_W} {\delta}_{q,N+2} \hat{\Phi}_{\mr{red},q}\bkt{0} +  Z_0 \hat{Q}_{\mr{red},q}\bkt{0} + 2  \sbkt{Z_0 \tilde Q_\mr{in}\bkt{\tilde s_p}}{\delta}_{N+2,q}}\right.\non\\
&\left.\cbkt{\tilde s_m \sum_{l}C_{\mr{red},n,l}\hat{\Phi}_{\mr{red},l}\bkt{0} +\frac{Z_0}{Z_W} {\delta}_{n,N+2} \hat{\Phi}_{\mr{red},n}\bkt{0} +  Z_0 \hat{Q}_{\mr{red},n}\bkt{0} + 2  \sbkt{Z_0 \tilde Q_\mr{in}\bkt{\tilde s_m}}{\delta}_{N+2,n}}:}
}
We note that the normal-ordered expectation values of the flux and charge variables for the initial state in consideration are given by  $\avg{:\hat \Phi _{\mr{red},p}(0)\hat \Phi _{\mr{red},m}(0) :} = \hbar Z_A \delta_{p,1} \delta_{m,1} $ and $\avg{:\hat Q _{\mr{red},p}(0)\hat Q_{\mr{red},m} (0):} = \frac{\hbar}{Z_A}  \delta_{p,1} \delta_{m,1} $. The auto-correlation function of the input noise quadratures are given by (see Appendix~\ref{App:noise} for  derivation):
\eqn{
\label{eq:q1q2}
\avg{:\hat Q_\mr{in}\bkt{{ t_1}} \hat Q_\mr{in}\bkt{ t_2}:}   =&\frac{\bkt{k_B T}^2}{2\pi \hbar Z_W }  \re\sbkt{\psi^{(1)}\bkt{1 - i \bkt{\frac{t_1 - t_2}{\beta}}}}\rightarrow
 \frac{ k_B T }{2Z_W } \delta \bkt{t_1 - t_2},
}
where $\beta = \frac{\hbar \Omega_0 }{k_BT}$, and  $\psi^{(1)}\bkt{z}$ represents trigamma function of $z\in \mathbb{Z}$ \cite{Carmichael}. We assume a  high temperature limit, such that the input noise from the transmission line can be approximated to be delta-correlated (Eq.~\eqref{Eq:deltanoise}), which can be justified for the parameters in consideration. 

Thus we can simplify Eq.~\eqref{Eq:Ycorr} as follows:
\eqn{
\avg{:\hat{Y}_{q}\bkt{\tilde s_p}\hat{Y}_{n}\bkt{\tilde s_m}:} = \hbar Z_A&\sbkt{\tilde s_p \tilde s_m  C_{\mr{red}, q,1} C_{\mr{red}, n,1} + \bkt{\frac{Z_0}{Z_A}}^2\delta_{q,1}\delta_{n,1} \right.\non\\
&\left.+  \frac{2k_B TZ_0^2}{\hbar\Omega_0 Z_AZ_W\bkt{\tilde s_p + \tilde s_m }} \delta_{q,N+2}\delta_{n,N+2}
 \bkt{e^{- \bkt{\tilde s_p + \tilde s_m }\tilde t} -1  }}.
}
We use the above correlation to simplify the excitation number at the atomic position in Eq.~\eqref{Eq:nat0} as follows:

\eqn{
&\avg{\hat n_A ^{(0)}\bkt{\tilde t}} = \frac{1}{2} \sum_{p,q,m,n} e^{\bkt{\tilde s_p + \tilde s_m}\tilde t} \sbkt{  \eta_{p,q}\bkt{\tilde s_p}  \eta_{m,n}\bkt{\tilde s_m}+\bkt{ \frac{Z_A}{Z_0}}^2\zeta_{p,q}\bkt{\tilde s_p}  \zeta_{m,n}\bkt{\tilde s_m}}\non\\
&\sbkt{\tilde s_p \tilde s_m  C_{\mr{red}, q,1} C_{\mr{red}, n,1} + \bkt{\frac{Z_0}{Z_A}}^2\delta_{q,1}\delta_{n,1} +  \frac{2k_B TZ_0^2}{\hbar \Omega_0 Z_A Z_W\bkt{\tilde s_p + \tilde s_m }} \delta_{q,N+2}\delta_{n,N+2}
 \bkt{e^{- \bkt{\tilde s_p + \tilde s_m }\tilde t} -1  }}\non\\
& =  \frac{1}{2} \sum_{p,q,m,n} e^{\bkt{\tilde s_p + \tilde s_m}\tilde t} \sbkt{  \eta_{p,q}\bkt{\tilde s_p}  \eta_{m,n}\bkt{\tilde s_m}+\bkt{ \frac{Z_A}{Z_0}}^2\zeta_{p,q}\bkt{\tilde s_p}  \zeta_{m,n}\bkt{\tilde s_m}}\sbkt{\tilde s_p \tilde s_m  C_{\mr{red}, q,1} C_{\mr{red}, n,1} + \bkt{\frac{Z_0}{Z_A}}^2\delta_{q,1}\delta_{n,1} }\non\\
&+\frac{k_B TZ_0^2}{\hbar \Omega_0 Z_AZ_W}\sum_{p,m}\frac{ e^{\bkt{\tilde s_p + \tilde s_m}\tilde t} }{\bkt{\tilde s_p + \tilde s_m }}\sbkt{  \eta_{p,N+2}\bkt{\tilde s_p}  \eta_{m,N+2}\bkt{\tilde s_m}+\bkt{ \frac{Z_A}{Z_0}}^2\zeta_{p,N+2}\bkt{\tilde s_p}  \zeta_{m,N+2}\bkt{\tilde s_m}} 
 \bkt{e^{- \bkt{\tilde s_p + \tilde s_m }\tilde t} -1  }
}
The above equation can be rewritten as Eq.~\eqref{Eq:nAt} and the $t\rightarrow\infty $ limit of the above expression  yields Eq.~\eqref{Eq:ninf}.

\section{Dynamics of artificial atom coupled to single open resonator}
\label{App:toy}

In this section we analyze the spontaneous emission dynamics of an artificial atom coupled to an open LC-resonator, as shown in Fig.~\ref{Fig:fft}~(a).  We can write the equations of motion for the reduced subspace of Atom+Coupler+Resonator+First waveguide node explicitly as follows (see Eqs.~\eqref{eq:dphiexdt} and \eqref{eq:dqexdt}):


\eqn{&
\der{}{t}\bkt{\begin{array}{c}
     \hat\Phi_A  \\
     \hat\Phi_C  \\
     \hat\Phi_R\\
     \hat\Phi_0^W\\
     Z_0 \hat Q_A\\
     Z_0 \hat Q_C\\
     Z_0\hat Q_R\\
     Z_0\hat Q_0^W
\end{array}} =\non\\
&\Omega_0\bkt{\begin{array}{cccccccc}
     0&0&0&0& \bkt{\dbar{C}_\mr{red}^{-1}}_{11}& \bkt{\dbar{C}_\mr{red}^{-1}}_{12}& \bkt{\dbar{C}_\mr{red}^{-1}}_{13}& \bkt{\dbar{C}_\mr{red}^{-1}}_{14}  \\
    0&0&0&0& \bkt{\dbar{C}_\mr{red}^{-1}}_{21}& \bkt{\dbar{C}_\mr{red}^{-1}}_{22}& \bkt{\dbar{C}_\mr{red}^{-1}}_{23}& \bkt{\dbar{C}_\mr{red}^{-1}}_{24}\\
    0&0&0&0& \bkt{\dbar{C}_\mr{red}^{-1}}_{31}& \bkt{\dbar{C}_\mr{red}^{-1}}_{32}& \bkt{\dbar{C}_\mr{red}^{-1}}_{33}& \bkt{\dbar{C}_\mr{red}^{-1}}_{34}\\
    0&0&0&0& \bkt{\dbar{C}_\mr{red}^{-1}}_{41}& \bkt{\dbar{C}_\mr{red}^{-1}}_{42}& \bkt{\dbar{C}_\mr{red}^{-1}}_{43}& \bkt{\dbar{C}_\mr{red}^{-1}}_{44}\\
    \bkt{\dbar{L}_\mr{red}^{-1}}_{11} &  \bkt{\dbar{L}_\mr{red}^{-1}}_{12}&  \bkt{\dbar{L}_\mr{red}^{-1}}_{13}  & 0 &0&0 &0&0\\
    \bkt{\dbar{L}_\mr{red}^{-1}}_{21} &  \bkt{\dbar{L}_\mr{red}^{-1}}_{22}&\bkt{\dbar{L}_\mr{red}^{-1}}_{23}  & 0 &0&0 &0&0\\
    \bkt{\dbar{L}_\mr{red}^{-1}}_{31} &  \bkt{\dbar{L}_\mr{red}^{-1}}_{32}&\bkt{\dbar{L}_\mr{red}^{-1}}_{33}  & 0 &0&0 &0&0\\
    0 & 0 & 0&0& -\frac{Z_0}{Z_W}{ \bkt{\dbar{C}_\mr{red}^{-1}}_{41}}&-\frac{Z_0}{Z_W}{ \bkt{\dbar{C}_\mr{red}^{-1}}_{42}}&-\frac{Z_0}{Z_W}{ \bkt{\dbar{C}_\mr{red}^{-1}}_{43}}&-\frac{Z_0}{Z_W}{ \bkt{\dbar{C}_\mr{red}^{-1}}_{44}}
\end{array}}\bkt{\begin{array}{c}
     \hat\Phi_A  \\
     \hat\Phi_C\\
     \hat\Phi_R\\
     \hat\Phi_0^W\\
     Z_0 \hat Q_A\\
     Z_0 \hat Q_C\\
     Z_0\hat Q_R\\
     Z_0\hat Q_0^W
\end{array}}\non\\
& +\Omega_0 \bkt{\begin{array}{c}
     0  \\
     0  \\
     0  \\
     0\\
     0\\
     0\\
     0\\
     2\frac{1}{\Omega_0} Z_0\hat Q_\mr{in}
\end{array}},
}
where  $\bkt{\dbar C_\mr{red}^{-1}}_{jk}$ and $\bkt{\dbar L_\mr{red}^{-1}}_{jk}$ indicates the $\cbkt{j,k}$ element of the inverse capacitance and inductance matrices in the reduced subspace as defined in Eqs.~\eqref{Cex} and \eqref{Linex}.

Taking the Laplace transform of the above and rearranging terms, we get:
\eqn{
 \bkt{\begin{array}{c}
     \tilde\Phi_A  \\
     \tilde\Phi_C \\
     \tilde\Phi_R\\
     \tilde\Phi_0^W\\
     Z_0 \tilde Q_A\\
     Z_0 \tilde Q_C\\
     Z_0\tilde Q_R\\
     Z_0\tilde Q_0^W
\end{array}} =\frac{1}{\Omega_0}\tilde G \bkt{\begin{array}{c}
     \hat\Phi_A\bkt{0}  \\
     \hat\Phi_C\bkt{0}  \\
     \hat\Phi_R\bkt{0}\\
     \hat\Phi_W\bkt{0}\\
     Z_0 \hat Q_A\bkt{0}\\
     Z_0 \hat Q_C\bkt{0}\\
     Z_0\hat Q_R\bkt{0}\\
     Z_0\hat Q_0^W\bkt{0}+     2Z_0 \tilde Q_\mr{in}
\end{array}},
\label{Eq:ToyLap}
}
where we have defined Laplace transform of the fluxes at various nodes as
$ \tilde{\mc{O}}\bkt{s} = \int_0 ^\infty \dd t~ e^{-st}\hat{\mc{O}}(t)$. The propagator matrix $\tilde G$ is defined as:
\eqn{&\tilde G\equiv\bkt{\begin{array}{cccccccc}
     \tilde s&0&0&0&- \bkt{\dbar{C}_\mr{red}^{-1}}_{11}&- \bkt{\dbar{C}_\mr{red}^{-1}}_{12}&- \bkt{\dbar{C}_\mr{red}^{-1}}_{13}&- \bkt{\dbar{C}_\mr{red}^{-1}}_{14}  \\
    0&\tilde s&0&0&- \bkt{\dbar{C}_\mr{red}^{-1}}_{21}&- \bkt{\dbar{C}_\mr{red}^{-1}}_{22}&- \bkt{\dbar{C}_\mr{red}^{-1}}_{23}&- \bkt{\dbar{C}_\mr{red}^{-1}}_{24}\\
    0&0&\tilde s&0&- \bkt{\dbar{C}_\mr{red}^{-1}}_{31}&- \bkt{\dbar{C}_\mr{red}^{-1}}_{32}&- \bkt{\dbar{C}_\mr{red}^{-1}}_{33}&- \bkt{\dbar{C}_\mr{red}^{-1}}_{34}\\
    0&0&0&\tilde s&- \bkt{\dbar{C}_\mr{red}^{-1}}_{41}&- \bkt{\dbar{C}_\mr{red}^{-1}}_{42}&- \bkt{\dbar{C}_\mr{red}^{-1}}_{43}&- \bkt{\dbar{C}_\mr{red}^{-1}}_{44}\\
    \bkt{\dbar{L}_\mr{red}^{-1}}_{11} &  \bkt{\dbar{L}_\mr{red}^{-1}}_{12}&\bkt{\dbar{L}_\mr{red}^{-1}}_{13}  & 0 &\tilde s&0 &0&0\\
    \bkt{\dbar{L}_\mr{red}^{-1}}_{21} &  \bkt{\dbar{L}_\mr{red}^{-1}}_{22} &  \bkt{\dbar{L}_\mr{red}^{-1}}_{23}  & 0 &0&\tilde s &0&0\\
    \bkt{\dbar{L}_\mr{red}^{-1}}_{31} &  \bkt{\dbar{L}_\mr{red}^{-1}}_{32} &  \bkt{\dbar{L}_\mr{red}^{-1}}_{33}  & 0 &0 &0&\tilde s&0\\
    0&0 & 0 & 0& \frac{Z_0}{Z_W}{ \bkt{\dbar{C}_\mr{red}^{-1}}_{41}}&\frac{Z_0}{Z_W}{ \bkt{\dbar{C}_\mr{red}^{-1}}_{42}}&\frac{Z_0}{Z_W}{ \bkt{\dbar{C}_\mr{red}^{-1}}_{43}}&\tilde s+\frac{Z_0}{Z_W}{ \bkt{\dbar{C}_\mr{red}^{-1}}_{44}}
\end{array}}^{-1}
}
with $\tilde s\equiv s/\Omega_0 $.

We can take the inverse Laplace transform of Eq.~\eqref{Eq:ToyLap} to obtain the dynamics of the atomic observables as follows:

\eqn{
\label{Eq:phiat}
&\hat\Phi_A \bkt{\tilde t} = G_{AA}^{\Phi\Phi}\bkt{\tilde t}\hat\Phi_A \bkt{0} +G_{AC}^{\Phi\Phi}\bkt{\tilde t}\hat\Phi_C \bkt{0} + G_{AR}^{\Phi\Phi}\bkt{\tilde t}\hat\Phi_R \bkt{0} + G_{AW}^{\Phi\Phi}\bkt{\tilde t}\hat\Phi_W\bkt{0}+ G_{AA}^{\Phi Q}\bkt{\tilde t} \sbkt{Z_0\hat Q_A \bkt{0}} \non\\
&+ G_{AC}^{\Phi Q}\bkt{\tilde t}\sbkt{Z_0\hat Q_C \bkt{0}}+ G_{AR}^{\Phi Q}\bkt{\tilde t}\sbkt{Z_0\hat Q_R \bkt{0}}+ G_{AW}^{\Phi Q}\bkt{\tilde t}\sbkt{Z_0\hat Q_W\bkt{0} } + \int_0 ^{\tilde t}\dd\tilde\tau~ G_{AW}^{\Phi Q}\bkt{\tilde t - \tilde \tau} \sbkt{\frac{2}{\Omega_0 }Z_0 \hat Q_\mr{in}\bkt{\frac{\tilde \tau}{\Omega_0}}}\\
\label{Eq:qat}
&Z_0\hat Q_A \bkt{\tilde t} = G_{AA}^{Q\Phi}\bkt{\tilde t}\hat\Phi_A \bkt{0} + G_{AC}^{Q\Phi}\bkt{\tilde t}\hat\Phi_C\bkt{0}+ G_{AR}^{Q\Phi}\bkt{\tilde t}\hat\Phi_R \bkt{0} + G_{AW}^{Q\Phi}\bkt{\tilde t}\hat\Phi_W\bkt{0}+ G_{AA}^{Q Q}\bkt{\tilde t}\sbkt{Z_0\hat Q_A \bkt{0} }\non\\
&+ G_{AC}^{Q Q}\bkt{\tilde t}\sbkt{Z_0\hat Q_C \bkt{0}} + G_{AR}^{Q Q}\bkt{\tilde t}\sbkt{Z_0\hat Q_R \bkt{0}}  + G_{AW}^{Q Q}\bkt{ \tilde t}\sbkt{Z_0\hat Q_W\bkt{0}}+ \int_0 ^{\tilde t}\dd\tilde \tau G_{AW}^{Q Q}\bkt{\tilde t -\tilde  \tau} \sbkt{\frac{2}{\Omega_0}Z_0\hat Q_\mr{in}\bkt{\frac{\tilde \tau}{\Omega_0}}},
}
where we have defined $\tilde t \equiv \Omega_0 t$ and the inverse Laplace transform of the propagator matrix $\tilde G$ as:
\eqn{\frac{1}{2\pi i } \int_{-i\infty} ^{i\infty} \dd s ~e^{s t} \tilde G\bkt{s} \equiv \bkt{\begin{array}{cccccccc}
G_{AA}^{\Phi\Phi}\bkt{t}&G_{AC}^{\Phi\Phi}\bkt{t}&G_{AR}^{\Phi\Phi}\bkt{t} & G_{AW}^{\Phi\Phi}\bkt{t}&G_{AA}^{\Phi Q}\bkt{t}&G_{AC}^{\Phi Q}\bkt{t}&G_{AR}^{\Phi Q}\bkt{t} & G_{AW}^{\Phi Q}\bkt{t}\\
G_{AA}^{Q\Phi}\bkt{t}&G_{AC}^{Q\Phi}\bkt{t}&G_{AR}^{Q\Phi}\bkt{t} & G_{AW}^{Q\Phi}\bkt{t}&G_{AA}^{Q Q}\bkt{t}&G_{AC}^{Q Q}\bkt{t}&G_{AR}^{Q Q}\bkt{t} & G_{AW}^{Q Q}\bkt{t}\\
G_{RA}^{\Phi\Phi}\bkt{t}&G_{RC}^{\Phi\Phi}\bkt{t}&G_{RR}^{\Phi\Phi}\bkt{t} & G_{RW}^{\Phi\Phi}\bkt{t}&G_{RA}^{\Phi Q}\bkt{t}&G_{RC}^{\Phi Q}\bkt{t}&G_{RR}^{\Phi Q}\bkt{t} & G_{RW}^{\Phi Q}\bkt{t}\\
G_{RA}^{Q\Phi}\bkt{t}&G_{RC}^{Q\Phi}\bkt{t}&G_{RR}^{Q\Phi}\bkt{t} & G_{RW}^{Q\Phi}\bkt{t}&G_{RA}^{Q Q}\bkt{t}&G_{RC}^{Q Q}\bkt{t}&G_{RR}^{Q Q}\bkt{t} & G_{RW}^{Q Q}\bkt{t}\\
G_{WA}^{\Phi\Phi}\bkt{t}&G_{WC}^{\Phi\Phi}\bkt{t}&G_{WR}^{\Phi\Phi}\bkt{t} & G_{WW}^{\Phi\Phi}\bkt{t}&G_{WA}^{\Phi Q}\bkt{t}&G_{WC}^{\Phi Q}\bkt{t}&G_{WR}^{\Phi Q}\bkt{t} & G_{WW}^{\Phi Q}\bkt{t}\\
G_{WA}^{Q\Phi}\bkt{t}&G_{WC}^{Q\Phi}\bkt{t}&G_{WR}^{Q\Phi}\bkt{t} & G_{WW}^{Q\Phi}\bkt{t}&G_{WA}^{Q Q}\bkt{t}&G_{WC}^{Q Q}\bkt{t}&G_{WR}^{Q Q}\bkt{t} & G_{WW}^{Q Q}\bkt{t}
\end{array}}.
}

We obtain the normal ordered expectation values of $\hat \Phi_A^2\bkt{t}$ and $\hat Q_A^2\bkt{t}$ as follows:

\eqn{
\label{Eq:phia2}
\avg {:\hat\Phi_A ^2\bkt{\tilde t} :} =& \hbar Z_A \sbkt{\bkt{G_{AA}^{\Phi\Phi}\bkt{\tilde t}}^2+\frac{Z_0^2}{Z_A^2}\bkt{G_{AA}^{\Phi Q}\bkt{\tilde t}}^2}  \avg{\hat n_A\bkt{0}}  \non\\
&  + \frac{4 Z_0^2}{ \Omega_0^2 } \int_0^{\tilde t }\dd\tilde \tau_1\int_0^{\tilde t }\dd\tilde \tau_2 G_{AW}^{\Phi Q}\bkt{\tilde t - \tilde \tau_1} G_{AW}^{\Phi Q}\bkt{\tilde t - \tilde \tau_2} \avg{:\hat Q_\mr{in}\bkt{\frac{\tilde \tau_1}{\Omega_0}} \hat Q_\mr{in}\bkt{\frac{\tilde \tau_2}{\Omega_0}}:},\\
\label{Eq:qa2}
\avg {:Z_0^2\hat Q_A ^2\bkt{\tilde t} :} =& {\hbar Z_A} \sbkt{\bkt{G_{AA}^{Q\Phi}\bkt{\tilde t}}^2+\frac{Z_0^2}{Z_A^2}\bkt{G_{AA}^{Q Q}\bkt{\tilde t}}^2}  \avg{\hat n_A\bkt{0}}  \non\\
&  + \frac{4 Z_0^2}{ \Omega_0^2 } \int_0^{\tilde t }\dd\tilde \tau_1\int_0^{\tilde t }\dd\tilde \tau_2 G_{AW}^{Q Q}\bkt{\tilde t - \tilde \tau_1} G_{AW}^{Q Q}\bkt{\tilde t - \tilde \tau_2} \avg{:\hat Q_\mr{in}\bkt{\frac{\tilde \tau_1}{\Omega_0}} \hat Q_\mr{in}\bkt{\frac{\tilde \tau_2}{\Omega_0}}:},
}
where we have assumed that the total system is initially in a vacuum state for all nodes except the atomic node.

 Substituting the noise correlation  in the high temperature limit in the above (Eq.~\eqref{Eq:deltanoise}), we obtain:
\eqn{
\label{Eq:phia22}
\avg {:\hat\Phi_A ^2\bkt{\tilde t} :} =& {\hbar Z_A} \sbkt{\bkt{G_{AA}^{\Phi\Phi}\bkt{\tilde t}}^2+\frac{Z_0^2}{Z_A^2}\bkt{G_{AA}^{\Phi Q}\bkt{\tilde t}}^2}  \avg{\hat n_A\bkt{0}}   + \frac{2 k_B T Z_0^2}{ \Omega_0 Z_W } \int_0^{\tilde t }\dd\tilde \tau\bkt{ G_{AW}^{\Phi Q}\bkt{\tilde t - \tilde \tau}}^2 ,\\
\label{Eq:qa22}
\avg {:Z_A^2\hat Q_A ^2\bkt{\tilde t} :} =& {\hbar Z_A} \sbkt{\bkt{G_{AA}^{Q\Phi}\bkt{\tilde t}}^2+ \frac{Z_0^2}{Z_A^2}\bkt{G_{AA}^{Q Q}\bkt{\tilde t}}^2}  \avg{\hat  n_A\bkt{0}}  + \frac{2 k_B T Z_0^2}{ \Omega_0 Z_W } \int_0^{\tilde t }\dd\tilde \tau\bkt{ G_{AW}^{Q Q}\bkt{\tilde t - \tilde \tau}}^2 .
}

\section{Including the atomic nonlinearity perturbatively}
\label{App:NL}

In this Appendix, we evaluate the perturbative corrections due to nonlinearity  considering a cubic form of the nonlinear potential  $\mathcal{U}_A(\Phi_A) = - E_A c_3 \bkt{\frac{\Phi_A}{\phi_0}}^3\equiv \Lambda \Phi_A^3$. We simplify the  following two constituent terms in the nonlinear corrections to the number expectation value  Eq.~\eqref{nat1} separately:
\eqn{\label{naphi1}
\avg{\hat n_{A,\Phi} ^{(1)}\bkt{\tilde t}} \equiv&\frac{\lambda^2}{2\hbar Z_A} \avg{: \sbkt{\sum_{p,q} e^{\tilde s_p \tilde t}  \frac{1}{\bkt{\frac{\delta\gamma_{pp} \bkt{\tilde s_p}}{\delta \tilde s}}}\beta^{-1}_{1,p}\bkt{\tilde s_p}  \alpha^{-1}_{p,q}\bkt{\tilde s_p} \hat{Y}_{q}^{(1)}\bkt{\tilde s_p}}\right.\non\\
&\left.\sbkt{\sum_{m,n} e^{\tilde s_m \tilde t}  \frac{1}{\bkt{\frac{\delta\gamma_{mm} \bkt{\tilde s_m}}{\delta \tilde s}}}\beta^{-1}_{1,m}\bkt{\tilde s_m}  \alpha^{-1}_{m,n}\bkt{\tilde s_m} \hat{Y}_{n}^{(1)}\bkt{\tilde s_m}} }\\
\label{naq1}
\avg{\hat n_{A,Q}^{(1)}\bkt{\tilde t} }\equiv &\frac{\lambda^2 Z_A}{2\hbar} \avg{\sbkt{\sum_{p,q,r} \tilde s_p e^{\tilde s_p \tilde t}  \frac{1}{\bkt{\frac{\delta\gamma_{pp} \bkt{\tilde s_p}}{\delta \tilde s}}}C_{\mr{red},1,q}\beta^{-1}_{q,p}\bkt{\tilde s_p}  \alpha^{-1}_{p,r}\bkt{\tilde s_p} \hat{Y}_{r}^{(1)}\bkt{\tilde s_p}}\right.\non\\
&\left.\sbkt{\sum_{l,m,n} \tilde s_l e^{\tilde s_l \tilde t}  \frac{1}{\bkt{\frac{\delta\gamma_{ll} \bkt{\tilde s_l}}{\delta \tilde s}}}C_{\mr{red},1,m}\beta^{-1}_{m,l}\bkt{\tilde s_l}  \alpha^{-1}_{l,n}\bkt{\tilde s_l} \hat{Y}_{n}^{(1)}\bkt{\tilde s_l}}:}.
}

\begin{figure}[t]
    \centering
    \includegraphics[width = 3.5 in]{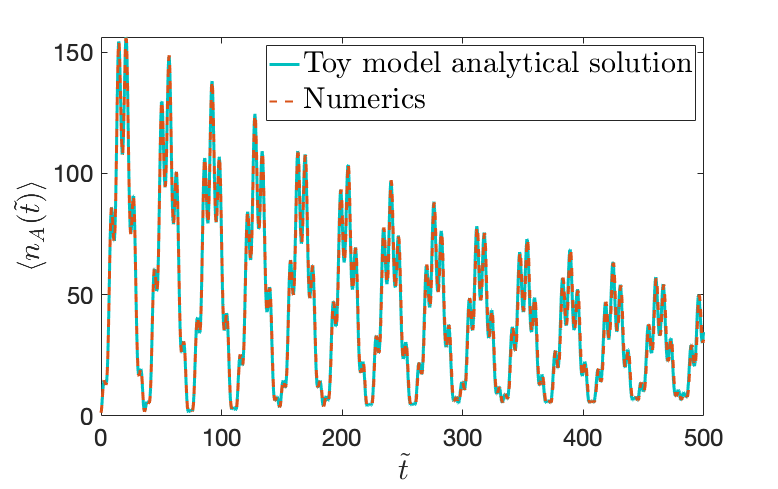}
    \caption{Comparison of the  atomic excitation number dynamics for the toy model of an atom coupled to an open resonator  (see  Fig.~\ref{Fig:fft}(a)) for  $\chi = 1$, $T = 50$~mK, and $\omega_A/(2\pi) \approx 5$~GHz. The analytical solution is obtained from substituting Eqs.~\eqref{Eq:phia22} and \eqref{Eq:qa22} in Eq.~\eqref{Eq:nat}, and the numerical solution in matrix representation is obtained from substituting Eqs.~\eqref{Eq:phiext} and \eqref{Eq:qext}  in Eq.~\eqref{Eq:nat}.}
    \label{Fig:comparison}
\end{figure}
The nonlinear source term for the case of cubic nonlinear potential  is given as:
\eqn{
\hat{\mb{Y}}^{(1)}\bkt{\tilde s} \equiv \int \dd\tilde t~ e^{- \tilde s \tilde t}\sbkt{Z_0 \Lambda \bkt{\dbar{\delta}_1\hat{\mb{\Phi}}_\mr{red}^{(0)}\bkt{\tilde t}}^2}.
}
Substituting the linear solution (Eq.~\eqref{Eq:phiext}) in the above we  obtain the nonlinear source term as follows:
\eqn{  
\hat{Y}_{j}^{(1)}\bkt{\tilde s} = &\int \dd\tilde t~ e^{- \tilde s \tilde t} \sbkt{\frac{Z_0 \Lambda}{\lambda} \delta_{j1} \cbkt{ \sum_{p,q} e^{\tilde s_p \tilde t}  \frac{1}{\bkt{\frac{\delta\gamma_{pp} \bkt{\tilde s_p}}{\delta \tilde s}}}\beta^{-1}_{1,p}\bkt{\tilde s_p}  \alpha^{-1}_{p,q}\bkt{\tilde s_p} \hat{Y}_{q}\bkt{\tilde s_p}}\right.\non\\
&\left.\cbkt{\sum_{m,n} e^{\tilde s_m \tilde t}  \frac{1}{\bkt{\frac{\delta\gamma_{mm} \bkt{\tilde s_m}}{\delta \tilde s}}}\beta^{-1}_{1,m}\bkt{\tilde s_m}  \alpha^{-1}_{m,n}\bkt{\tilde s_m} \hat{Y}_{n}\bkt{\tilde s_m}}}
}
We now substitute the nonlinear source term in  Eqs.~\eqref{naphi1} and \eqref{naq1}, which yields:
\eqn{
&\avg{\hat n_{A,\Phi} ^{(1)}\bkt{\tilde t}} \equiv\frac{\bkt{Z_0 \Lambda}^2} {2\hbar Z_A}\non\\
& \avg{: \sbkt{\sum_{p,q} e^{\tilde s_p \tilde t}  \frac{1}{\bkt{\frac{\delta\gamma_{pp} \bkt{\tilde s_p}}{\delta \tilde s}}}\beta^{-1}_{1,p}\bkt{\tilde s_p}  \alpha^{-1}_{p,1}\bkt{\tilde s_p} \int \dd\tilde t_1~ e^{- \tilde s_p \tilde t_1} \sbkt{ \cbkt{ \sum_{p_1,q_1} e^{\tilde s_{p_1} \tilde t_1}  \frac{1}{\bkt{\frac{\delta\gamma_{p_1p_1} \bkt{\tilde s_{p_1}}}{\delta \tilde s}}}\beta^{-1}_{1,p_1}\bkt{\tilde s_{p_1}}  \alpha^{-1}_{p_1,q_1}\bkt{\tilde s_{p_1}} \hat{Y}_{q_1}\bkt{\tilde s_{p_1}}}\right.\right.\right.\non\\
&\left.\left.\left.\cbkt{\sum_{m_1,n_1} e^{\tilde s_{m_1} \tilde t_1}  \frac{1}{\bkt{\frac{\delta\gamma_{m_1m_1} \bkt{\tilde s_{m_1}}}{\delta \tilde s}}}\beta^{-1}_{1,{m_1}}\bkt{\tilde s_{m_1}}  \alpha^{-1}_{m_1,n_1}\bkt{\tilde s_{m_1}} \hat{Y}_{n_1}\bkt{\tilde s_{m_1}}}}}\right.\non\\
&\left.\sbkt{\sum_{m} e^{\tilde s_m \tilde t}  \frac{1}{\bkt{\frac{\delta\gamma_{mm} \bkt{\tilde s_m}}{\delta \tilde s}}}\beta^{-1}_{1,m}\bkt{\tilde s_m}  \alpha^{-1}_{m,1}\bkt{\tilde s_m} \int \dd\tilde t_2~ e^{- \tilde s_m \tilde t_2} \sbkt{ \cbkt{ \sum_{p_2,q_2} e^{\tilde s_{p_2} \tilde t_2}  \frac{1}{\bkt{\frac{\delta\gamma_{p_2p_2} \bkt{\tilde s_{p_2}}}{\delta \tilde s}}}\beta^{-1}_{1,p_2}\bkt{\tilde s_{p_2}}  \alpha^{-1}_{p_2,q_2}\bkt{\tilde s_{p_2}} \hat{Y}_{q_2}\bkt{\tilde s_{p_2}}}\right.\right.\right.\non\\
&\left.\left.\left.\cbkt{\sum_{m_2,n_2} e^{\tilde s_{m_2} \tilde t_2}  \frac{1}{\bkt{\frac{\delta\gamma_{m_2m_2} \bkt{\tilde s_{m_2}}}{\delta \tilde s}}}\beta^{-1}_{1,{m_2}}\bkt{\tilde s_{m_2}}  \alpha^{-1}_{m_2,n_2}\bkt{\tilde s_{m_2}} \hat{Y}_{n_2}\bkt{\tilde s_{m_2}}}} 
} :}
}
\eqn{
= & \frac{\bkt{Z_0 \Lambda}^2} {2\hbar Z_A}  \sbkt{\sum_{p,m}\sum_{p_1,q_1}\sum_{m_1,n_1}\sum_{p_2,q_2}\sum_{m_2,n_2} \avg{ : \hat{Y}_{q_1}\bkt{\tilde s_{p_1}} \hat{Y}_{n_1}\bkt{\tilde s_{m_1}} \hat{Y}_{q_2}\bkt{\tilde s_{p_2}}\hat{Y}_{n_2}\bkt{\tilde s_{m_2} } :} \frac{ e^{\bkt{\tilde s_p + \tilde s_m}\tilde t} }{\bkt{-\tilde s_p  + \tilde s_{p_1} + \tilde s_{m_1}}\bkt{-\tilde s_m + \tilde s_{p_2} + \tilde s_{m_2}}  }\right.\non\\
&\left. \bkt{e^{-\bkt{\tilde s _p -\tilde s _{p_1}  - \tilde s _{m_1}   } \tilde t}-1} \bkt{e^{-\bkt{\tilde s _m -\tilde s _{p_2}  - \tilde s _{m_2}   } \tilde t}-1} \cbkt{ \frac{1}{\bkt{\frac{\delta\gamma_{pp} \bkt{\tilde s_p}}{\delta \tilde s}}}\beta^{-1}_{1,p}\bkt{\tilde s_p} \alpha^{-1}_{p,1}\bkt{\tilde s_p} }\cbkt{ \frac{1}{\bkt{\frac{\delta\gamma_{mm} \bkt{\tilde s_m}}{\delta \tilde s}}}\beta^{-1}_{1,m}\bkt{\tilde s_m}  \alpha^{-1}_{m,1}\bkt{\tilde s_m}}    \right.\non\\
&\left. \cbkt{  \frac{1}{\bkt{\frac{\delta\gamma_{p_1 p_1} \bkt{\tilde s_{p_1}}}{\delta \tilde s}}}\beta^{-1}_{1,p_1}\bkt{\tilde s_{p_1}}  \alpha^{-1}_{p_1,q_1}\bkt{\tilde s_{p_1}} } \cbkt{ \frac{1}{\bkt{\frac{\delta\gamma_{m_1m_1} \bkt{\tilde s_{m_1}}}{\delta \tilde s}}}\beta^{-1}_{1,{m_1}}\bkt{\tilde s_{m_1}}  \alpha^{-1}_{m_1,n_1}\bkt{\tilde s_{m_1}}}\right.\non\\
&\left. \cbkt{  \frac{1}{\bkt{\frac{\delta\gamma_{p_2p_2} \bkt{\tilde s_{p_2}}}{\delta \tilde s}}}\beta^{-1}_{1,p_2}\bkt{\tilde s_{p_2}}  \alpha^{-1}_{p_2,q_2}\bkt{\tilde s_{p_2}} } \cbkt{ \frac{1}{\bkt{\frac{\delta\gamma_{m_2m_2} \bkt{\tilde s_{m_2}}}{\delta \tilde s}}}\beta^{-1}_{1,{m_2}}\bkt{\tilde s_{m_2}}  \alpha^{-1}_{m_2,n_2}\bkt{\tilde s_{m_2}} } } 
\label{naphi11}
.
}

\eqn{
&\avg{\hat n_{A,Q} ^{(1)}\bkt{\tilde t}}\equiv\frac{\bkt{Z_0 \Lambda}^2 Z_A} {2\hbar }\non\\
&\sbkt{\sum_{p,q,l,m}\sum_{p_1,q_1}\sum_{m_1,n_1}\sum_{p_2,q_2}\sum_{m_2,n_2} \avg{ : \hat{Y}_{q_1}\bkt{\tilde s_{p_1}} \hat{Y}_{n_1}\bkt{\tilde s_{m_1}} \hat{Y}_{q_2}\bkt{\tilde s_{p_2}}\hat{Y}_{n_2}\bkt{\tilde s_{m_2} }: } \frac{ \tilde s_p \tilde s_l e^{\bkt{\tilde s_p + \tilde s_l}\tilde t} }{\bkt{-\tilde s_p  + \tilde s_{p_1} + \tilde s_{m_1}}\bkt{-\tilde s_l + \tilde s_{p_2} + \tilde s_{m_2}}  }\right.\non\\
&\left. \bkt{e^{-\bkt{\tilde s _p -\tilde s _{p_1}  - \tilde s _{m_1}   } \tilde t}-1} \bkt{e^{-\bkt{\tilde s _l -\tilde s _{p_2}  - \tilde s _{m_2}   } \tilde t}-1} \right.\non\\
& \left.\cbkt{ \frac{1}{\bkt{\frac{\delta\gamma_{pp} \bkt{\tilde s_p}}{\delta \tilde s}}} C_{\mr{red},1,p}\beta^{-1}_{p,q}\bkt{\tilde s_p} \alpha^{-1}_{q,1}\bkt{\tilde s_p} }\cbkt{ \frac{1}{\bkt{\frac{\delta\gamma_{mm} \bkt{\tilde s_m}}{\delta \tilde s}}} C_{\mr{red},1,m}\beta^{-1}_{m,n}\bkt{\tilde s_m}  \alpha^{-1}_{n,1}\bkt{\tilde s_m}}    \right.\non\\
&\left. \cbkt{  \frac{1}{\bkt{\frac{\delta\gamma_{p_1 p_1} \bkt{\tilde s_{p_1}}}{\delta \tilde s}}}\beta^{-1}_{1,p_1}\bkt{\tilde s_{p_1}}  \alpha^{-1}_{p_1,q_1}\bkt{\tilde s_{p_1}} } \cbkt{ \frac{1}{\bkt{\frac{\delta\gamma_{m_1m_1} \bkt{\tilde s_{m_1}}}{\delta \tilde s}}}\beta^{-1}_{1,{m_1}}\bkt{\tilde s_{m_1}}  \alpha^{-1}_{m_1,n_1}\bkt{\tilde s_{m_1}}}\right.\non\\
&\left. \cbkt{  \frac{1}{\bkt{\frac{\delta\gamma_{p_2p_2} \bkt{\tilde s_{p_2}}}{\delta \tilde s}}}\beta^{-1}_{1,p_2}\bkt{\tilde s_{p_2}}  \alpha^{-1}_{p_2,q_2}\bkt{\tilde s_{p_2}} } \cbkt{ \frac{1}{\bkt{\frac{\delta\gamma_{m_2m_2} \bkt{\tilde s_{m_2}}}{\delta \tilde s}}}\beta^{-1}_{1,{m_2}}\bkt{\tilde s_{m_2}}  \alpha^{-1}_{m_2,n_2}\bkt{\tilde s_{m_2}} }}. 
}
The above expressions for $ \avg{n_{A,\Phi}^{(1)}\bkt{\tilde t}} $ and $ \avg{n_{A,Q}^{(1)}\bkt{\tilde t}} $ can be simplified by substituting the linear solution Eq.~\eqref{eq:Ys}  to obtain the fourth order correlation function $\avg{ : \hat{Y}_{q_1}\bkt{\tilde s_{p_1}} \hat{Y}_{n_1}\bkt{\tilde s_{m_1}} \hat{Y}_{q_2}\bkt{\tilde s_{p_2}}\hat{Y}_{n_2}\bkt{\tilde s_{m_2} }: }$.

\end{widetext}

\bibliography{draft.bib}

\end{document}